\documentclass[a4paper,fleqn,usenatbib,useAMS]{mnras}

\usepackage{graphicx}	
\usepackage{amsmath}	
\usepackage{amssymb}	

\usepackage[T1]{fontenc}
\usepackage{ae,aecompl}

\usepackage{newtxtext,newtxmath}

\newcommand*\hess{H.E.S.S.}
\newcommand*\fermi{\textit{Fermi}}
\newcommand*\ctools{\texttt{ctools}}

\title[Prospects for the detection of HE \textit{Fermi} pulsars with CTA]{Prospects for the detection of high-energy ($E$$>$25 GeV) \textit{Fermi} pulsars with the Cherenkov Telescope Array}

\author[Burtovoi et al.]{A. Burtovoi$^{1,2}$\thanks{E-mail: \href{mailto:aleksandr.burtovoi90@gmail.com}{aleksandr.burtovoi90@gmail.com}}, T. Y. Saito$^{3,4}$, L. Zampieri$^{2}$\thanks{E-mail: \href{mailto:luca.zampieri@oapd.inaf.it}{luca.zampieri@oapd.inaf.it}} and T. Hassan$^{5}$
\\
$^{1}$Centre of Studies and Activities for Space (CISAS) ``G. Colombo'', University of Padova, Via Venezia 15, I-35131 Padova, Italy\\
$^{2}$INAF - Astronomical Observatory of Padova, vicolo dell' Osservatorio 5, I-35122 Padova, Italy\\
$^{3}$Division of Physics and Astronomy, Kyoto University, 606-8502 Kyoto, Japan\\
$^{4}$The Hakubi Center for Advanced Research, Kyoto University, 606-8501 Kyoto, Japan\\
$^{5}$Institut de Fisica d'Altes Energies (IFAE), The Barcelona Institute of Science and Technology, Campus UAB, 08193 Bellaterra (Barcelona) Spain
}

\pubyear{2016}

\begin{document}
\label{firstpage}
\pagerange{\pageref{firstpage}--\pageref{lastpage}}
\maketitle

\begin{abstract} 
Around 160 gamma-ray pulsars were discovered by the \fermi~Large Area Telescope (LAT) since 2008. The most energetic of them, 12 objects with emission above 25 GeV, are suitable candidates for the detection with the current and future Imaging Atmospheric Cherenkov Telescopes above few tens of GeV. We perform an analysis of the \fermi-LAT data of these high-energy pulsars in order to determine if such objects can be detected with the Cherenkov Telescope Array (CTA). Our goal is to forecast the significance of their point source detection with CTA. We analyze 5 years of the \fermi-LAT data fitting the spectra of each pulsar at energies $E$$>$10 GeV with a power-law function. Assuming no spectral cut-off, we extrapolate the resulting spectra to the very high energy range (VHE, $E$$>$0.1 TeV) and simulate CTA observations of all 12 pulsars with the \ctools~software package. Using different analysis tools, individual CTA sensitivity curves are independently calculated for each pulsar and cross-checked with the \ctools~results. Our simulations result in significant CTA detections of up to 8 pulsars in 50 h. Observations of the most energetic \fermi~pulsars with CTA will shed light on the nature of the high-energy emission of pulsars, clarifying whether the VHE emission detected in the Crab pulsar spectrum is present also in other gamma-ray pulsars.
\end{abstract}


\begingroup
\let\clearpage\relax
\endgroup
\newpage

\section{Introduction}\label{sec:1}
Pulsars are energetic neutron stars, which produce radiation in a broad energy range: from radio up to extremely energetic gamma-rays (see e.g. \citealt{Becker1997,Manchester2005,Mignani2011,Abdo2013, Ansoldi2016}). These compact objects have strong magnetic fields of about $B=10^9-10^{15}$ G for young isolated pulsars and $B<10^9$ G for millisecond pulsars. The rotational energy of pulsars is lost mainly in the form of a Poynting flux. However, a small fraction of this energy is converted into kinetic energy of magnetospheric electrons and positrons. These particles can be accelerated inside or outside the magnetosphere (see e.g. \citealt{Arons1983,Muslimov2003} for the slot gap model; \citealt{Cheng1986_1,Romani1996} for the outer gap model; \citealt{Coroniti1990} for the striped wind model) producing then gamma-ray emission.

The light curves (or phaseograms) of pulsars in gamma-rays have on- and off- peak intervals and, in some cases, a bridge interval in phase. The on-peak phase interval can be characterized by one (or a few) prominent and sharp emission peak(s) (also called ``on-pulse emission''). The off-peak (OP) emission is usually dominated by the background from the surrounding nebula and/or the pulsar itself. The average level of the bridge emission (if present) between two peaks is higher than that of the off-peak phase interval. One of the most recent investigations of the gamma-ray pulsars light curves was presented by \citet{Pierbattista2015}. Gamma-ray observations showed that the peaks of the pulsars light curves become sharper with increasing energy (see e.g \citealt{Abdo2010,Aliu2011,Aleksic2012,Ackermann2013,Aleksic2014_2,Ansoldi2016}). 

About 160 gamma-ray pulsars \citep{Acero2015} were detected above 100 MeV with the Large Area Telescope (LAT, \citealt{Atwood2009}) aboard the \textit{Fermi Gamma-ray Space Telescope}. It was found that 12 of them, associated with known sources, show significant pulsations at energies $E$$>$25 GeV \citep{Ackermann2013}. They are: PSRs J0007$+$7303, J0534$+$2201 (Crab), J0614$-$3329, J0633$+$1746 (Geminga), J0835$-$4510 (Vela), J1028$-$5819, J1048$-$5832, J1413$-$6205, J1809$-$2332, J1836$+$5925, J2021$+$3651, J2229$+$6114. These pulsars (HE \fermi~pulsars; see Sect. \ref{subs:1.1}) have been searched with the currently operating Imaging Atmospheric Cherenkov Telescopes (IACTs, see e.g. \citealt{Weekes2005arx}), such as the Major Atmospheric Gamma-Ray Imaging Cherenkov telescope (MAGIC,  $E$$>$50 GeV, \citealt{Aleksic2016}), the Very Energetic Radiation Imaging Telescope Array System (VERITAS, $E$$>$85 GeV, \citealt{Park2015arx}) and the High Energy Stereoscopic System (\hess, $E$$>$20 GeV, \citealt{Hinton2004,Hofverberg2013arx}). However, so far only two pulsars -- the Crab pulsar and the Vela pulsar -- were detected with IACTs at very high energies (VHE, $E$$>$100 GeV, for details see Sect. \ref{subs:1.2}).

Considering their very energetic pulsed emission, these 12 HE \fermi~pulsars are suitable targets for IACT investigations above few tens of GeV among the whole population of gamma-ray pulsars.

\subsection{\fermi-LAT investigations of HE pulsars}\label{subs:1.1}
HE \fermi~pulsars are included in different \fermi-LAT catalogs such as, e.g. the Third source catalog (3FGL, \citealt{Acero2015}), the Second catalog of gamma-ray pulsars (2PC, \citealt{Abdo2013}), the First high-energy catalog (1FHL, \citealt{Ackermann2013}) and the Third high-energy catalog (3FHL, \citealt{3FHLarx}). PSR J0835$-$4510 is also included in the Second high-energy catalog (2FHL, \citealt{Ackermann2016}).

The spectral properties of their gamma-ray emission were investigated in a number of previous studies. Phase-resolved spectral analyses were performed for PSRs J0007$+$7303 \citep{Abdo2012,Li2016}, J0534$+$2201 \citep{Abdo2010,Buehler2012}, J0633$+$1746 \citep{Abdo2010_5}, J0835$-$4510 \citet{Abdo2010_3} and J1836$+$5925 \citep{Abdo2010_4}. Phase-averaged spectra were obtained in J0614$-$3329 \citep{Ransom2011}, J0835$-$4510 (for $E$$>$50 GeV, \citealt{Leung2014}), J1028$-$5819 \citep{Abdo2009}, J1048$-$5832 \citep{Abdo2009_1} and J2229$+$6114 \citep{Abdo2009_1}, J1413$-$6205 \citep{Saz2010} and J2021$+$3651 \citep{Abdo2009_2}.

When studying the high-energy pulsed emission from pulsars it is important to account for the OP emission. It can originate in the magnetosphere of the pulsar (with a spectral cut-off) and/or in the surrounding pulsar wind nebula (PWN, without a spectral cut-off, \citealt{Abdo2013}). The OP emission in the Crab pulsar is dominated by the surrounding PWN (Crab nebula). In the \fermi-LAT energy range it can be modeled as a sum of two power laws \citep{Abdo2010} or as a sum of a power-law and a smoothly broken power-law component \citep{Buehler2012}. The OP emission in the Vela pulsar is dominated by the extended PWN associated with the Vela X source. \citet{Abdo2010_2} fitted the \fermi-LAT data of this PWN with a power-law model. A more detailed analysis of the Vela X region \citep{Grondin2013} shows that a broken power law describes its gamma-ray spectrum better than a simple power law. In addition, PSRs J0007$+$7303 and J1413$-$6205 show prominent OP point source emission in gamma-rays \citep[see][]{Abdo2012,Li2016,Saz2010}. The spectra of these OP components can be modeled with a power law (for PSR J0007$+$7303) and a power law with an exponential cut-off (for both PSRs J0007$+$7303 and J1413$-$6205). The OP emission in PSRs J0633$+$1746 and J1836$+$5925 most probably is the own pulsar emission \citep{Abdo2010_5,Abdo2010_4}. A spectral analysis of the off-peak emission performed within the framework of 2PC confirmed significant OP emission also in PSRs J0007$+$7303, J0534$+$2201, J0633$+$1746, J0835$-$4510, J1809$-$2332 and J1836$+$5925 \citep{Abdo2013}.

\subsection{IACTs investigations of HE \fermi~pulsars}\label{subs:1.2}
VERITAS observations \citep{Aliu2013} discovered TeV emission from the supernova remnant CTA 1\footnote{The CTA 1 supernova remnant refers to the first radio source in the first Caltech catalog, discovered by \citet{Harris1960}.}, which is situated near PSR J0007$+$7303 ($5^{\prime}$ apart) and associated with TeV emission from the corresponding PWN. Both the Crab pulsar and PWN were detected with IACTs in the VHE range. The Crab nebula was detected already with Whipple \citep{Weekes1989} and HEGRA \citep{Aharonian2004} and then it was investigated with all three currently operating Cherenkov telescopes (MAGIC, \citealt{Aleksic2012_2,Aleksic2015}; VERITAS, \citealt{Aliu2014_3}; \hess, \citealt{Aharonian2006_2}). The discovery of new unexpected VHE gamma-ray emission (up to TeV energies) from the Crab pulsar \citep{Aliu2011,Aleksic2012,Aleksic2014_2,Ansoldi2016} put rather strong constrains on emission models. The existence of VHE photons requires an emission region located in the outer parts of the pulsars magnetosphere or even outside it. Different theoretical interpretations of the sub-TeV/TeV gamma-ray emission from the Crab pulsar are reported in e.g. \citet{Bogovalov2000,Lyutikov2012,Du2012,Bednarek2012,Aharonian2012,Lyutikov2013,Hirotani2015,Mochol2015}. The Vela X region was studied by the \hess~Collaboration in the 0.75--70 TeV energy range \citep{Aharonian2006_3,Abramowski2012}. The detection of pulsations at 20$-$120 GeV with \hess~has been recently reported also for the Vela pulsar\footnote{\url{http://www.mpg.de/8287998/velar-pulsar}}. With the current statistics its spectrum is compatible with a power-law model. The VHE source HESS J1026$-$582, which is coincident with PSR J1028$-$5819, was discovered with \hess~\citep{Abramowski2011_4}. PSR J2021$+$3651 can be one of the possible contributors to the VHE emission from the  bright extended source MGRO J2019$+$37, which was observed with VERITAS \citep{Aliu2014_2}. VHE emission was seen with VERITAS at the position of PSR J2229$+$6114 during an observation of G106.3$+$2.7 \citep{Acciari2009}. The upper limit estimates obtained for PSR J2229$+$6114 with MAGIC are reported in \citet{Reyes2009arx}. No significant pulsed emission from PSR J0633$+$1746 was observed with VERITAS above 100 GeV \citep{Aliu2015} and with  MAGIC above 50 GeV \citep{Ahnen2016}.

\subsection{HE \fermi~pulsars with the Cherenkov Telescope Array}\label{subs:1.3}
More detailed investigations of the point source emission from the 12 HE \fermi~pulsars in VHE gamma-rays will be possible with the future ground-based IACT installations as the Cherenkov Telescope Array (CTA, see e.g. \citet{Ona2013} for the prospects for observations of VHE pulsars and \citet{Burtovoi2016} for the VHE timing analysis of the Crab pulsar). This Cherenkov instrument will be a next-generation facility, consisting of two arrays in the northern and southern hemispheres (CTA-North and CTA-South) with more than one hundred telescopes in total \citep{CTA2011,Acharya2013}. CTA will observe the sky over the energy range from a few tens of GeV to more than 100 TeV. The declared sensitivity, which reaches $\sim3\times10^{-3}$ Crab Unit flux\footnote{1 Crab Unit = $2.79\times10^{-11} \times (E/\text{1 TeV})^{-2.57}$ cm$^{-2}$ s$^{-1}$ TeV$^{-1}$.} at 1 TeV \citep{Bernlohr2013}, is an order of magnitude better than that of the present IACTs (MAGIC, VERITAS, \hess).

In this work our goal is to estimate how many pulsars detected with the \fermi-LAT above 25 GeV would be detectable with CTA above 0.04 TeV and especially at VHE. We perform a power-law fitting of the spectra of 12 HE \fermi~pulsars in order to determine their spectral slope at energies $E$$>$10 GeV. Extrapolating the power-law spectra to VHE, we simulate CTA observations of these objects and estimate the significances of their point source detection with the \ctools~software \citep{Knodlseder2016}. The assumption concerning the power-law spectral behavior of pulsars at energies $>$100 GeV is motivated by the detection of the Crab pulsar at VHE. Therefore, in this respect, our estimates of the CTA detectability of the HE \fermi~pulsars may be considered somewhat optimistic. In addition to the pulsars simulations, we also calculate independently CTA sensitivity curves for each pulsar, comparing them with results of the \ctools~analysis. Our results are compared with similar estimates reported in 1FHL \citep{Ackermann2013} and in \citet{Ona2013}.

The outline of the paper is the following. In Sect. \ref{sec:2} we present the spectral analysis of the \fermi-LAT data. The simulation of CTA observations in the VHE band and the calculation of CTA sensitivities are described in Sect. \ref{sec:3}. The main results and a discussion can be found in Sects. \ref{sec:4} and \ref{sec:5}, respectively. Conclusions are reported in Sect. \ref{sec:6}.

\section{Analysis of the \fermi-LAT data}\label{sec:2}
We perform an analysis of 12 HE \fermi~pulsars using the data acquired during about 5 years of the \fermi-LAT sky survey. For each pulsar direction we extract events with energies from 10 up to 500 GeV, within the time interval shown in Table \ref{tab:psr_pars} and from a region of interest (ROI) of 2$^{\circ}$ centered on the position of the pulsar (Table \ref{tab:psr_pars}). With this ROI both an adequate coverage of the whole \fermi-LAT PSF (0.8$^{\circ}$ at 10 GeV for a 95\% containment radius\footnote{\url{http://www.slac.stanford.edu/exp/glast/groups/canda/lat_Performance.htm}}) and a proper treatment of the background diffuse emission are achieved \citep{Ackermann2013}. The time intervals used here correspond to the epochs within which the ephemerides of the pulsars are valid (see Sect. \ref{subs:2.1}). In our analysis we consider events with zenith angles of less than 90$^{\circ}$, in order to avoid including photons coming from the Earth limb.

Typical gamma-ray spectra of pulsars show a steep decline at $\sim$1-10 GeV (see e.g. \citealt{Abdo2013}). Conventionally it is described with a Power Law Exponential Cut-off (PLEC) model:
\begin{equation}
	F(E) = N_0 \left( \frac{E}{E_0} \right)^{-\gamma}\exp\left[ -\left(\frac{E}{E_c}\right)^b \right],
	\label{eq:3}
\end{equation}
where $N_0$ is the normalization factor, $\gamma$ the spectral index, $E_c$ the cut-off energy, $E_0$ a reference energy and $b$ an additional free parameter.

With the aim of achieving a more accurate characterization of the high-energy part of the pulsars spectra above their typical cut-off energy, in this work we perform a spectral fit of the pulsars spectra above 10 GeV using a power-law model:
\begin{equation}
	F(E) = N_0 \left( \frac{E}{E_0} \right)^{-\gamma}.
	\label{eq:1}
\end{equation}

For the spectral fit we use the binned likelihood analysis tool \texttt{gtlike} from the \fermi~Science Tools software package v10r0p5\footnote{\url{http://fermi.gsfc.nasa.gov/ssc/data/analysis/software/}}. We also accounted for the energy dispersion. The adopted instrument response (IRF) is \texttt{P8R2\_SOURCE\_V6}. We fit the \fermi-LAT data with a model consisting of background emission and non-background point-like and diffuse sources. The background component represents both the Galactic (\texttt{gll\_iem\_v06}) and isotropic diffuse emission (\texttt{iso\_P8R2\_SOURCE\_V6\_v06})\footnote{\url{http://fermi.gsfc.nasa.gov/ssc/data/access/lat/BackgroundModels.html}}. The latter sums up the contribution of the extragalactic and instrumental backgrounds.
We keep the normalizations of both background components as free parameters. 
Finally, in our spectral model we include all point and diffuse sources within a radius of $\sim$3$^\circ$ (from the center of the ROI) that are listed in the 3FGL catalog and have a value of Test Statistic\footnote{Test Statistic is defined as $\mathrm{TS}=-2\ln(L_0/L_1)$, where $L_0$ is the maximum likelihood value for a model without an additional source and $L_1$ is the maximum likelihood value for a model with the additional source at a specified location (from \url{http://fermi.gsfc.nasa.gov/ssc/data/analysis/documentation/Cicerone/Cicerone_Likelihood/Likelihood_overview.html}).} (TS) $\ge$25.

\subsection{Folded light curves}\label{subs:2.1}
We folded the light curves of each pulsar extracted from the ROI (Fig. \ref{fig:ph_interv}) using the \texttt{TEMPO2} software package \citep{Hobbs2006} and the ephemerides\footnote{\url{http://www.slac.stanford.edu/~kerrm/fermi_pulsar_timing/}} reported in \citet{Kerr2015}, which are valid within the epochs listed in Table \ref{tab:psr_pars}. We define the on- and off- peak phase intervals (Table \ref{tab:psr_phi}) using the Knuth's rule algorithm \citep{astroML,astroMLText}, which provides a good determination of the pulsars peaks. Only for the Crab pulsar we use the Bayesian Block algorithm that provides better estimates of the phase intervals for light curves with relatively narrow peaks \citep{astroML,astroMLText}. 10 out of 12 HE \fermi~pulsars show a pulse profile with at least two peaks and bridge emission. The two exceptions are PSRs J0007$+$7303 and J2229$+$6114. These pulsars (Figs. \ref{fig:ph_interv}a, \ref{fig:ph_interv}l) appear to have two closely spaced peaks or only one peak, respectively \citep{Abdo2012,Abdo2009_1}. PSR J0835$-$4510 (Vela pulsar) has a more complex pulse shape with 3 peaks and bridge emission overlapping with each other \citep{Abdo2010_3}. For this pulsar we consider a single on-peak interval comprising all peaks and any possible bridge emission (Fig. \ref{fig:ph_interv}e). It should be noticed that only two events are detected from PSR J1836$+$5925 at energies $E$$>$25 GeV (Fig. \ref{fig:ph_interv}j). As in \citet{Ackermann2013} these events were considered significant\footnote{See \citet[Sect. 4.5]{Ackermann2013} for details.}, we include PSR J1836$+$5925 in the present analysis.

\begin{table}
\caption{Pulsars with emission at energies $E$$>$25 GeV detected with the \fermi-LAT and investigated in this work. Right ascension (RA) and declination (DEC) of each pulsar, taken from 3FGL \citep{Acero2015}, are shown in the second and third columns$^a$. Data are taken from the time intervals in modified Julian days (MJD) listed in the fourth column.}
\label{tab:psr_pars}
\centering
\begin{tabular}{l c c c}
\hline\hline
Pulsar 			& RA (deg)	& DEC (deg)	& Time interval of						\\
				&			&			& ephemerides (MJD)$^{a}$	\\
\hline
J0007$+$7303	& 1.77		& 73.05 		& 54686--56583	\\
J0534$+$2201	& 83.64		& 22.02		& 54686--56583	\\
J0614$-$3329		& 93.54		& -33.50		& 54696--56572	\\
J0633$+$1746	& 98.48		& 17.77		& 54689--56579	\\
J0835$-$4510		& 128.84		& -45.18		& 54686--56583	\\
J1028$-$5819		& 157.12		& -58.32		& 54686--56583	\\
J1048$-$5832		& 162.07		& -58.54		& 54686--56583	\\
J1413$-$6205		& 213.37		& -62.09		& 54689--56579	\\
J1809$-$2332		& 272.46		& -23.54		& 54689--56579	\\
J1836$+$5925	& 279.06		& 59.43		& 54689--56579	\\
J2021$+$3651	& 305.28		& 36.86		& 54686--56583	\\
J2229$+$6114	& 337.27		& 61.24		& 54686--56583	\\
\hline
\multicolumn{4}{p{0.45\textwidth}}{\textbf{Notes.} $^{a}$ephemerides taken from \url{http://www.slac.stanford.edu/~kerrm/fermi_pulsar_timing/} (see \citealt{Kerr2015} for details).}
\end{tabular}
\end{table}

\begin{table*}
\caption{On-/off- peak and bridge phase intervals of our sample of pulsars at energies above 10 GeV. P1 and P2 (if present) correspond to the first and second peaks, respectively (see text for details).}
\label{tab:psr_phi}
\centering
\begin{tabular}{l l l l l l}
\hline\hline
Pulsar 			& \multicolumn{2}{c}{$\phi_{\rm on}$}				& $\Delta\phi_{\rm on}$	& $\phi_{\rm off}$	& $\phi_{\rm bridge}$	\\
				& \multicolumn{1}{c}{P1} &  \multicolumn{1}{c}{P2}	& 					&				&					\\
\hline
J0007$+$7303	& [0.09;0.42]		& $-$			&  0.33	& [0.42;1.09]		& $-$		\\
J0534$+$2201	& [-0.014;0.013] 	& [0.324;0.423]	&  0.126	& [0.423;0.986]	& [0.013;0.324]\\
J0614$-$3329		& [0.07;0.14] 		& [0.63;0.75]		&  0.19	& [0.14;0.63]		& [0.75;1.07]	\\
J0633$+$1746	& [0.05;0.12] 	 	& [0.51;0.68]		&  0.24	& [0.68;1.05]		& [0.12;0.51]	\\
J0835$-$4510		& [0.11;0.62]		& $-$			&  0.51	& [0.62;1.11]		& $-$		\\
J1028$-$5819		& [0.13;0.25]  		& [0.62;0.75]		&  0.25	& [0.75;1.13]		& [0.25;0.62]	\\
J1048$-$5832		& [0.13;0.25] 	 	& [0.5;0.63]		&  0.25	& [0.63;1.13]		& [0.25;0.5]	\\
J1413$-$6205		& [0.11;0.22] 	 	& [0.44;0.55]		&  0.22	& [0.55;1.11]		& [0.22;0.44]	\\
J1809$-$2332		& [0.08;0.14]  		& [0.39;0.45]		&  0.12	& [0.45;1.08]		& [0.14;0.39]	\\
J1836$+$5925	& [0.0;0.23]	  	& [0.50;0.71]		&  0.44	& [0.71;1.0]		& [0.23;0.50]	\\
J2021$+$3651	& [0.08;0.17]	  	& [0.49;0.59]		&  0.19	& [0.59;1.08]		& [0.17;0.49]	\\
J2229$+$6114	& [0.37;0.52]		& $-$			&  0.15	& [0.52;1.37]		& $-$		\\
\hline
\end{tabular}
\end{table*}

\newcommand*\lcw{0.33}
\begin{figure*}
	\includegraphics[width=\lcw\textwidth]{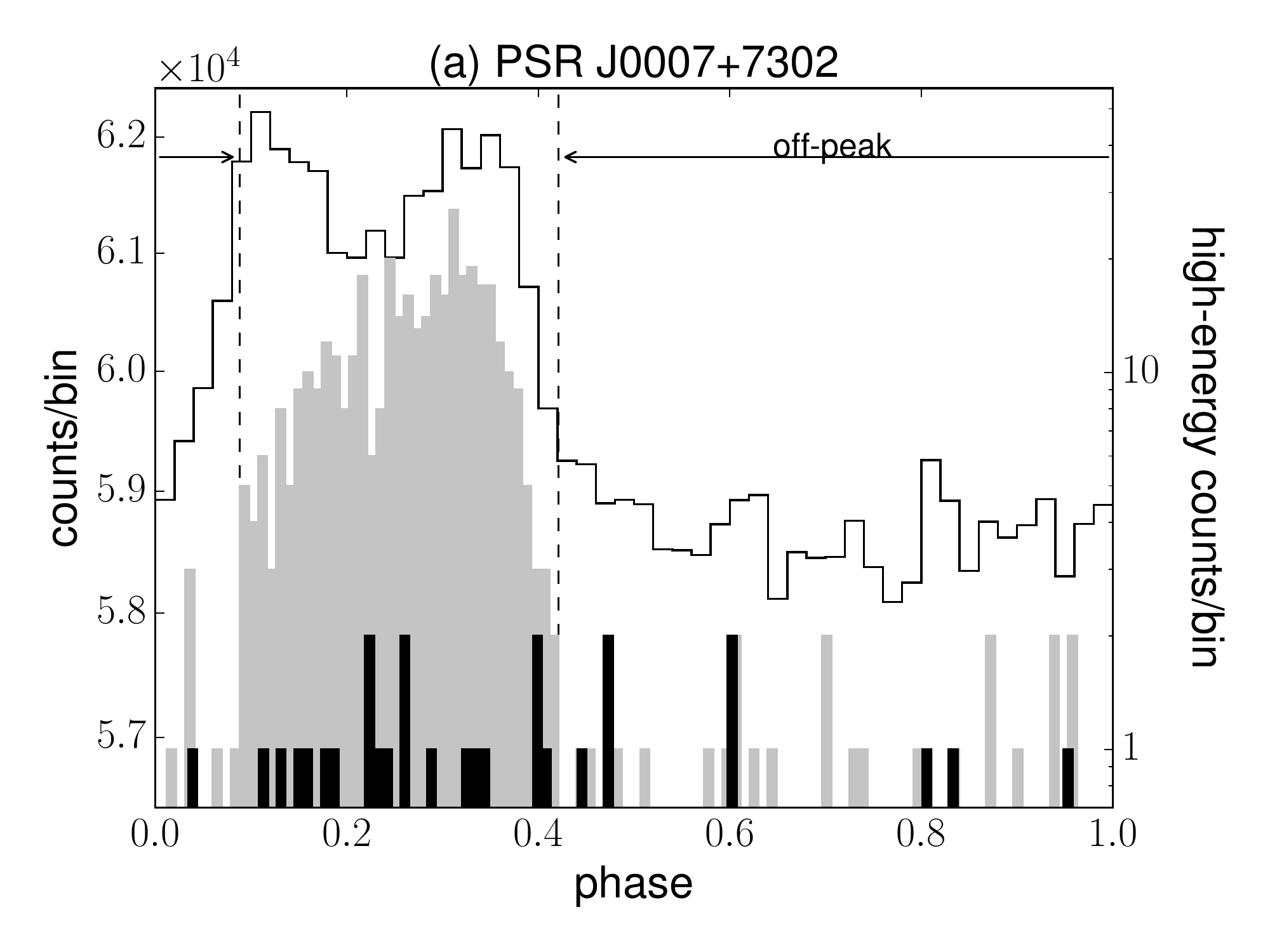}
	\hfill
	\includegraphics[width=\lcw\textwidth]{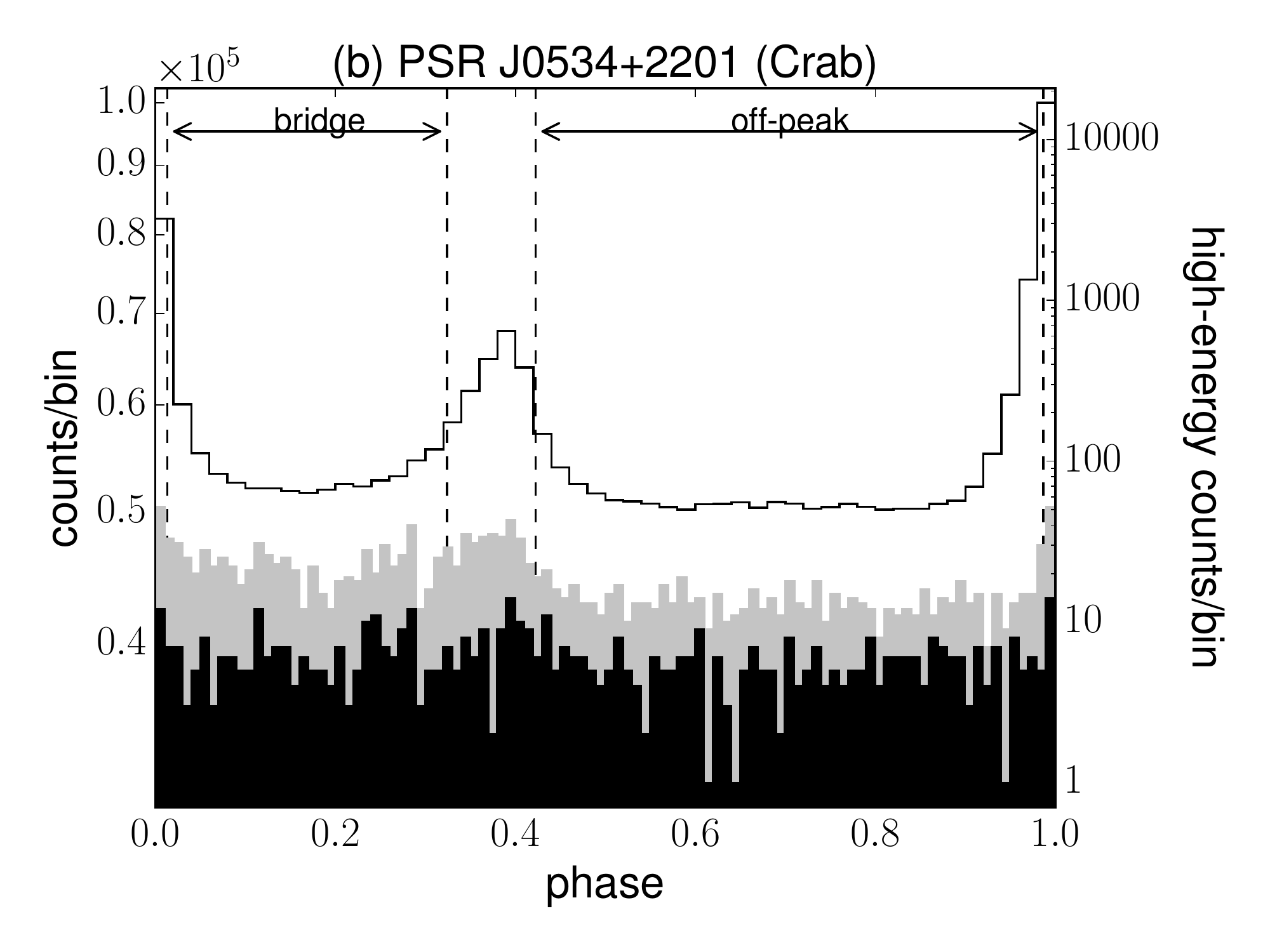}
	\hfill
	\includegraphics[width=\lcw\textwidth]{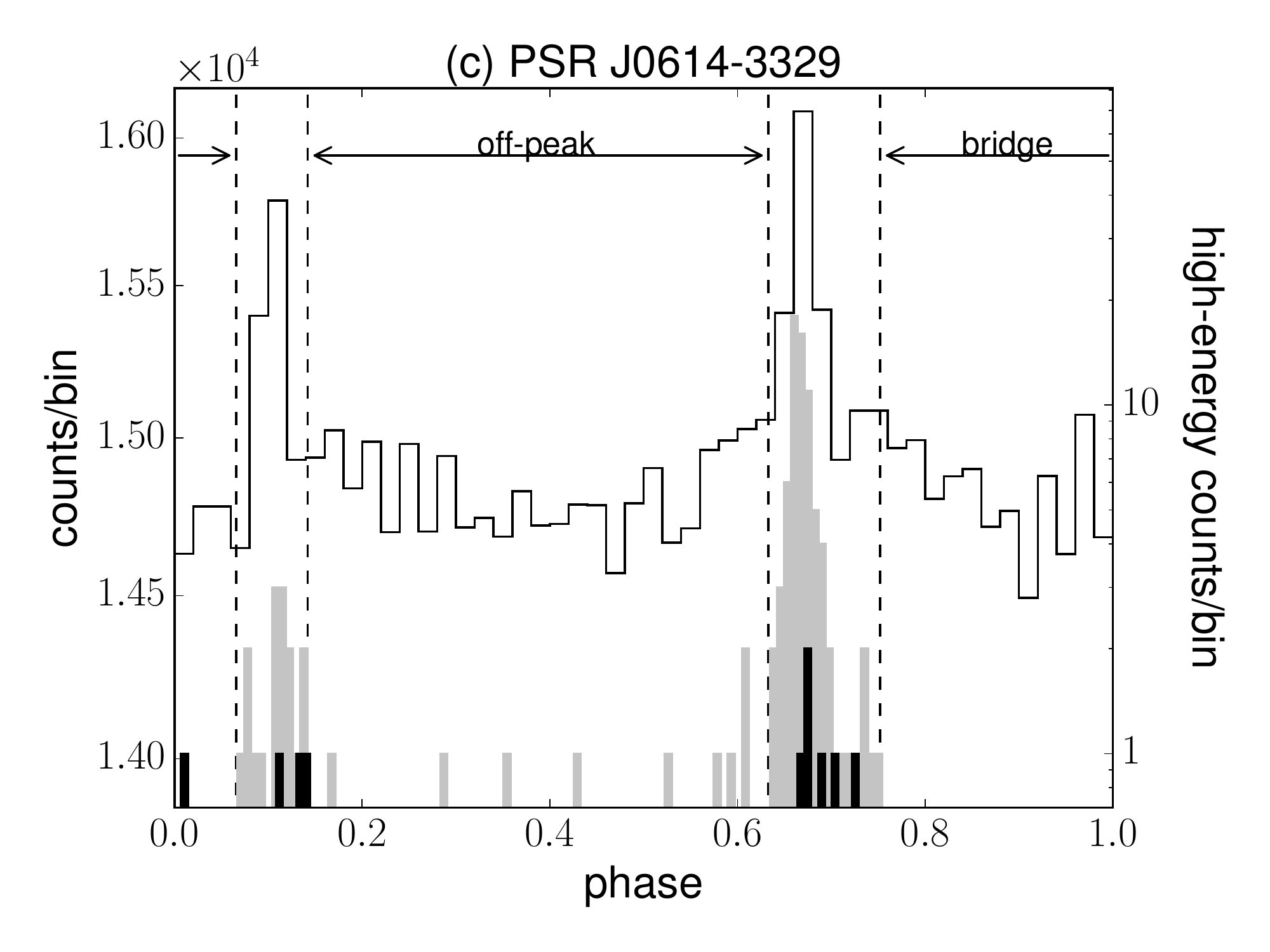}
\\
	\includegraphics[width=\lcw\textwidth]{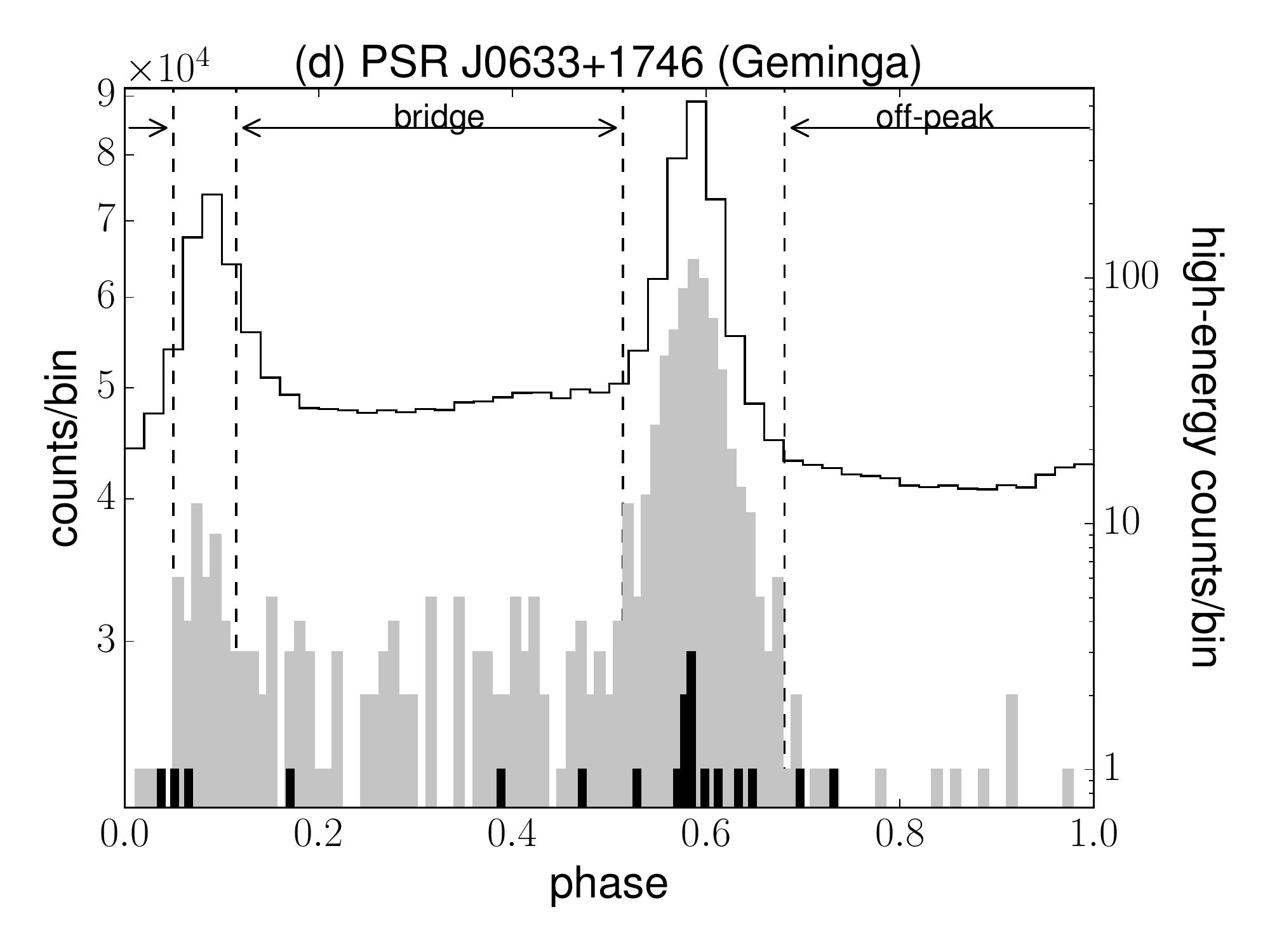} 
	\hfill
	\includegraphics[width=\lcw\textwidth]{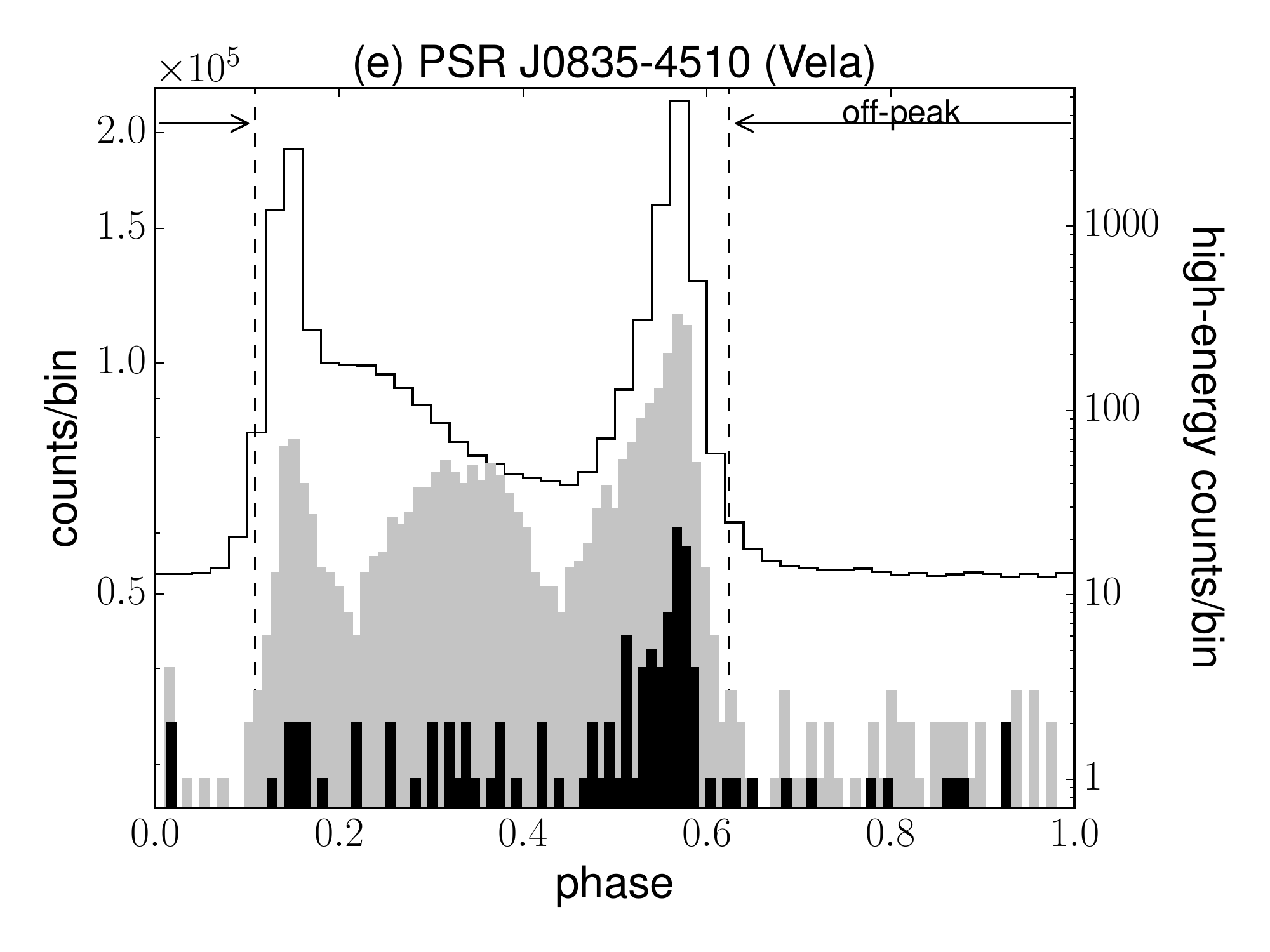}
	\hfill
	\includegraphics[width=\lcw\textwidth]{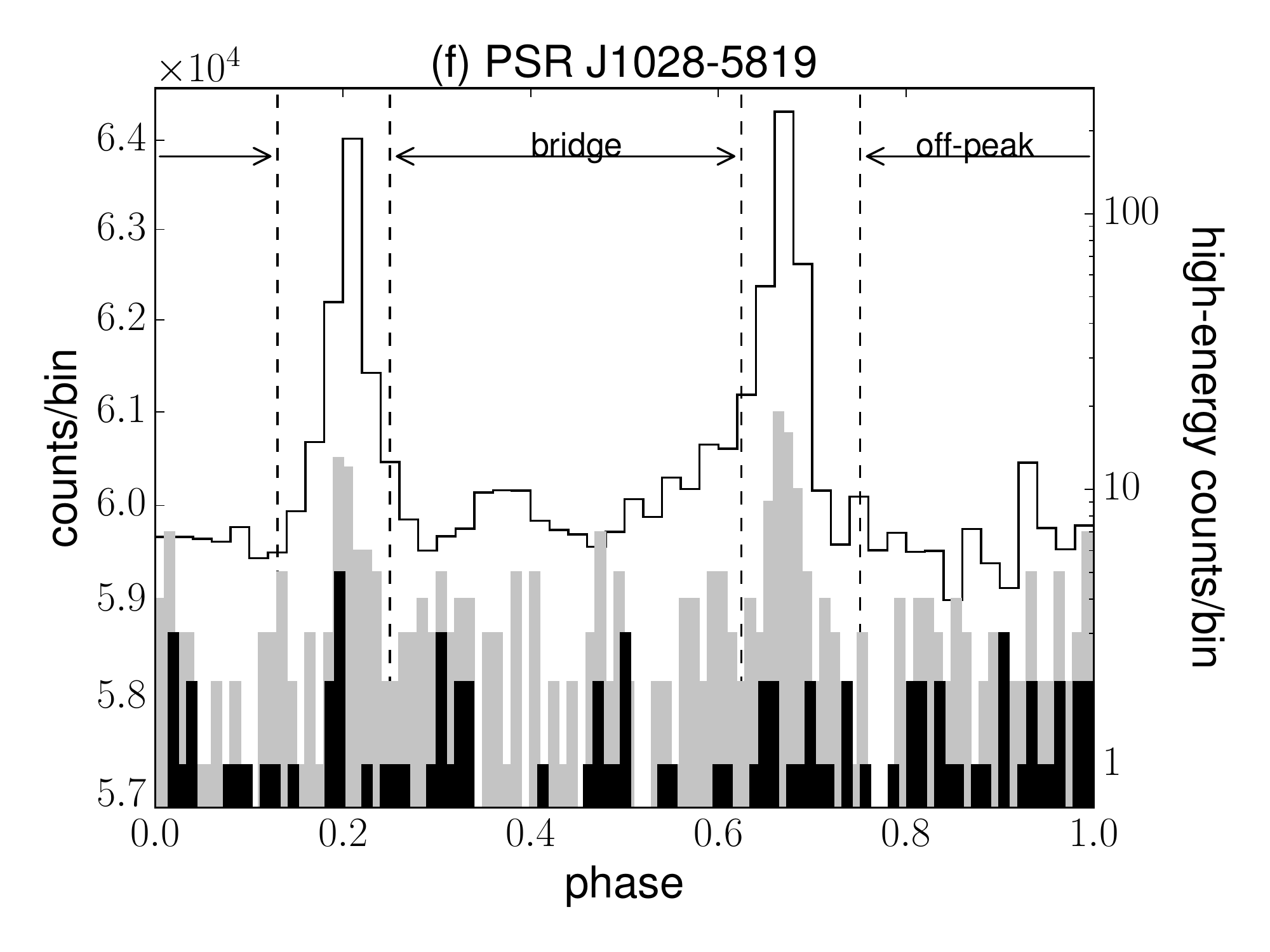}
\\
	\includegraphics[width=\lcw\textwidth]{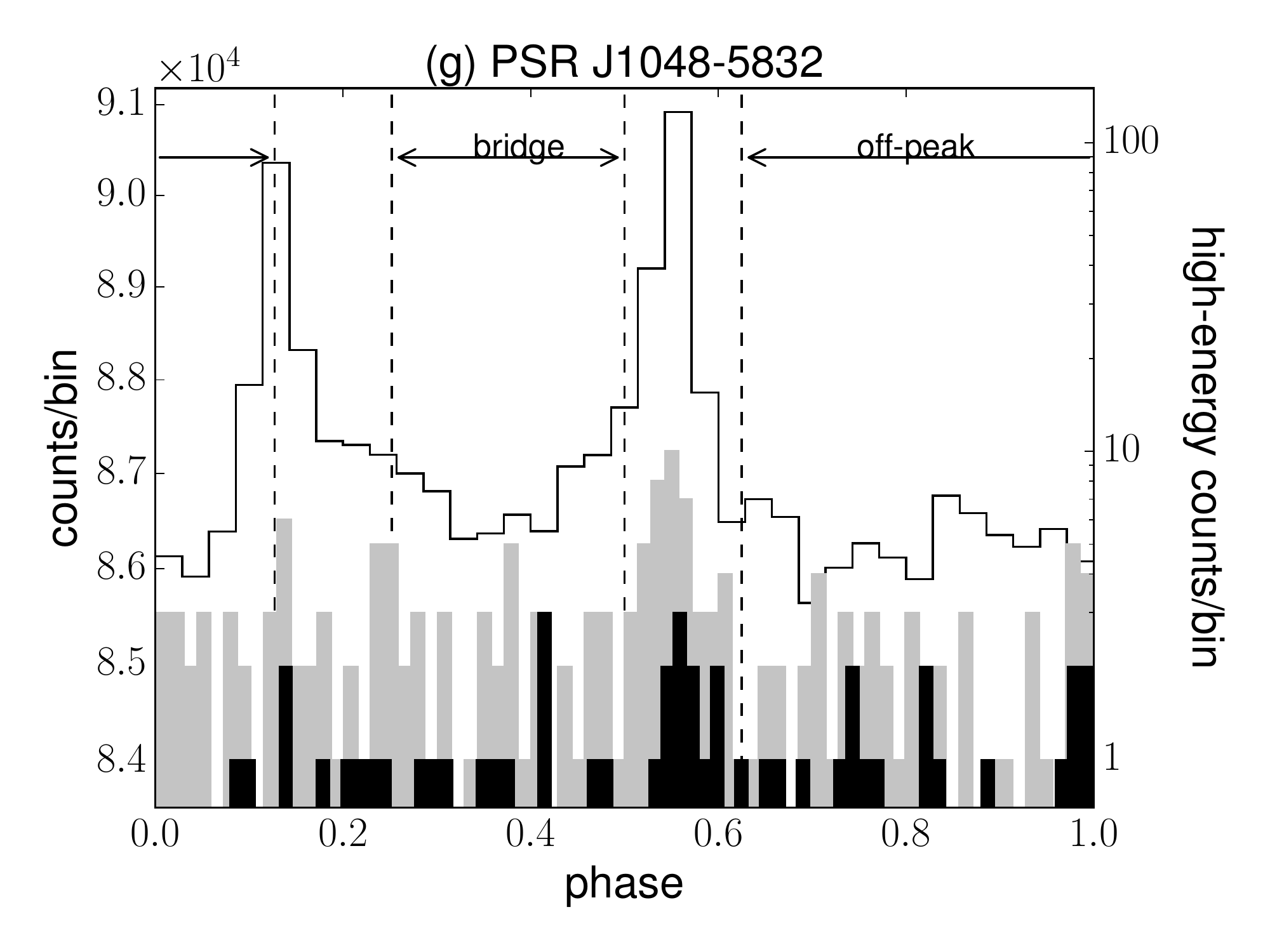}
	\hfill
	\includegraphics[width=\lcw\textwidth]{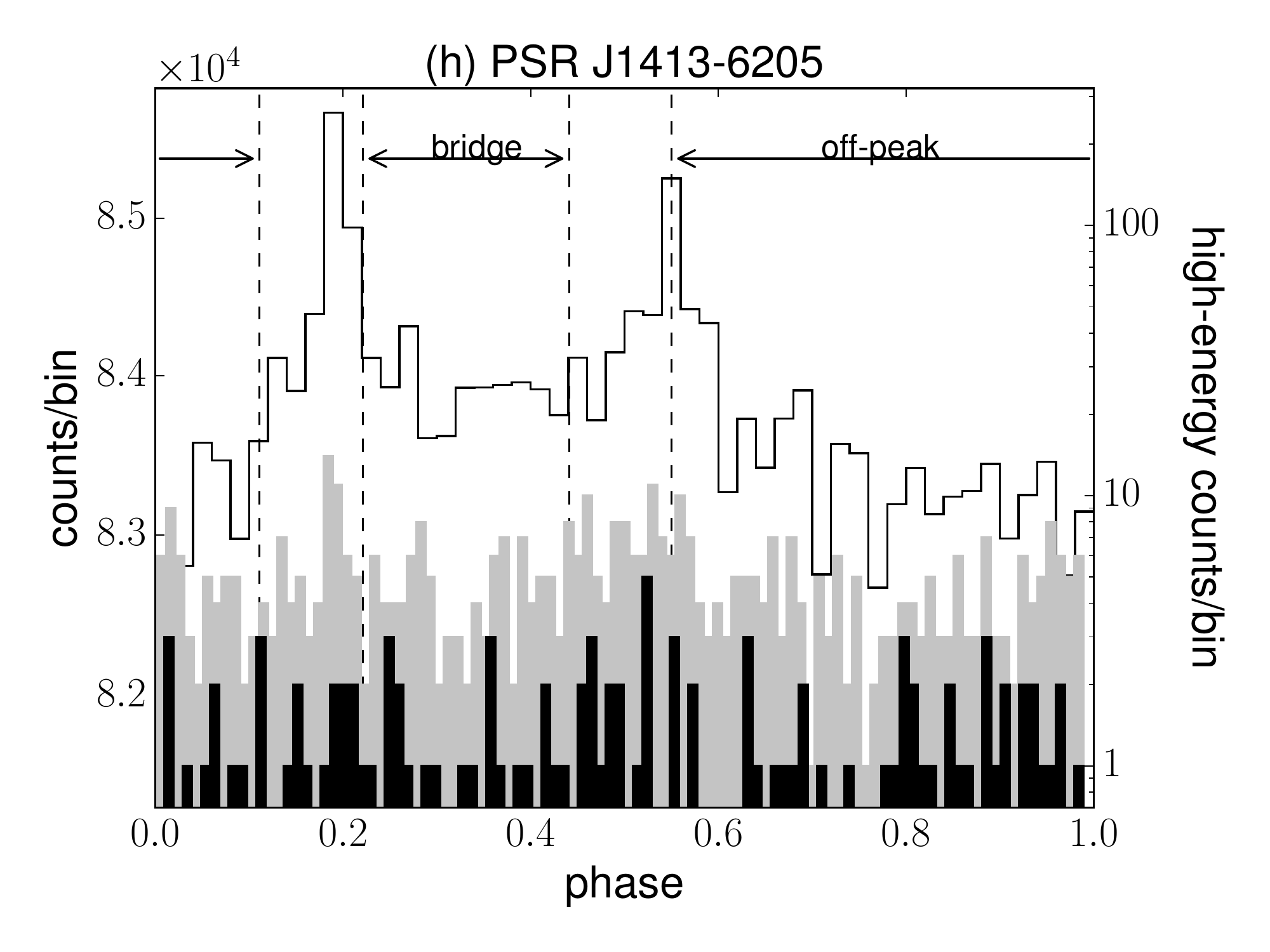} 
	\hfill
	\includegraphics[width=\lcw\textwidth]{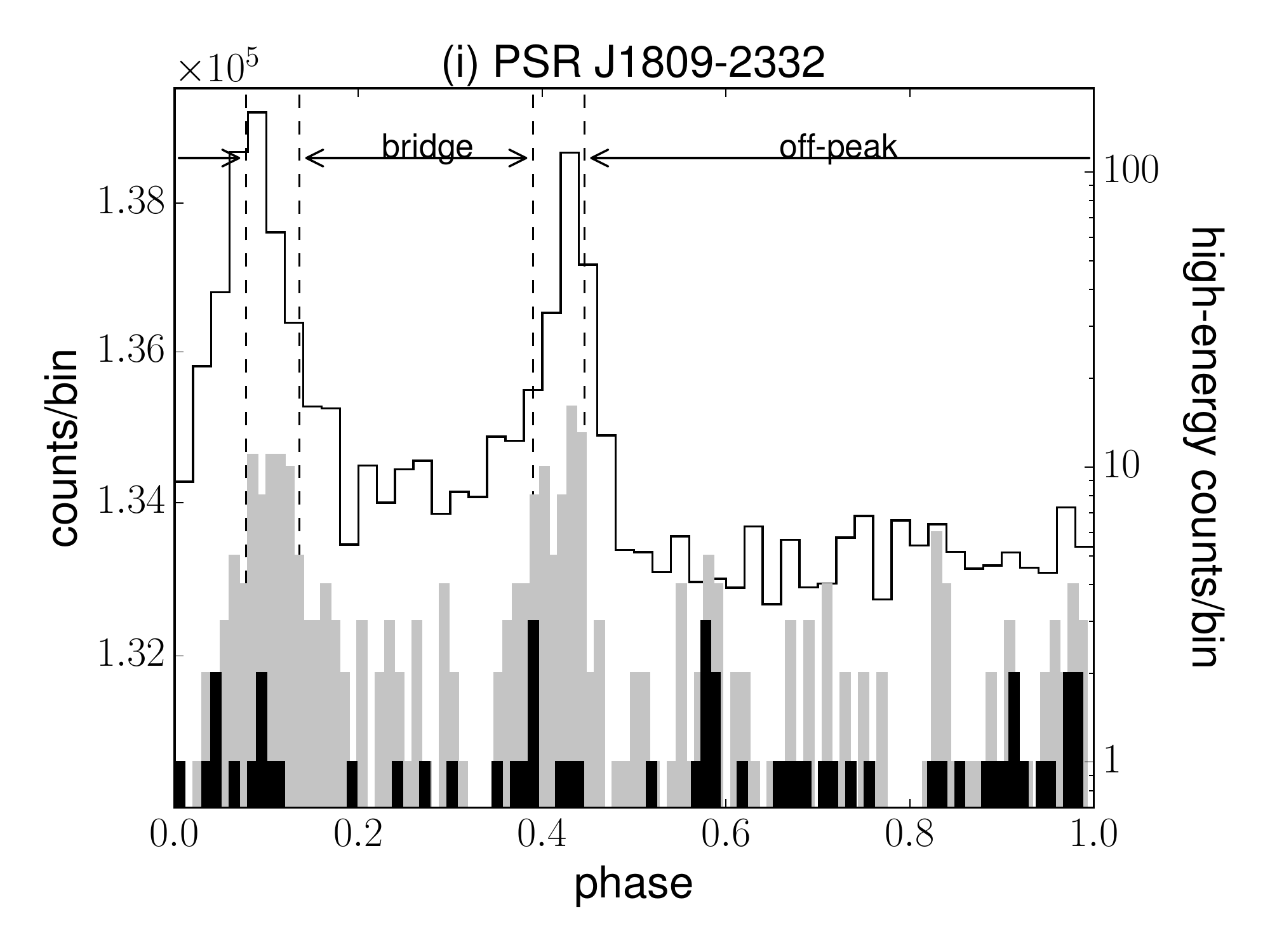}
\\
	\includegraphics[width=\lcw\textwidth]{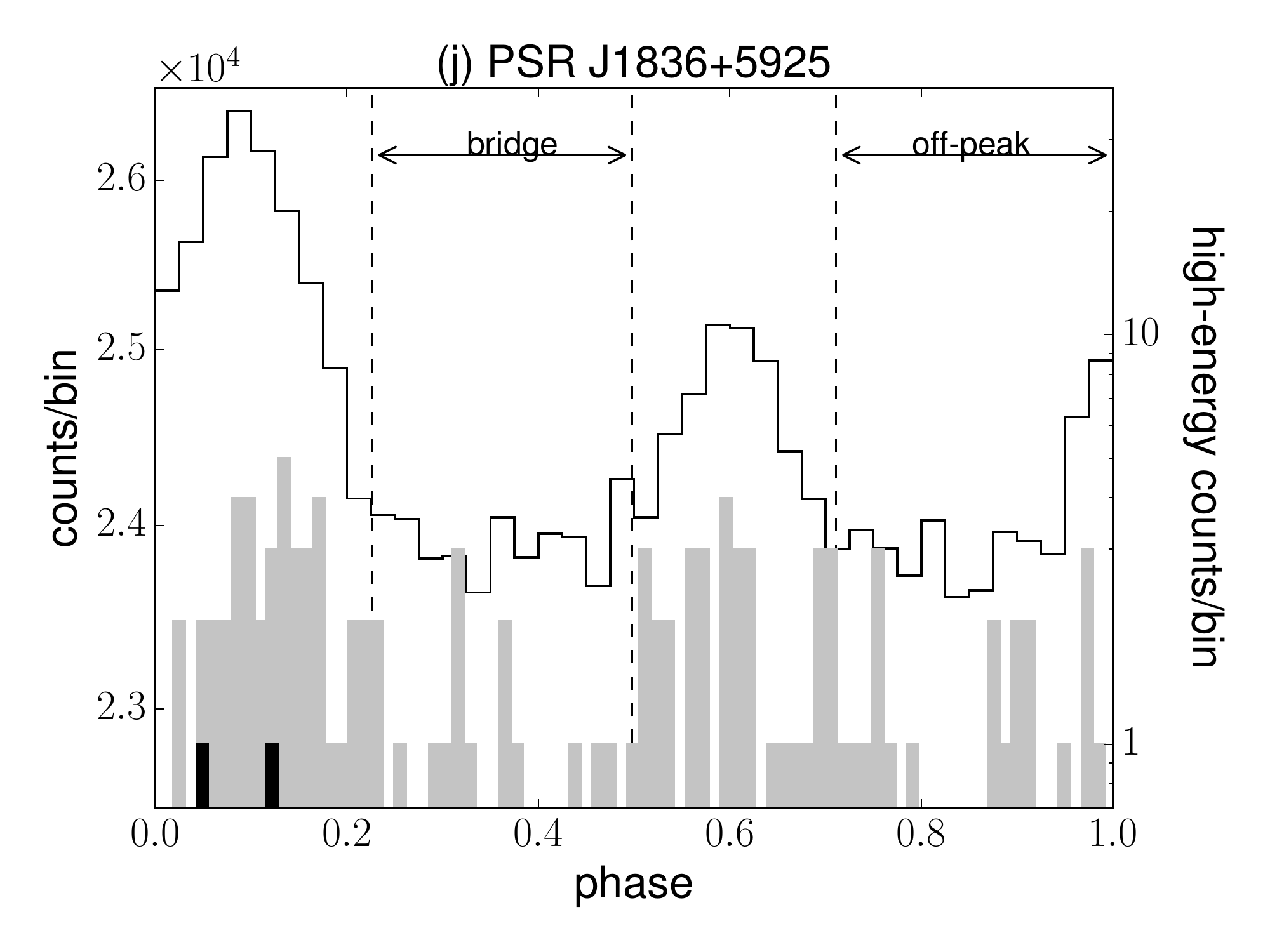}
	\hfill
	\includegraphics[width=\lcw\textwidth]{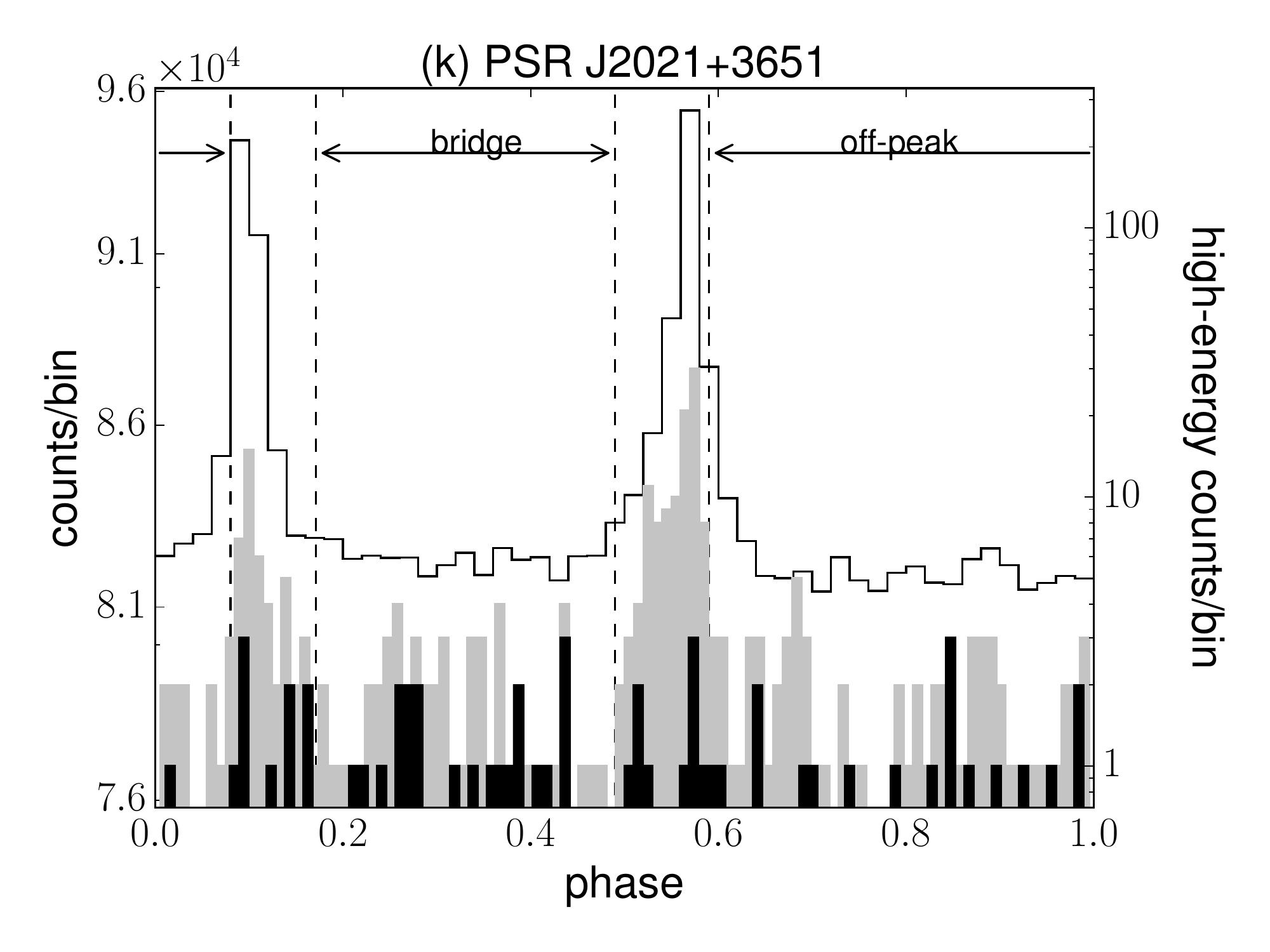}
	\hfill
	\includegraphics[width=\lcw\textwidth]{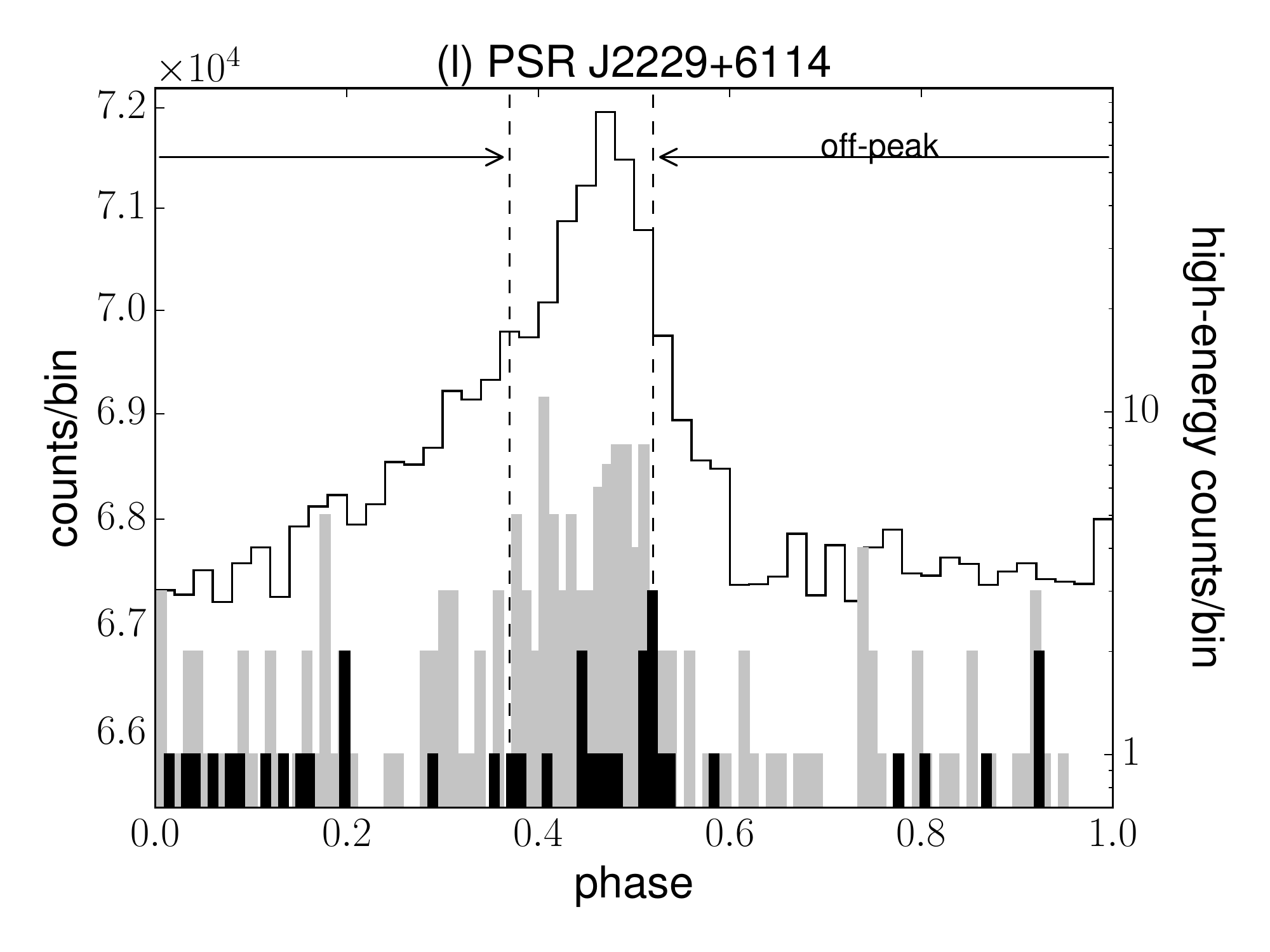}
\\
	\caption{Pulse profiles of 12 HE \fermi~pulsars, obtained folding $\sim$5 years of the \fermi-LAT data. Gray and black histograms ($y$-axis on the right) correspond to events with energies $E$$>$10 GeV and $E$$>$25 GeV, respectively. White histograms ($y$-axis on the left) correspond to energies $E$$>$100 MeV and are shown for comparison. Counts at energies $E$$>$100 MeV ($E$$>$10 GeV/25 GeV) are extracted from a region with a radius of 14$^{\circ}$ (0.8$^{\circ}$, 95\% containment radius). Vertical dashed lines mark the on-peak, off-peak and bridge phase intervals for energies above 10 GeV (see Table \ref{tab:psr_phi}).}
	\label{fig:ph_interv}
\end{figure*}

\subsection{Phase-folded spectral analysis at energies $E$$>$10 GeV}\label{subs:2.2}
We perform separate spectral fits of the off- and on- peak intervals to achieve a better background subtraction. The treatment of the OP emission follows the method described in \citet{Abdo2010_1PC}. Data from the off-peak intervals were fitted with a model in which the pulsar is replaced by a point-like source with OP emission. A power-law model with a free normalization parameter and spectral index was used to describe the high-energy ($E$$>$10 GeV) OP emission in all pulsars.

In case the Galactic background, the extragalactic background and/or the OP component are significant in the off-peak spectral fit (TS$\ge$9), we include them in the on-peak spectral analysis, renormalizing their spectra according to the ratio ($\Delta\phi_{\rm on}/\Delta\phi_{\rm off}$) of the on/off-peak phase intervals (see Table \ref{tab:psr_phi}) and fixing their spectral parameters. In case of non-significant detection of these components in the off-peak spectral fit, the background is still considered in the on-peak analysis with free normalization parameters, while the OP is not included.

After dealing with the background and OP emission as described above, we fitted the on-peak peak emission (without the bridge) from the central pulsar using a power-law model (equation \ref{eq:1}).

Unlike other pulsars, the OP emission of the Vela pulsar is produced by a source which extends beyond the \fermi-LAT PSF ($\sim$0.8$^{\circ}$ at 10 GeV for a 95\% containment radius) -- the Vela PWN. We model this source using a disk spatial template with the parameters reported in 3FGL (uniform disk with radius of 0.88$^\circ$ centered on the coordinates $\alpha = 128.29^\circ$ and $\delta=-45.19^\circ$). To achieve a better accuracy on the determination of the Vela PWN spectrum we perform a phase-averaged spectral analysis, considering all phases and fitting simultaneously the pulsar and PWN emission.

Systematic uncertainties on the spectral parameters of pulsars are estimated using the method described in \citet{Ackermann2012}. The systematics at high energies are dominated by the uncertainty in the effective area of the \fermi-LAT. Following the general caveats on the Pass 8 analysis, we redo the analysis with two different energy-dependent effective areas, which bracket the total uncertainty in the effective area\footnote{We use energy-dependent effective areas, which linearly connect differences of $\pm3\%$, $\pm3\%$, $\pm15\%$ at $\log(E/{\rm MeV}$ of 4, 5, 6, respectively, from the nominal curve (see \url{http://fermi.gsfc.nasa.gov/ssc/data/analysis/LAT_caveats.html} and \url{http://fermi.gsfc.nasa.gov/ssc/data/analysis/scitools/Aeff_Systematics.html\#bracketing}).}. The corresponding spectral fits bracket the uncertainty in the fitted parameters. For the spectral index an energy-dependent function is usually multiplied by the two aforementioned (piece-wise continuous) functions to maximize the effect of the uncertainty on it \citep[see][equation 29]{Ackermann2012}.

We also calculated the spectral energy distribution of HE \fermi~pulsars and corresponding OPs, dividing the energy range from 10 to 500 GeV into eight energy bins (see Figs. \ref{fig:Fermi_Specs} and \ref{fig:PWN_Specs}). For all pulsars, we use a power-law model with free normalization factor and fixed spectral index ($\gamma=2.0$). We also left free to vary normalizations of all other sources inside 3$^\circ$ considered in our analysis, since different slopes and spectral shapes have a small influence on the resulting value of the flux in such narrow energy bins. If TS$_i$ of the $i$-th bin is $<$9, we consider an upper limit to the pulsar flux in that bin, provided by the \fermi~Science Tools.

\section{Simulation of CTA observations}\label{sec:3}
 In this work we use two methods to study the detectability of HE \fermi~pulsars in the VHE range. Sect. \ref{subs:3.1} describes the analysis carried out using \ctools\footnote{\texttt{ctools-1.0.1}, which is built on top of the \texttt{gammalib-1.0.1} package, see \url{http://cta.irap.omp.eu/ctools/}}, a prototype software package developed for the scientific analysis of CTA data \citep{Knodlseder2016}. Sect. \ref{subs:3.2} describes how to use the CTA standard (non-pulsed) IRFs to calculate pulsed CTA sensitivities for each of the simulated pulsars, taking into account both their duty cycle and nebula contribution. 

\subsection{\ctools~analysis}\label{subs:3.1}
Assuming a power-law spectral behavior at VHE, we extrapolated the high-energy ($E$$>$10 GeV) spectral fits of the 12 HE \fermi~pulsars obtained in Sect. \ref{sec:2} to the VHE range and then simulated observations of these sources with CTA using the \ctools~software, tool \texttt{ctobssim}. Because of the different position on the sky, some pulsars, like PSRs J0007$+$7303, J0534$+$2201, J0633$+$1746, J1836$+$5925, J2021$+$3651 and J2229$+$6114, will be better observable from the northern hemisphere, i.e. with CTA-North, while others (PSRs J0614$-$3329, J0835$-$4510, J1028$-$5819, J1048$-$5832, J1413$-$6205 and J1809$-$2332) with the CTA-South installation. We take this into account by using CTA-North and CTA-South configurations\footnote{Public available Monte Carlo Production 2 configurations (files ``CTA-Performance-South-20150511'' and ``CTA-Performance-North-20150511'' taken from \url{https://portal.cta-observatory.org/Pages/CTA-Performance.aspx}).} in our simulations. The CTA-South array will consist of 4 large-size telescopes (LSTs, 24-meter diameter), 24 medium-size telescopes (MSTs, 12-meter dish) and 72 small-size telescopes (SSTs, 4-meter diameter)\footnote{It should be noted, that the CTA-South array can be possibly extended with a few dozens of 10-meter Schwarzchild-Couder dual-mirror Telescopes (SCTs), which are not implemented in current configurations.}, whereas CTA-North -- of 4 LSTs and 15 MSTs \citep{Hassan2015arx}. IRFs of these configurations were derived from detailed Monte Carlo simulations of 50 h observations of a point source with 1 Crab Unit flux, observed at a 20$^{\circ}$ zenith angle and located at the center of the field of view. We assumed that the northern CTA installation will be located in La Palma, Spain (28.76$^\circ$ N, 17.89$^\circ$ W) and the southern installation -- in Paranal, Chile (24.63$^\circ$ S, 70.40$^\circ$ W)\footnote{see \url{https://portal.cta-observatory.org/Pages/Preparatory-Phase.aspx}}.

We note, that observations at zenith angles below 20$^{\circ}$ with CTA-North and CTA-South are possible only for some of the HE pulsars (J0534$+$2201, J0614$-$3329, J0633$+$1746, J0835$-$4510, J1809$-$2332 and J2021$+$3651, see Table \ref{tab:culm}). Other pulsars culminate at zenith angles above 30$^{\circ}$, which will lead to larger energy thresholds (up to $\sim$100 GeV). Comparing CTA sensitivity curves calculated for 20$^\circ$ and 40$^\circ$ zenith angles (see \citealt{Hassan2017arx}), we estimate that the sensitivity of CTA during observations at high zenith angles (40$^{\circ}$) will be $\sim$1.2 times worse than that during observations at 20$^{\circ}$ zenith angle in the energy range 0.1$-$0.2 TeV.

\begin{table}
\caption{Culmination zenith angles (min z.a.) of pulsars observed with different CTA installations. CTA-North is assumed to be located in La Palma, Spain (28.76$^\circ$ N, 17.89$^\circ$ W) and CTA-South -- in Paranal, Chile (24.63$^\circ$ S, 70.40$^\circ$ W).}
\label{tab:culm}
\centering
\begin{tabular}{l l c}
\hline\hline
Pulsar\hspace{1.2cm} & Array\hspace{0.8cm} & min z.a. \\
\hline
\textbf{J0007$+$7303}	& CTA-North 	& 44$^\circ$	\\
J0534$+$2201		& CTA-North 	& 7$^\circ$	\\
J0614$-$3329			& CTA-South 	& 9$^\circ$	\\
J0633$+$1746		& CTA-North 	& 11$^\circ$	\\
J0835$-$4510			& CTA-South 	& 21$^\circ$	\\
\textbf{J1028$-$5819}	& CTA-South 	& 34$^\circ$	\\
\textbf{J1048$-$5832}	& CTA-South 	& 34$^\circ$	\\
\textbf{J1413$-$6205}	& CTA-South 	& 37$^\circ$	\\
J1809$-$2332			& CTA-South 	& 1$^\circ$	\\
\textbf{J1836$+$5925}	& CTA-North 	& 31$^\circ$	\\
J2021$+$3651		& CTA-North 	& 8$^\circ$	\\
\textbf{J2229$+$6114}	& CTA-North 	& 32$^\circ$	\\
\hline
\multicolumn{3}{p{0.35\textwidth}}{\textbf{Notes.} Pulsars which will be observed at high zenith angles (above 30$^\circ$) are in \textbf{bold face}.}
\end{tabular}
\end{table}

Simulations are performed with a ROI of $3^{\circ} \times 3^{\circ}$, centered on the pulsar position. The duration of the observations is 50 h. The input spectral model to simulate consists of the pulsar spectrum obtained from the \fermi-LAT data and extrapolated to VHE, and the isotropic cosmic-ray background of CTA. In case of significant OP detection during the high-energy off-peak \fermi-analysis, we add the corresponding extrapolated power-law component into the VHE model, re-normalized to the total phase interval.

We also simulate the contribution from the Galactic diffuse background (GDB), assuming that its spatial distribution near a pulsar can be approximated as uniform and that its spectrum is an extrapolation of the power-law model obtained from the \fermi-LAT data above 10 GeV. To determine the spectral slope of the GDB we perform an additional spectral fit of the off-peak \fermi-LAT data ($E$$>$10 GeV). For this, we fix all components with the spectral parameters resulting from the high-energy off-peak analysis described  in Sect. \ref{sec:2} and replace the ``standard'' \fermi~Galactic background (\texttt{gll\_iem\_v06} template) with a model consisting of a uniform spatial distribution and a power-law spectrum\footnote{Normalization factor $N_0$ and spectral index $\gamma$ of the GDB are left free to vary.}.

As mentioned above, the OP emission in the Crab and Vela pulsars has been investigated in detail at VHE with the currently operating IACTs. We perform VHE simulations adopting a log-parabola spectral model for the Crab nebula at 0.05$-$30 TeV \citep{Aleksic2015} and a power-law model with an exponential cut-off for the Vela PWN at 0.75-70 TeV \citep{Abramowski2012}. We extrapolated these spectra down to 0.04 TeV (dot-dashed lines in Figs. \ref{fig:PWN_Specs}a and \ref{fig:PWN_Specs}c). In order to simulate the extended Vela X emission region, we use a spatial model which corresponds to a uniform disk with a radius of 1.2$^\circ$, centered at the coordinates $\alpha = 128.75^\circ$ and $\delta=-45.6^\circ$. This new template differs from the one used in our \fermi-analysis. It covers an area on the sky where VHE emission from Vela X has been detected with \hess~\citep{Abramowski2012}.

We carry out analogous simulations with reduced background and reduced OP contamination to mimic the results expected from an analysis of the sole on-peak phase intervals (hereafter we refer to this as the reduced-background analysis). With this it is possible to take into account different duty cycles of the pulsars. We renormalize the background/OP contribution to the high-energy on-peak phase interval ($\Delta\phi_{\rm on}$), without changing the pulsars spectra. 

We fit simulated data and estimate the significance $S$ of pulsars detection in four energy bins ($E$$>$0.04, $E$$>$0.1, $E$$>$0.25 and $E$$>$1 TeV) using the tool \texttt{ctlike}. This tool performs  the unbinned maximum likelihood analysis of data simulated with \texttt{ctobssim}. The square root of the output parameter TS roughly corresponds to the detection significance in Gaussian sigmas. In our analysis, the source is considered to be significantly detected, if $S\equiv\sqrt{TS}\ge5$.

An optimistic/pessimistic value of the significance $S$ for each pulsar is estimated in the following way, propagating the statistical uncertainties on the high-energy power-law spectrum (equation \ref{eq:1}). Using the covariance matrix obtained in the \fermi~likelihood analysis, we can write:
\begin{equation}
	F_{\rm opt,pes} = F(E) \pm \sqrt{\left(\frac{\partial F}{\partial N_0}\right)^2\sigma_{N_0}^2 + \left(\frac{\partial F}{\partial \gamma}\right)^2\sigma_{\gamma}^2 + 2\frac{\partial F}{\partial N_0}\frac{\partial F}{\partial \gamma}\sigma_{N_0\gamma}} \,,
	\label{eq:7}
\end{equation}
where $F_{\rm opt}$/$F_{\rm pes}$ are considered as optimistic/pessimistic approximations of the pulsar spectrum. In equation (\ref{eq:7}) $\partial F/\partial N_0$ and $\partial F/\partial \gamma$ correspond to the partial derivatives of the pulsar spectrum $F$ with respect to $N_0$ and $\gamma$, respectively. $\sigma_{N_0}^2$, $\sigma_{\gamma}^2$ and $\sigma_{N_0\gamma}$ are diagonal and non-diagonal elements of the corresponding covariance matrix, respectively. We repeat VHE simulations for all pulsars with $F_{\rm opt}$ and $F_{\rm pes}$ spectra, obtaining a range of significances. Similar simulations were performed to estimate how systematic uncertainties propagate on $S$.

\subsection{Calculation of CTA sensitivity to gamma-ray pulsars}\label{subs:3.2}
In order to provide a cross-check for the forecast described in the previous section, an independent analysis is performed to calculate the achieved 50 h sensitivity on each of the studied pulsars, directly from CTA IRFs. In order to do so, we need to take into account the main differences between standard and pulsed IACT analyses. To reduce the overwhelming population of cosmic-ray background, two different cuts are applied, optimizing the multiplicity, background rejection parameter (hadronness) and direction ($\theta^2$) of each event:
\begin{enumerate}
	\item Image parameters are used to calculate the likelihood of an event to represent a gamma-ray or cosmic-ray shower. This information is then used to apply a quality cut, selecting only the reconstructed events producing shower images similar to the ones expected from gamma-rays.
	\item Making use of the reconstructed arrival direction of each event, only those which agree with the location of the expected gamma-ray emitting source are selected.
\end{enumerate}

In this analysis we calculate sensitivity curves requiring a significance of 5 sigma in each energy bin. In addition, the number of excess events in each bin is required to be larger than 10 and also larger than 5\% of the background rate (see \citealt{Bernlohr2013,Hassan2015arx}). We note that here the significance is calculated using equation 17 from \citet{LiMa}, whereas in the \ctools~simulations (see Sect. \ref{subs:3.1}) it is assumed to be equal to the square root of the TS parameter.

In the case of pulsars analysis, gamma-ray periodic emission is expected in specific time periods (while cosmic-ray background is uniform in time), allowing additional background suppression by selecting the events with timestamps coherent with each pulsar ephemerides. The efficiency of this additional suppression is proportional to the pulsar duty cycle: the narrower the gamma-ray peak is, the larger percent of background events that will be suppressed.

These considerations are added to CTA sensitivity calculations \citep{Hassan2015arx} by decreasing background rate proportionally to the duty cycle of each pulsar (see the values of on-peak phase intervals $\Delta\phi_{\rm on}$ in Table \ref{tab:psr_phi}). In the case of pulsars located within gamma-ray nebulae, we added another background component in order to account for the PWN contribution to the background. For the Crab and Vela pulsars we used the spectra of the corresponding PWNe from \citet{Aleksic2015} and \citet{Abramowski2012}, respectively. Although the extended TeV emission around other HE \fermi~pulsars has been detected with IACTs (see \citealt{Acciari2009,Abramowski2011_4,Aliu2013,Aliu2014_2}), the CTA (point-source) sensitivity for all pulsars except Crab and Vela is calculated assuming that the background contribution from the surrounding nebula emission is negligible below 1 TeV.

\section{Results}\label{sec:4}
\subsection{High-energy spectral analysis above 10 GeV}\label{sec:4.2}
Results of the spectral analysis (off- and on- intervals) of the \fermi-LAT data at energies $E$$>$10 GeV are listed in Table \ref{tab:res}. In Fig. \ref{fig:Fermi_Specs} the best fitting high-energy spectra of 12 HE \fermi~pulsars are shown. We compared the spectra with those from different Fermi catalogs (3FGL, 2PC and 1FHL). In 2PC and 3FGL power-law models with an exponential (PLEC with $b$ fixed to 1) and sub-exponential (PLEC with $b<1$) cut-off were used to fit the spectra of different pulsars in the full \fermi-LAT energy range (0.1-100 GeV, see Fig. \ref{fig:Fermi_Specs}). The fit of the high-energy emission of the pulsars, reported in 1FHL, was performed with a simple power-law model. The 1FHL spectrum of the Crab pulsar is not shown in Fig. \ref{fig:Fermi_Specs}b since the spectral analysis in 1FHL was performed without removing the Crab nebula, which is dominating above 10 GeV. For comparison, in Fig. \ref{fig:Fermi_Specs}b we show the results of power-law spectral fits of both the \fermi-LAT and MAGIC data of the Crab pulsar, performed by \citet{Ansoldi2016}. We also checked that our spectra are consistent with those from the recently published high-energy \fermi~catalogs 2FHL and 3FHL (not shown in Fig. \ref{fig:Fermi_Specs}).

We estimated the systematic uncertainties on the spectral parameters of pulsars (see Sect. \ref{subs:2.2}) and found that the relative systematic variation of the normalization factor $\delta N_0/N_0$ is about 3\% and the absolute variation of the spectral index $\delta\gamma$ is about 0.05.

Significant OP emission (TS$\gtrsim$25) was detected in PSRs J0534$+$2201, J0633$+$1746, J0835$-$4510, J1809$-$2332 and J1836$+$5925. Their spectral parameters, renormalized to the total phase interval, are shown in Table \ref{tab:res}, together with corresponding TS-values (TS$_{\rm OP}$). For all other pulsars no significant OP component was found above 10 GeV during the off-peak high-energy analysis of the \fermi-LAT data. In Fig. \ref{fig:PWN_Specs} we compare the OP spectra with those from the literature (3FGL, 2PC, 2FHL; see references in Table \ref{tab:res}) and with those detected at VHE (hatched butterfly plots in Figs. \ref{fig:PWN_Specs}a, \ref{fig:PWN_Specs}c)\footnote{In 3FGL the emission of the Crab nebula is fitted with the sum of two power-law models (inverse-Compton and synchrotron components, see Fig. \ref{fig:PWN_Specs}a). Log-parabola and PLEC models were used to fit the MAGIC data of the Crab nebula \citep{Aleksic2015} and the \hess~data of Vela X \citep{Abramowski2012}, respectively.}. The spectrum of Vela X obtained from our analysis is consistent with that presented in 3FHL. In Fig. \ref{fig:PWN_Specs}d we show only the mean value of the 2PC spectrum of the OP emission in PSR J1809$-$2332, because of the relatively high uncertainties in the corresponding spectral parameters. 

In Fig. \ref{fig:PWN_Specs} we compare the OP spectra with publicly available CTA sensitivity curves\footnote{\url{https://portal.cta-observatory.org/Pages/CTA-Performance.aspx}}, calculated considering only the cosmic-ray background and taking the duty cycle equal to 1.

In order to include the GDB component in our simulations, we estimated its emission above 10 GeV around each pulsar assuming a power-law spectral shape and uniform spatial distribution. We detected significant emission from the GDB around all pulsars, except PSRs J0007$+$7303, J0633$+$1746, J1836$+$5925 and J2021+3651 (see Table \ref{tab:GDB_res}). However, the contribution of the GDB to the total VHE background around the pulsars of our sample is not higher than 1\% and, therefore, negligible in our simulations.

\begin{table*}
\caption{Results of the spectral fits at energies $E$$>$10 GeV for our sample of high-energy \fermi~pulsars. Best fit spectral parameters of the off-peak emission and the TS-values of the OP components are reported in the second and third columns, respectively. Normalization factors from the off-peak analysis are renormalized to the total phase interval. Power-law spectra of the pulsars obtained from the on-peak analysis and references to previous spectral analyses are reported in the last two columns.}
\label{tab:res}
\centering
\begin{tabular}{l l l l c}
\hline\hline
Pulsar\hspace{1cm} & off-peak analysis\hspace{0.3cm} & TS$_{\rm OP}$\hspace{0.5cm} & on-peak analysis\hspace{0.4cm} & Refs. \\

\hline
J0007$+$7303	& $-$					& $-$	& $N_0=2.7\pm0.2$		&  $[1]$,$[2]$	\\
				& 						&		& $\gamma=4.1\pm0.2$ 	&			\\

\hline
J0534$+$2201	& $N_0=15.9\pm0.6$		& 8532	& $N_0=3.5\pm0.3$		& $[3]$,$[4]$,$[5]$,	\\ 
(Crab)			& $\gamma=1.98\pm0.04$	&		& $\gamma=3.0\pm 0.2$	&  $[6]$,$[7]$		\\

\hline
J0614$-$3329		& $-$					& $-$	& $N_0=1.0\pm0.1$		&  $[8]$		\\
				& 						&		& $\gamma=3.6 \pm 0.3$	& 			\\ 

\hline
J0633$+$1746	& $N_0= 0.2\pm0.2$		& 22		& $N_0=3.6\pm0.3$		&  $[9]$		\\ 
(Geminga)		& $\gamma= 3.9\pm2.2$	&		& $\gamma=5.3 \pm 0.2$	& 			\\ 

\hline
J0835$-$4510$^{a}$	& $N_0=1.7 \pm 0.3$	      & 75	& $N_0=19.4 \pm 0.7$			&  $[10]$,$[11]$,$[12]$,	\\ 
(Vela) 				& $\gamma=2.2\pm0.2$ &		& $\gamma=4.45 \pm 0.07$		&  $[13]$,$[14]$		\\

\hline
J1028$-$5819		& $-$					& $-$	& $N_0=0.8\pm0.2$		&  $[15]$		\\
				&						&		& $\gamma=3.8\pm 0.4$	& 			\\

\hline
J1048$-$5832		& $-$					& $-$	& $N_0=0.25\pm0.09$		&  $[16]$		\\
				&						&		& $\gamma=4.4 \pm 0.7$	&			\\

\hline
J1413$-$6205		& $-$					& $-$	& $N_0=0.5\pm0.1$		& $[17]$		\\
				& 						&		& $\gamma=3.9 \pm 0.5$	&	 		\\

\hline
J1809$-$2332		& $N_0= 0.3\pm0.1$ 		& 27		& $N_0=0.9\pm0.2$		&  			\\
				& $\gamma= 3.1\pm0.7$ 	&		& $\gamma=4.2 \pm 0.4$	&  			\\

\hline
J1836$+$5925	& $N_0= 0.1\pm0.1$		& 121	& $N_0=0.4\pm0.1$		& $[18]$		\\
				& $\gamma= 7.3\pm1.6$ 	&		& $\gamma=4.6 \pm 0.4$	&  			\\

 \hline
J2021$+$3651	& $-$					& $-$	& $N_0=0.7\pm0.2$		&  $[19]$,$[20]$	\\
				& 						&		& $\gamma=4.7\pm0.5$		& 				\\

\hline
J2229$+$6114	& $-$					& $-$	& $N_0=0.6\pm0.1$		&  $[21]$,$[22]$ 	\\
				& 						&		& $\gamma=3.4 \pm 0.3$	& 				\\

\hline\hline
\multicolumn{5}{p{0.70\textwidth}}{\textbf{Notes.} $N_0$ is the normalization factor in units of $10^{-14}$ cm$^{-2}$ s$^{-1}$ MeV$^{-1}$ and $\gamma$ is the spectral index (see equation \ref{eq:1}). The reference energy $E_0$ used in the analysis is equal to 20 GeV. Errors include only statistical uncertainties. $^a$For the Vela pulsar we perform phase-averaged analysis, fitting the OP and pulsar emission simultaneously. The OP component dominated by extended emission from the Vela PWN was analyzed assuming a disk spatial template (from 3FGL) and a power-law spectral model (see text for details). References: [1] \citet{Abdo2012}; [2] \citet{Aliu2013}; [3] \citet{Abdo2010}; [4] \citet{Buehler2012}; [5] \citet{Aleksic2014_2}; [6] \citet{Aleksic2015}; [7] \citet{Ansoldi2016}; [8] \citet{Ransom2011}; [9] \citet{Abdo2010_5}; [10] \citet{Abdo2010_3}; [11] \citet{Leung2014}; [12] \citet{Abdo2010_2}; [13] \citet{Grondin2013}; [14] \citet{Abramowski2012}; [15] \citet{Abdo2009}; [16] \citet{Abdo2009_1}; [17] \citet{Saz2010}; [18] \citet{Abdo2010_4}; [19] \citet{Abdo2009_2}; [20] \citet{Aliu2014_2}; [21] \citet{Abdo2009_1}; [22] \citet{Acciari2009}.}
\end{tabular}
\end{table*}

\begin{table}

\caption{Results of the power-law fit of the diffuse Galactic background (GDB) emission above 10 GeV for the off-peak phases. Normalization factors $N_{\rm GDB}$ are renormalized to the total phase interval.}
\label{tab:GDB_res}
\centering
\begin{tabular}{l l}
\hline\hline
Pulsar\hspace{2.5cm} & GDB spectrum \\
\hline
J0534$+$2201	& $N_{\rm GDB}=4.3\pm0.8$	\\
				& $\gamma=2.4\pm0.3$		\\
\hline
J0614$-$3329		& $N_{\rm GDB}=0.9\pm0.5$	\\
				& $\gamma=5.0\pm1.0$		\\
\hline
J0835$-$4510		& $N_{\rm GDB}=8.7\pm1.7$	\\ 
				& $\gamma=2.5\pm0.3$		\\
\hline
J1028$-$5819		& $N_{\rm GDB}=6.6\pm2.5$	\\
				& $\gamma=1.9\pm0.3$		\\
\hline
J1048$-$5832		& $N_{\rm GDB}=26.2\pm1.8$\\
				& $\gamma=2.6\pm0.1$		\\
\hline
J1413$-$6205		& $N_{\rm GDB}=83\pm3$	\\
				& $\gamma=2.48\pm0.06$	\\
\hline
J1809$-$2332		& $N_{\rm GDB}=32.3\pm1.9$\\
				& $\gamma=2.47\pm0.09$	\\
\hline
J2229$+$6114	& $N_{\rm GDB}=6.7\pm0.9$	\\
				& $\gamma=2.4\pm0.2$		\\
\hline\hline
\multicolumn{2}{p{0.40\textwidth}}{\textbf{Notes.} $N_{\rm GDB}$ is the normalisation factor of the GDB power-law spectrum in units of $10^{-12}$ cm$^{-2}$ s$^{-1}$ MeV$^{-1}$ sr$^{-1}$ and $\gamma$ is its spectral index (see equation \ref{eq:1}). The reference energy $E_0$ used in the analysis is equal to 20 GeV. Errors include only statistical uncertainties.}
\end{tabular}
\end{table}

\newcommand*\spw{0.33}
\begin{figure*}
	\includegraphics[width=\spw\textwidth]{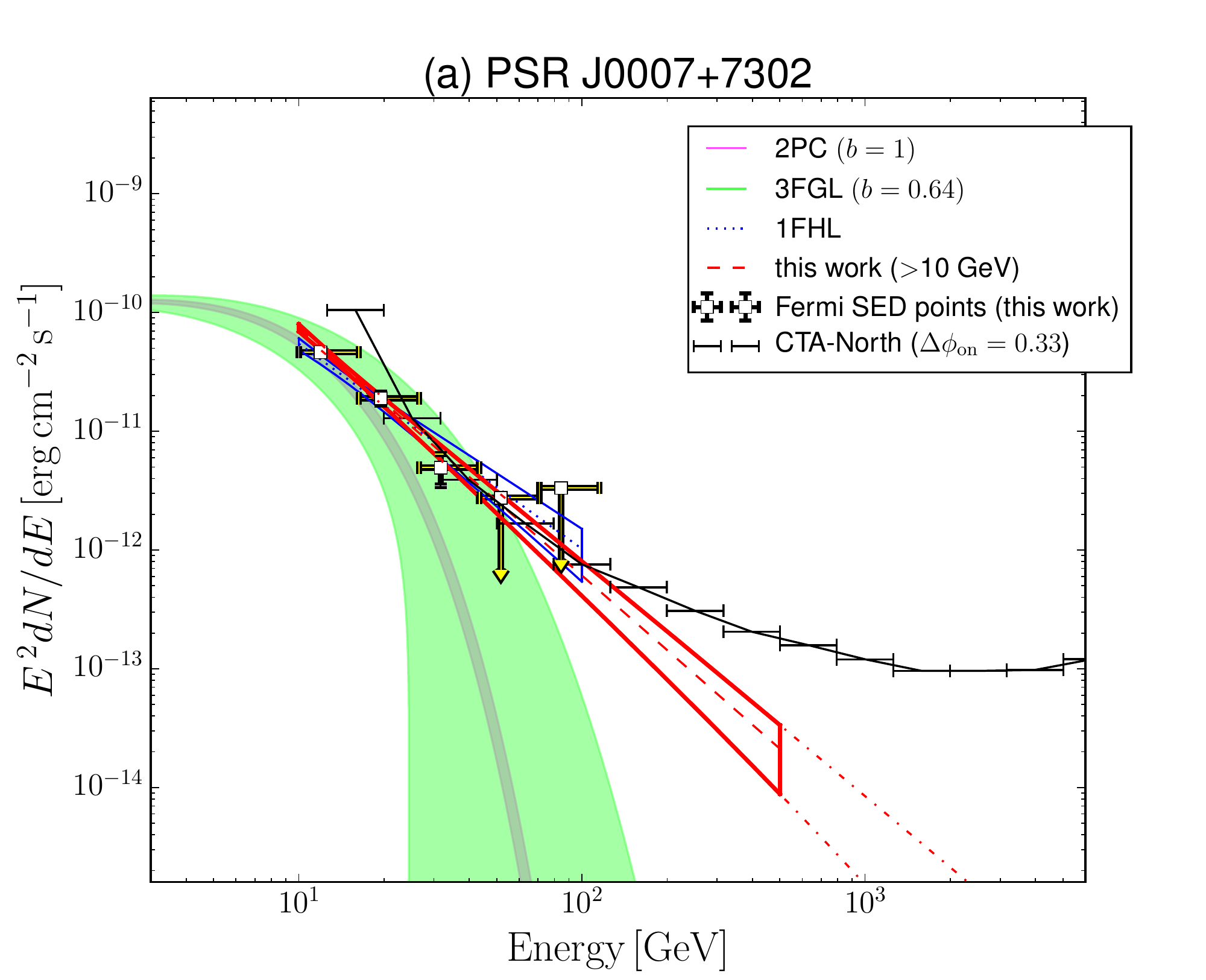}
	\hfill
	\includegraphics[width=\spw\textwidth]{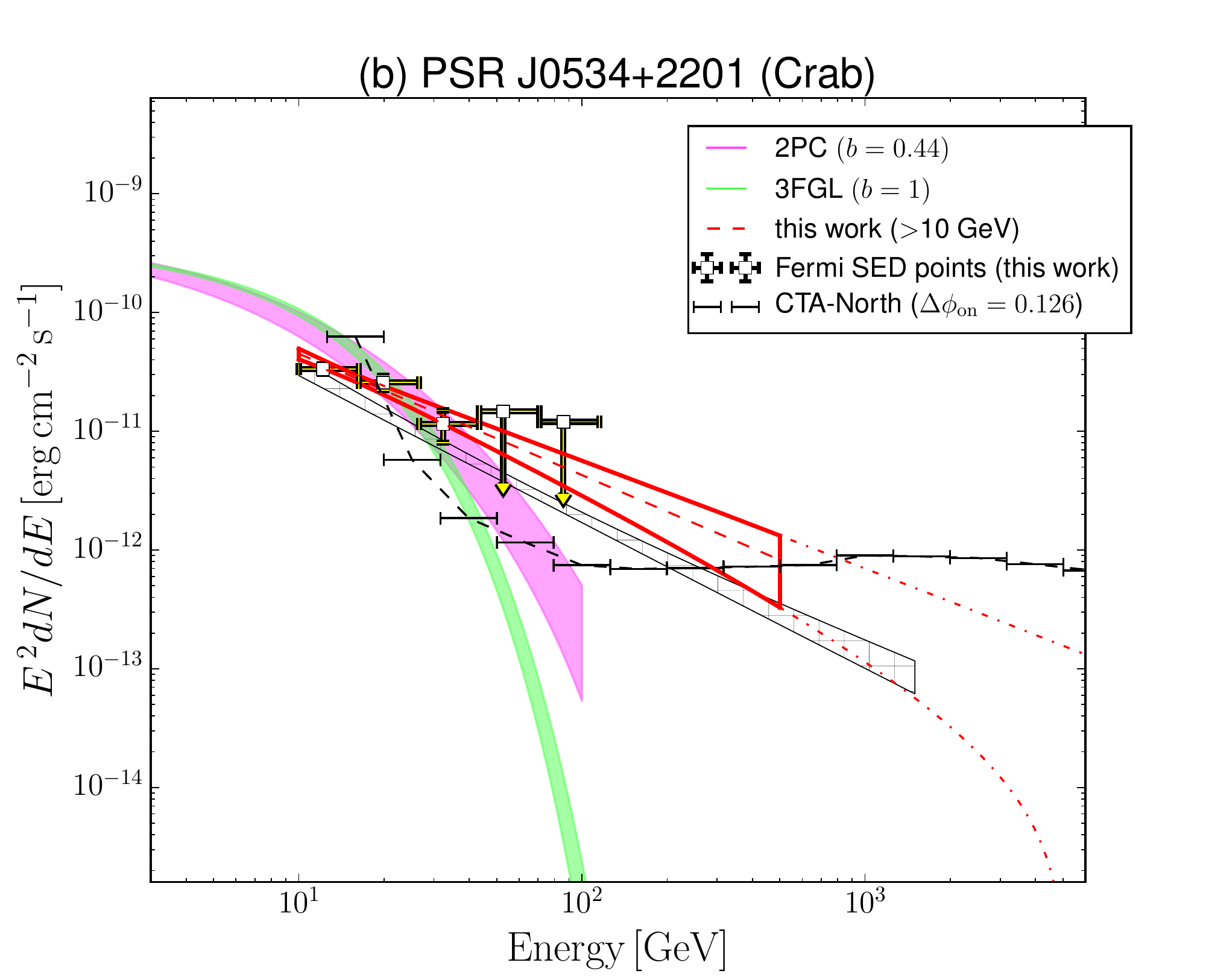}
	\hfill
	\includegraphics[width=\spw\textwidth]{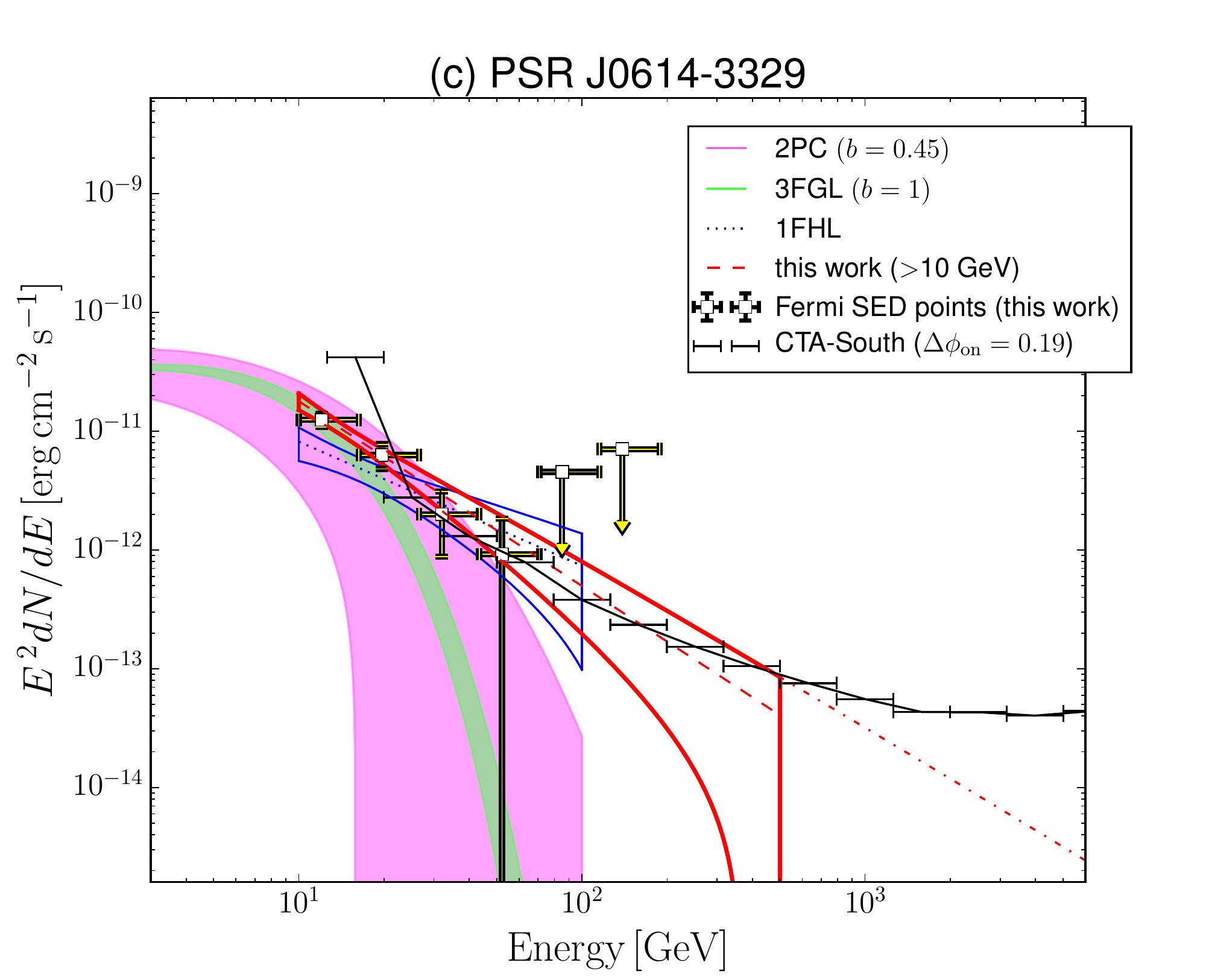}
\\
	\includegraphics[width=\spw\textwidth]{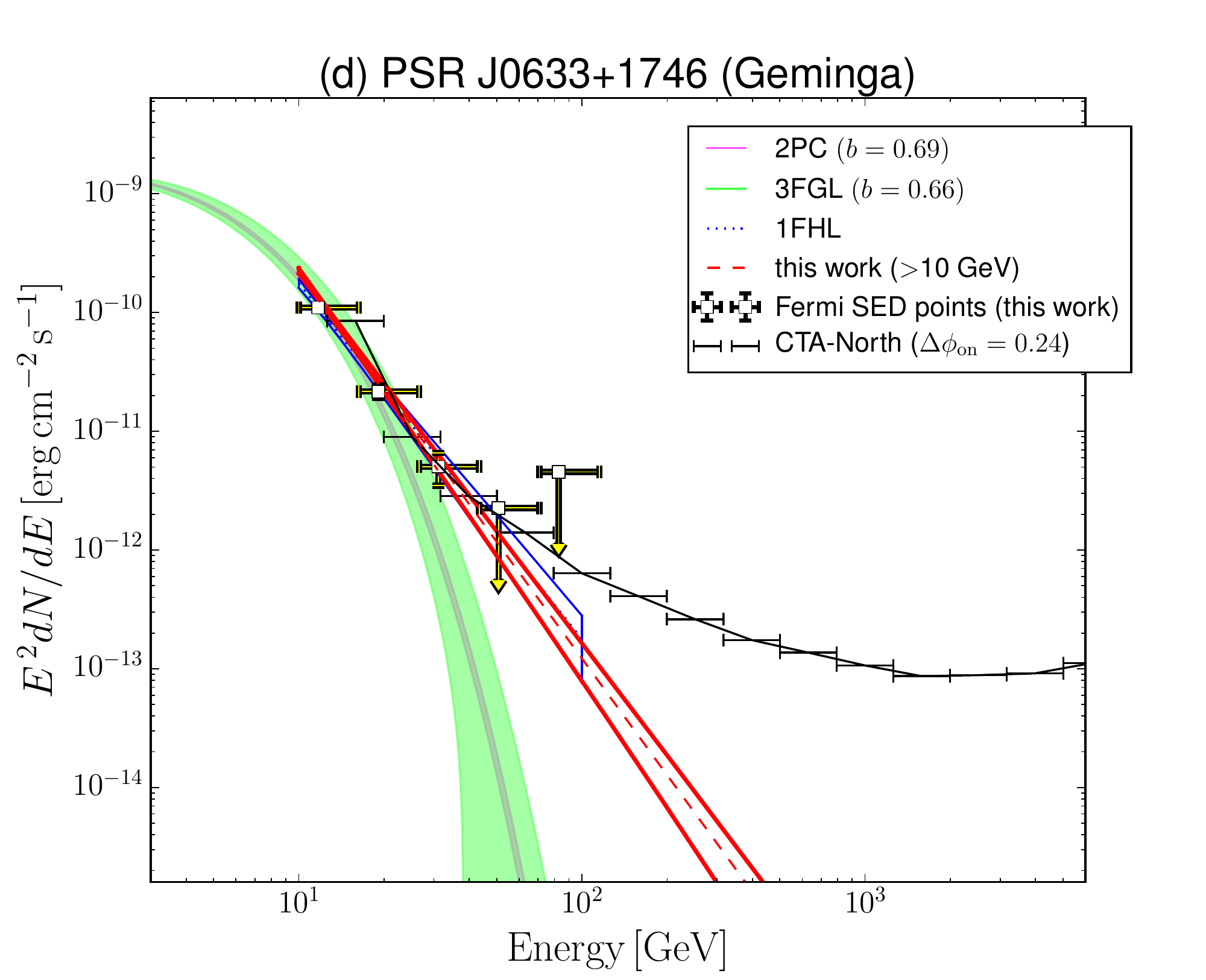}
	\hfill
	\includegraphics[width=\spw\textwidth]{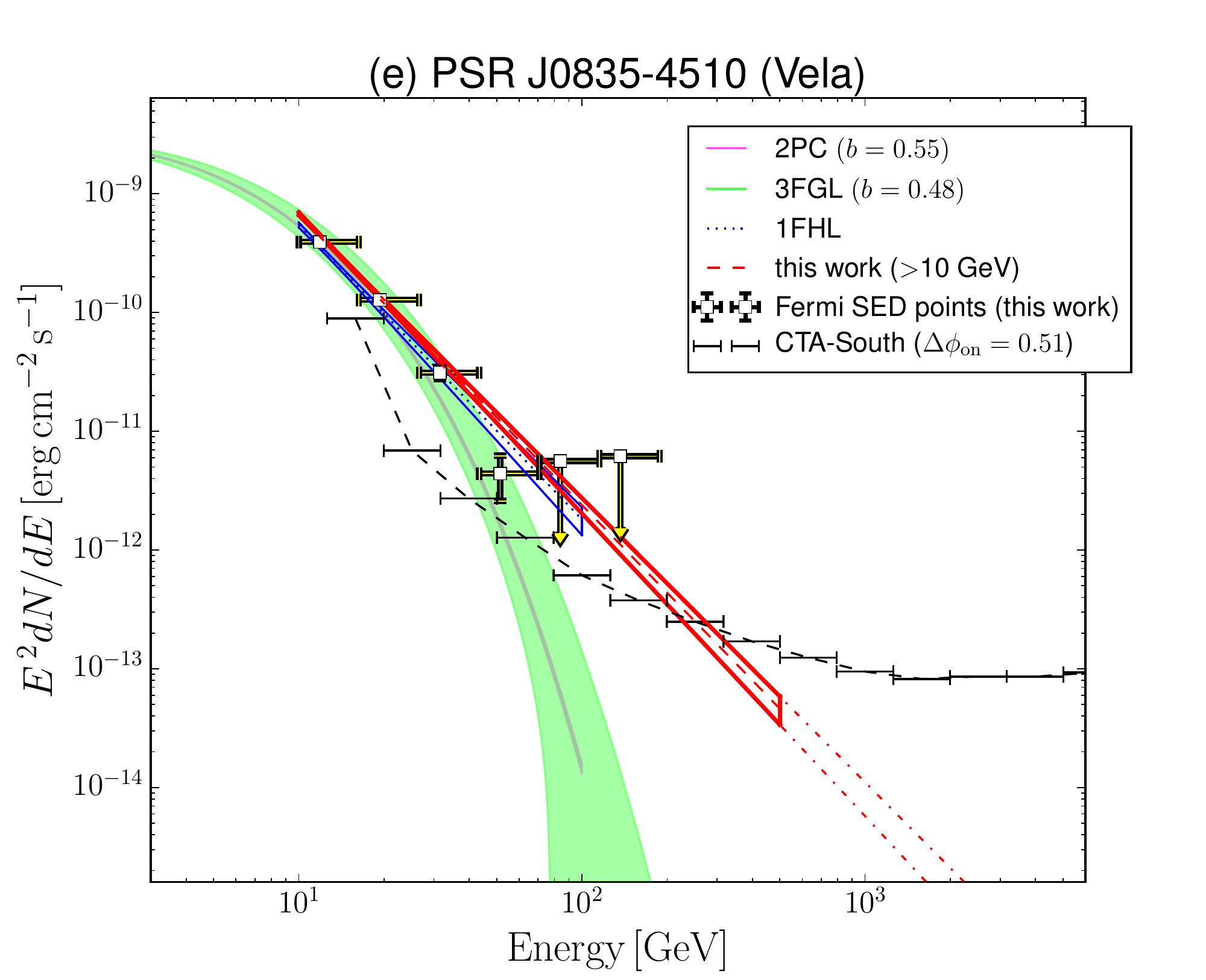}
	\hfill
	\includegraphics[width=\spw\textwidth]{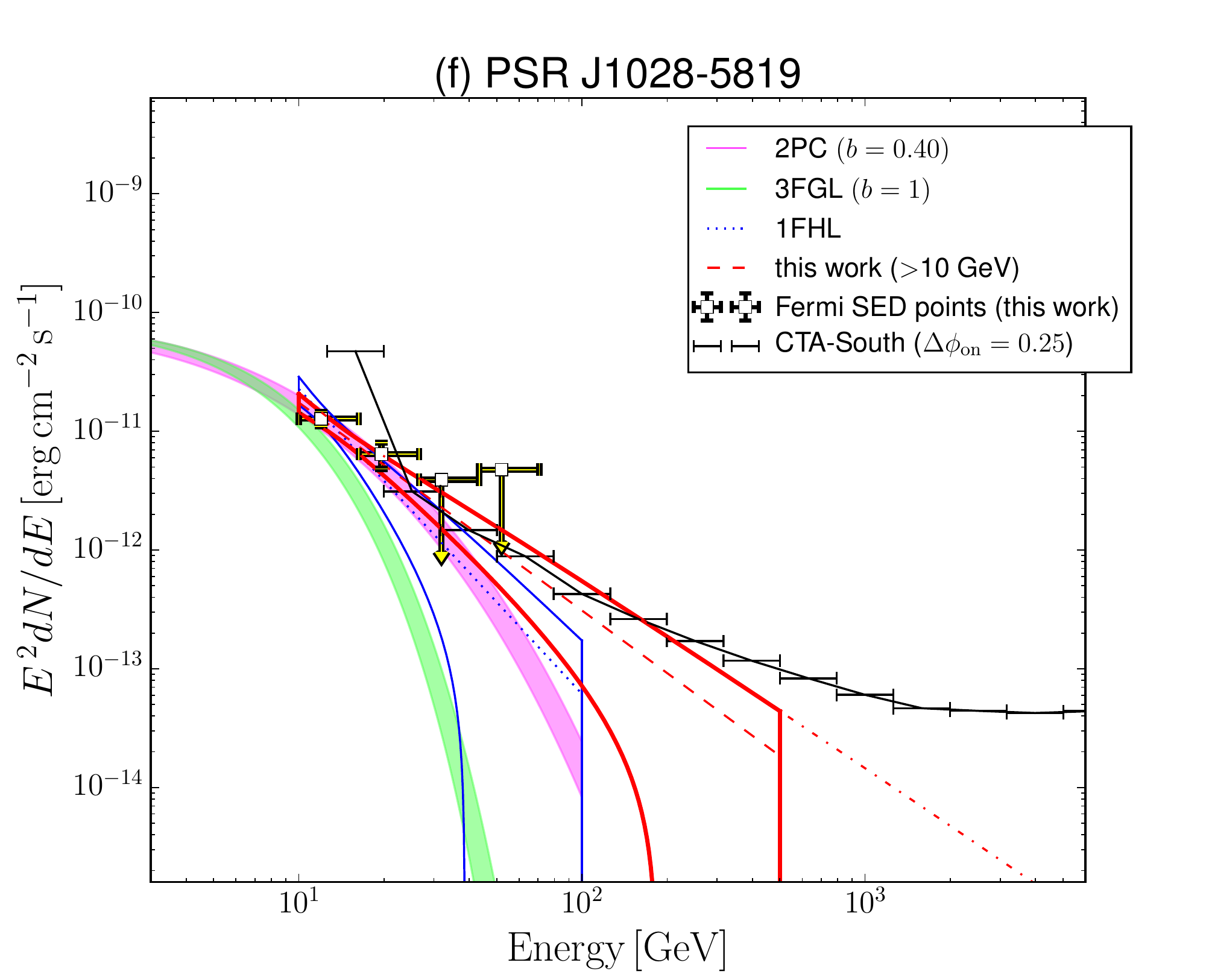}
\\
	\includegraphics[width=\spw\textwidth]{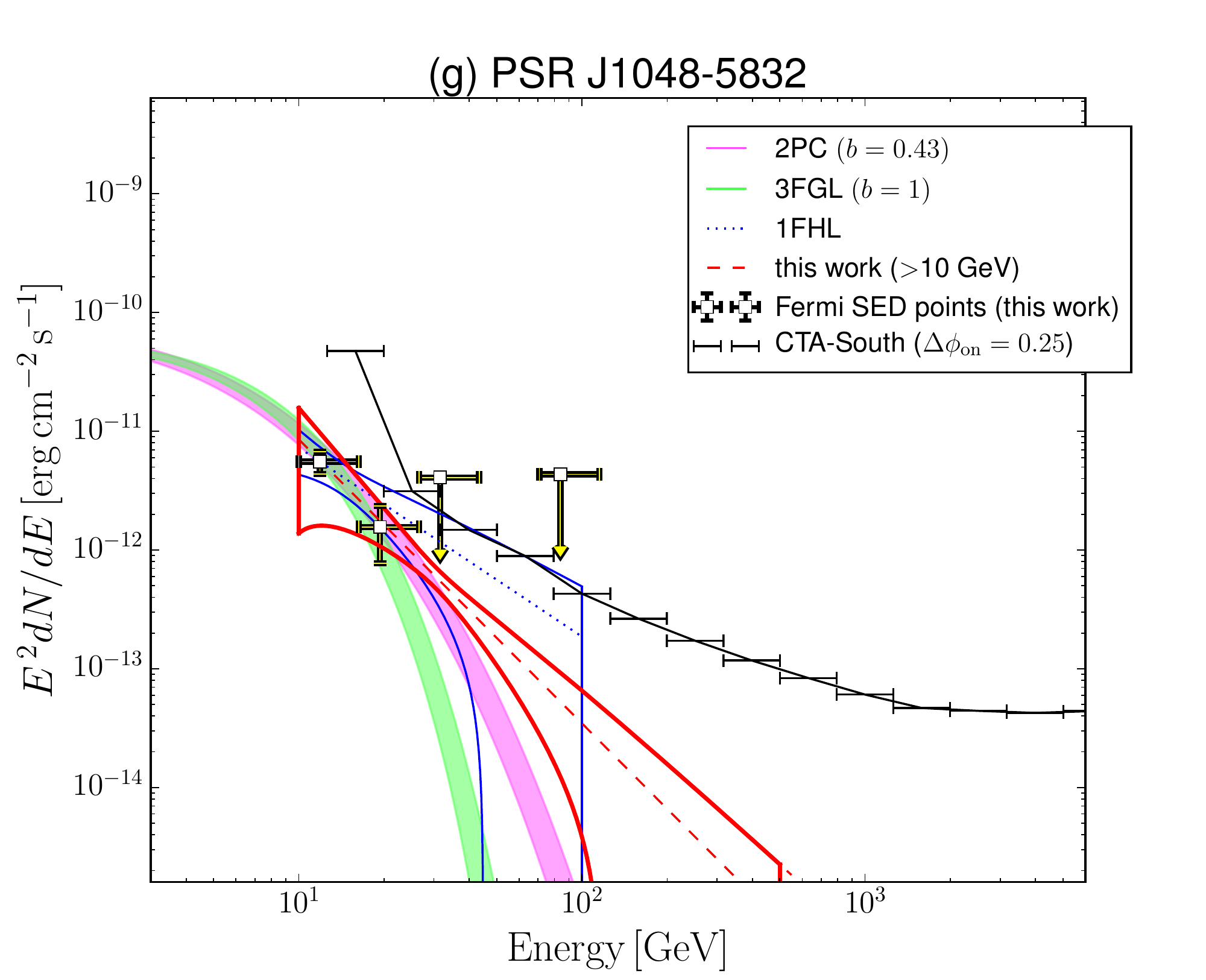}
	\hfill
	\includegraphics[width=\spw\textwidth]{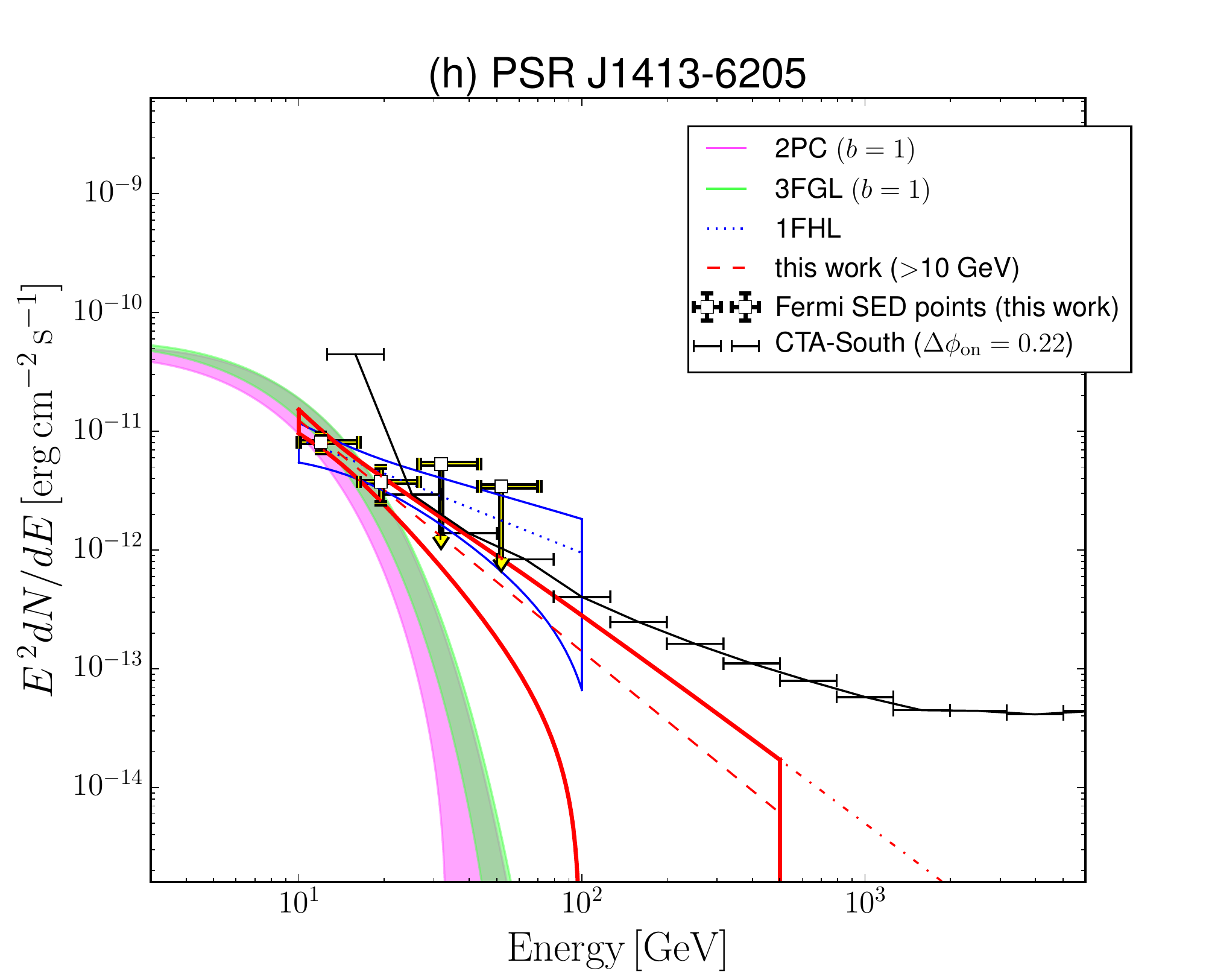}
	\hfill
	\includegraphics[width=\spw\textwidth]{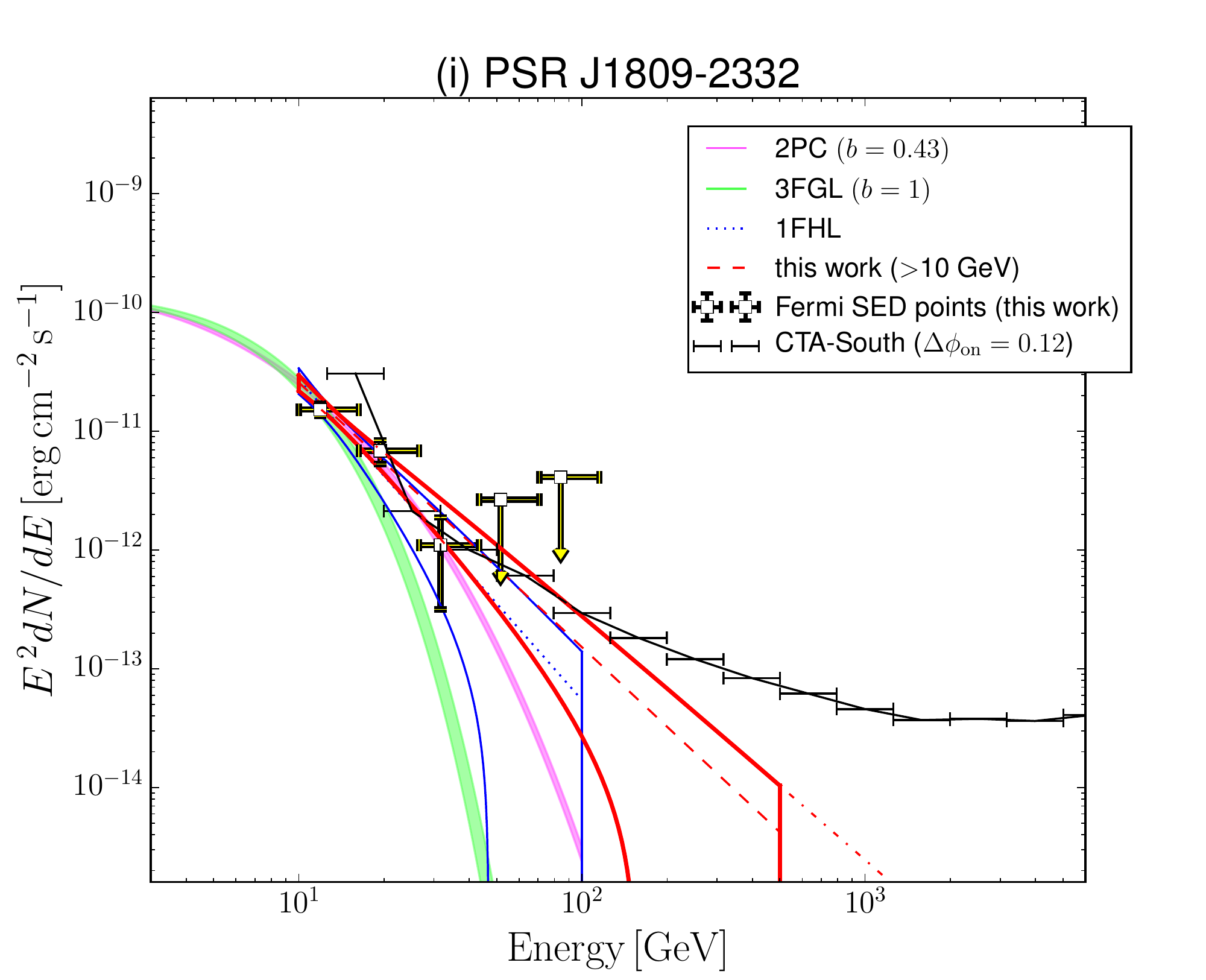}
\\
	\includegraphics[width=\spw\textwidth]{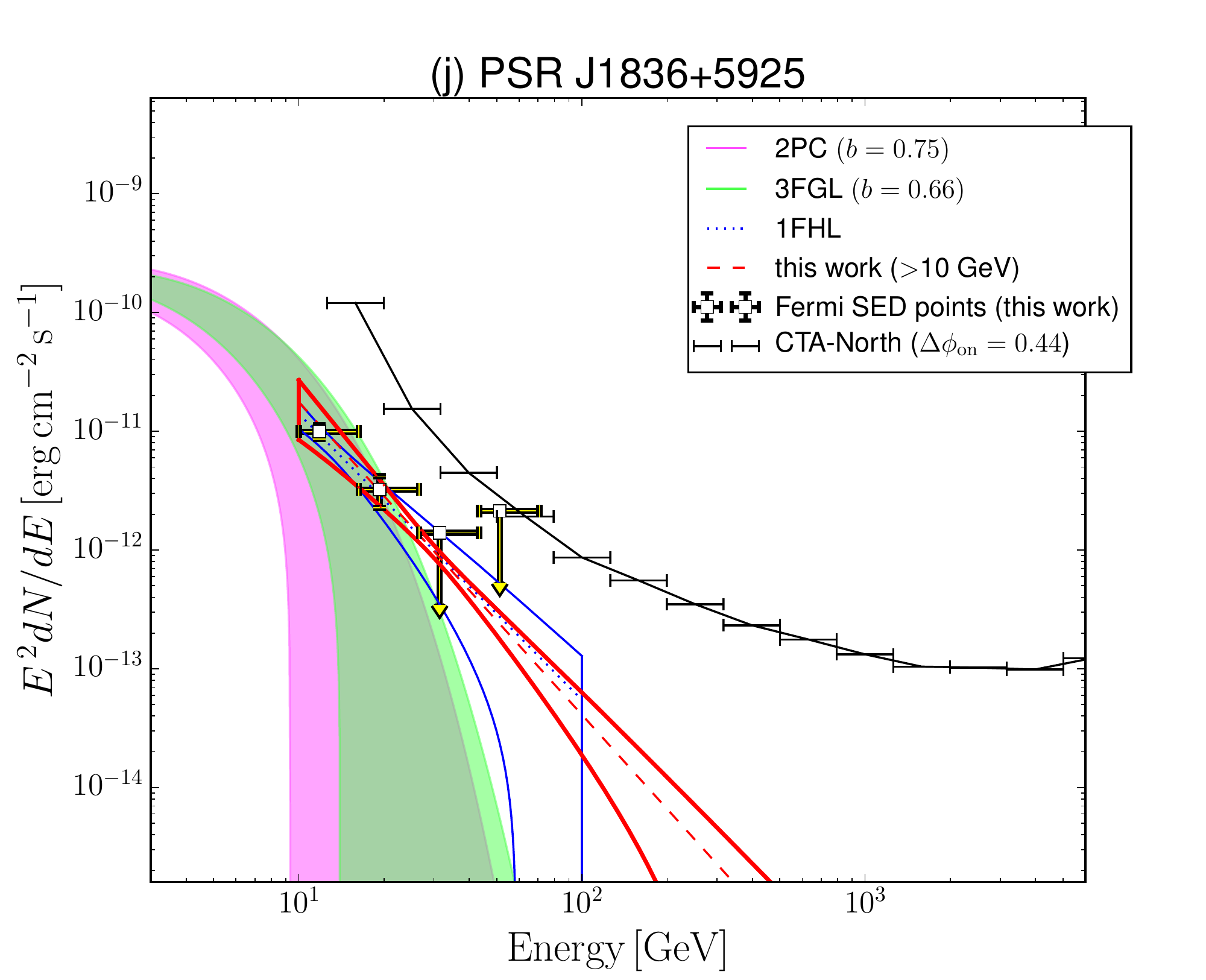}
	\hfill
	\includegraphics[width=\spw\textwidth]{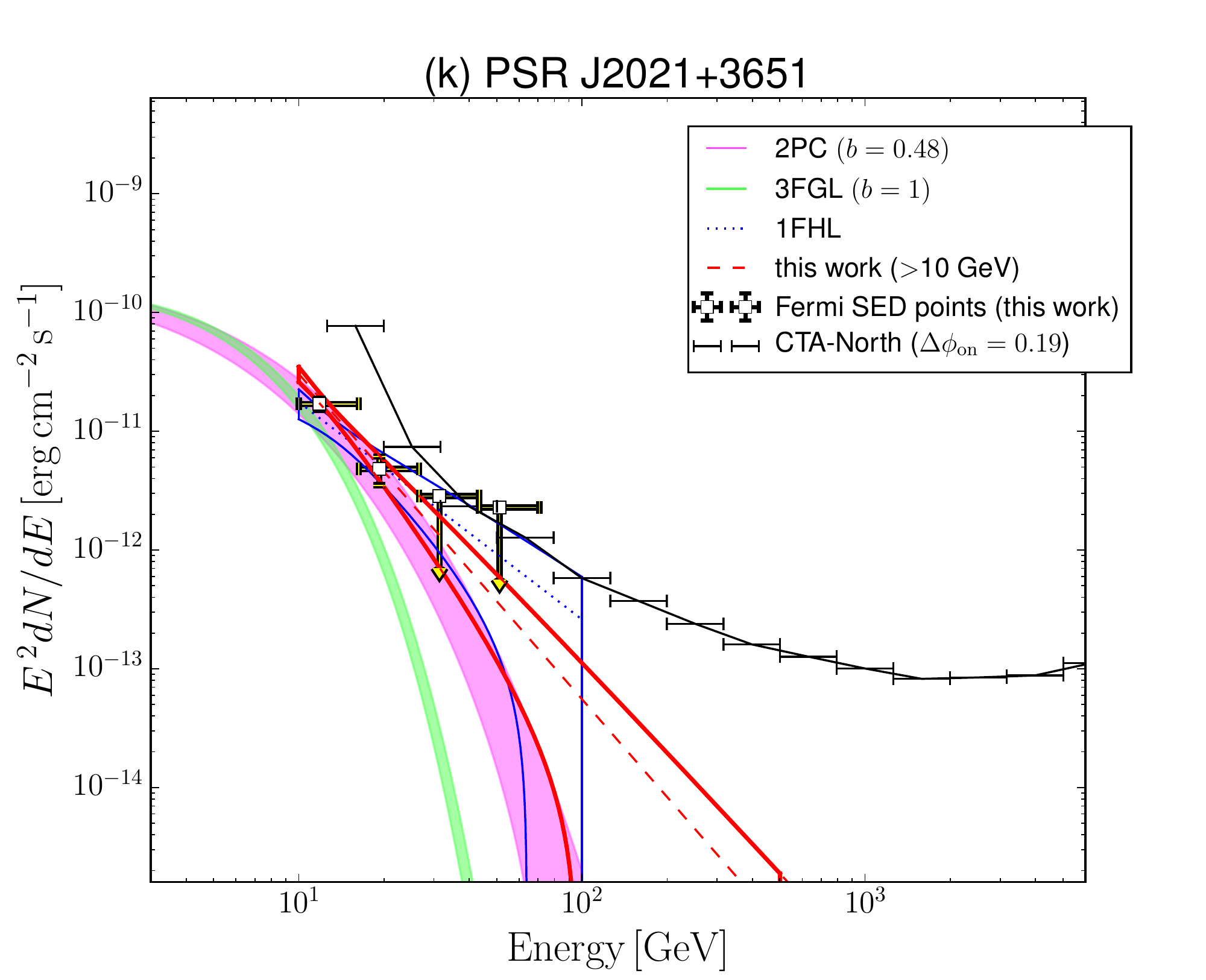}
	\hfill
	\includegraphics[width=\spw\textwidth]{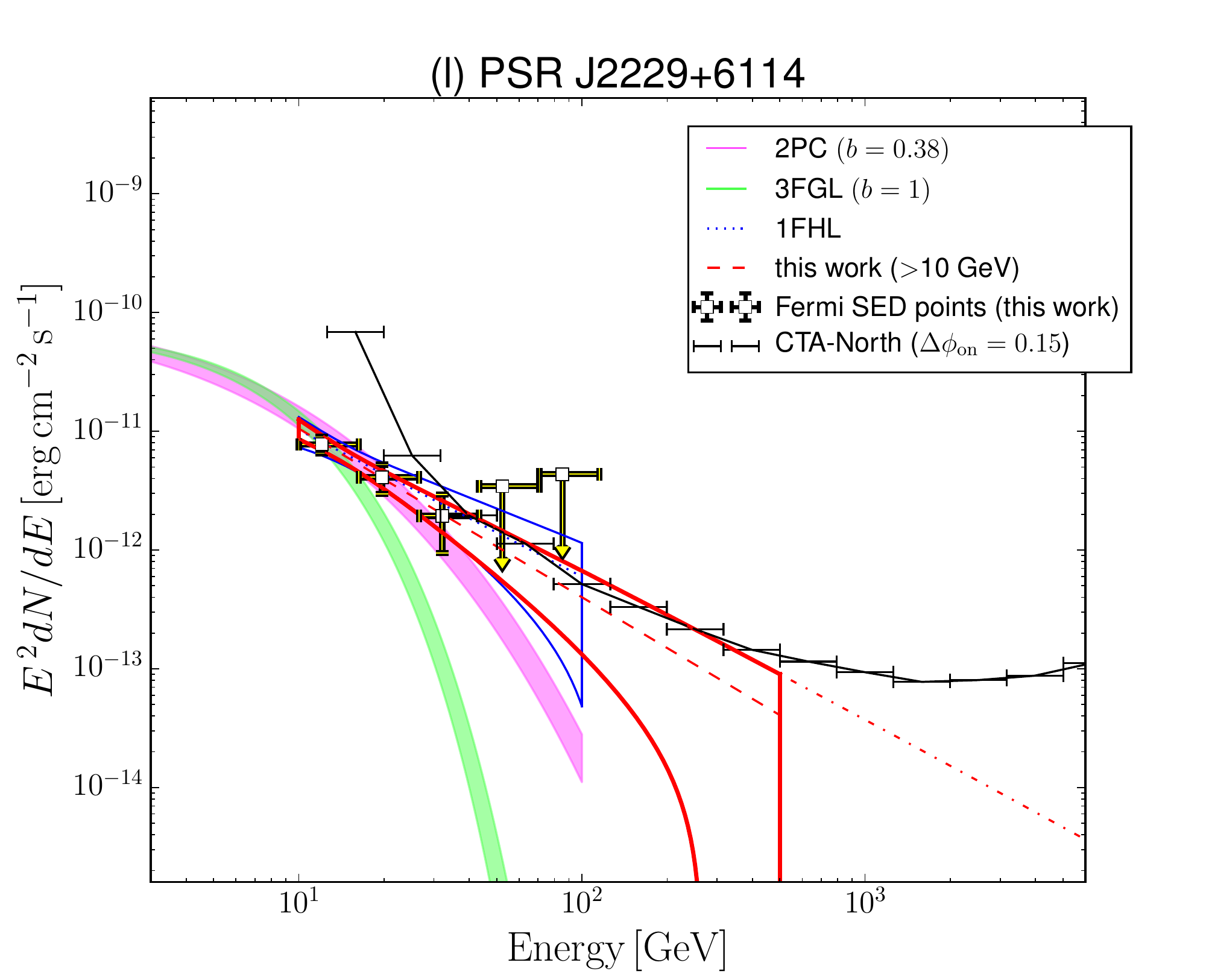} 

	\caption{Spectral energy distributions of our sample of pulsars above 10 GeV (red), compared with those from 3FGL (green, \citealt{Acero2015}), 2PC (magenta, \citealt{Abdo2013}) and 1FHL (blue, \citealt{Ackermann2013}). The value of the parameter $b$ of the PLEC model used in 3FGL and 2PC is shown in brackets. The red dashed and blue dotted lines are the mean values of spectra from the high-energy spectral fits ($E$$>$10 GeV). The red dot-dashed lines show an extrapolation of the butterfly plots to the VHE range. The hatched area in Fig. \ref{fig:Fermi_Specs}b corresponds to the joint \fermi-LAT and MAGIC power-law spectral fit found by \citet{Ansoldi2016}. Black dots show the the \fermi-LAT spectral energy distribution (SED) in different energy bins (upper limits correspond to TS$_i<9$, see text for details). Only statistical errors are considered in all butterfly plots and spectral energy distribution measurements. Individual CTA 50 h sensitivity curves calculated for each pulsar, with its own duty cycles ($\Delta\phi_{\rm on}$, marked as solid lines) and, if present, OP emission (marked as dashed lines), are shown in black.}
	\label{fig:Fermi_Specs}
\end{figure*}

\newcommand*\pwn{0.33}
\newcommand*\pwnn{0.471428571}
\begin{figure*}
	\includegraphics[width=\pwn\textwidth]{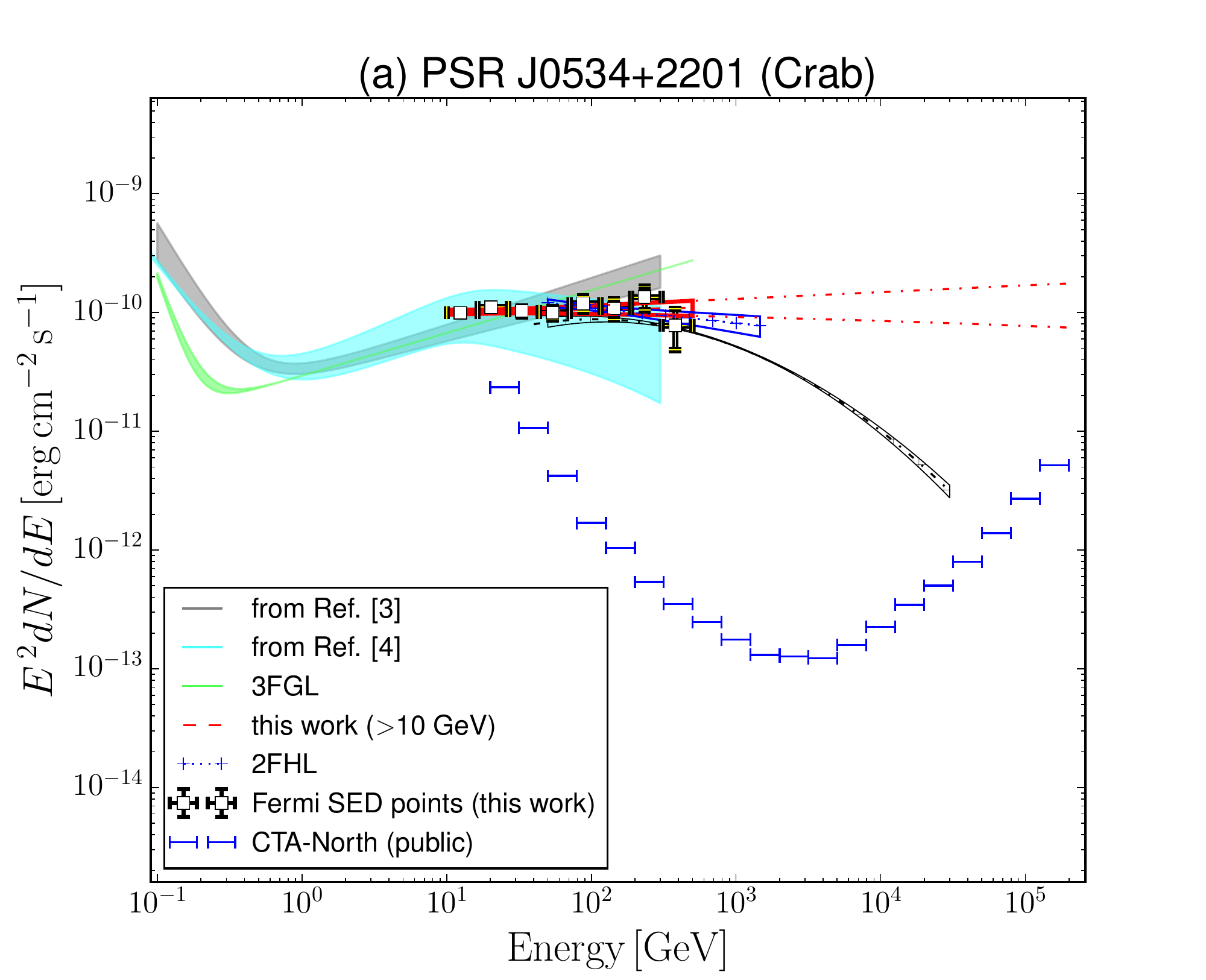}
	\hfill
	\includegraphics[width=\pwn\textwidth]{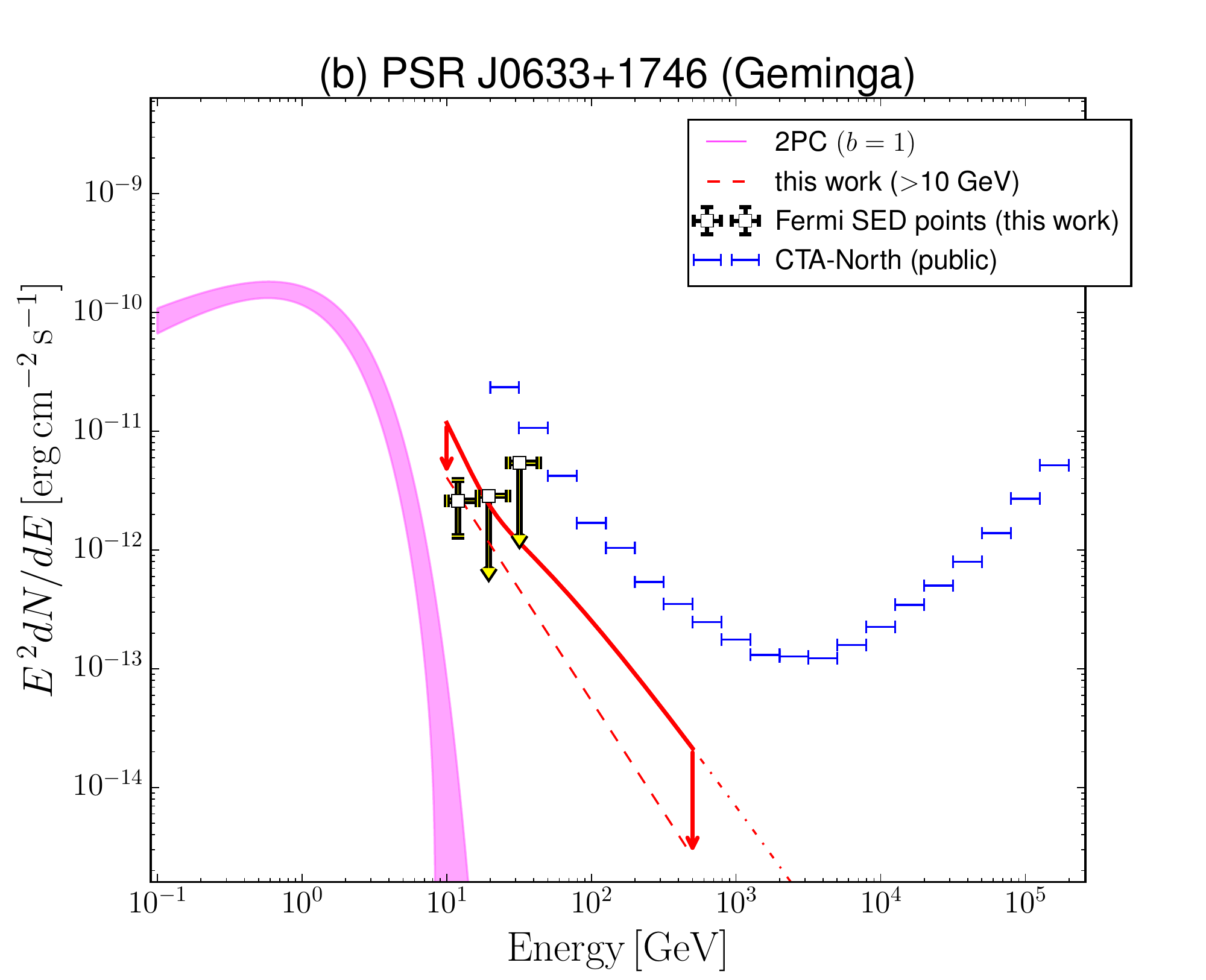}
	\hfill
	\includegraphics[width=\pwn\textwidth]{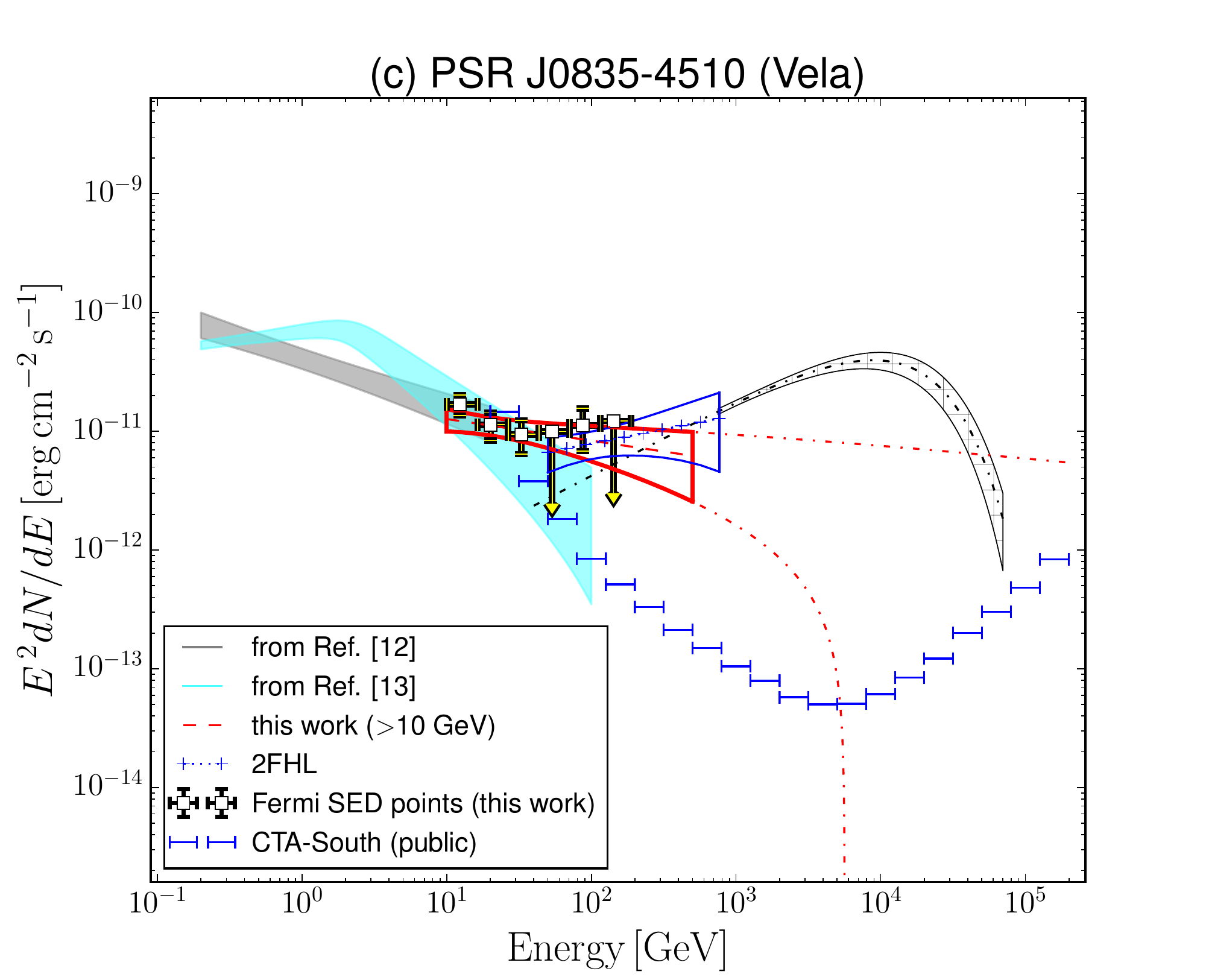}
\\
	\centering
	\begin{minipage}{.7\textwidth}
		\centering
		\includegraphics[width=\pwnn\textwidth]{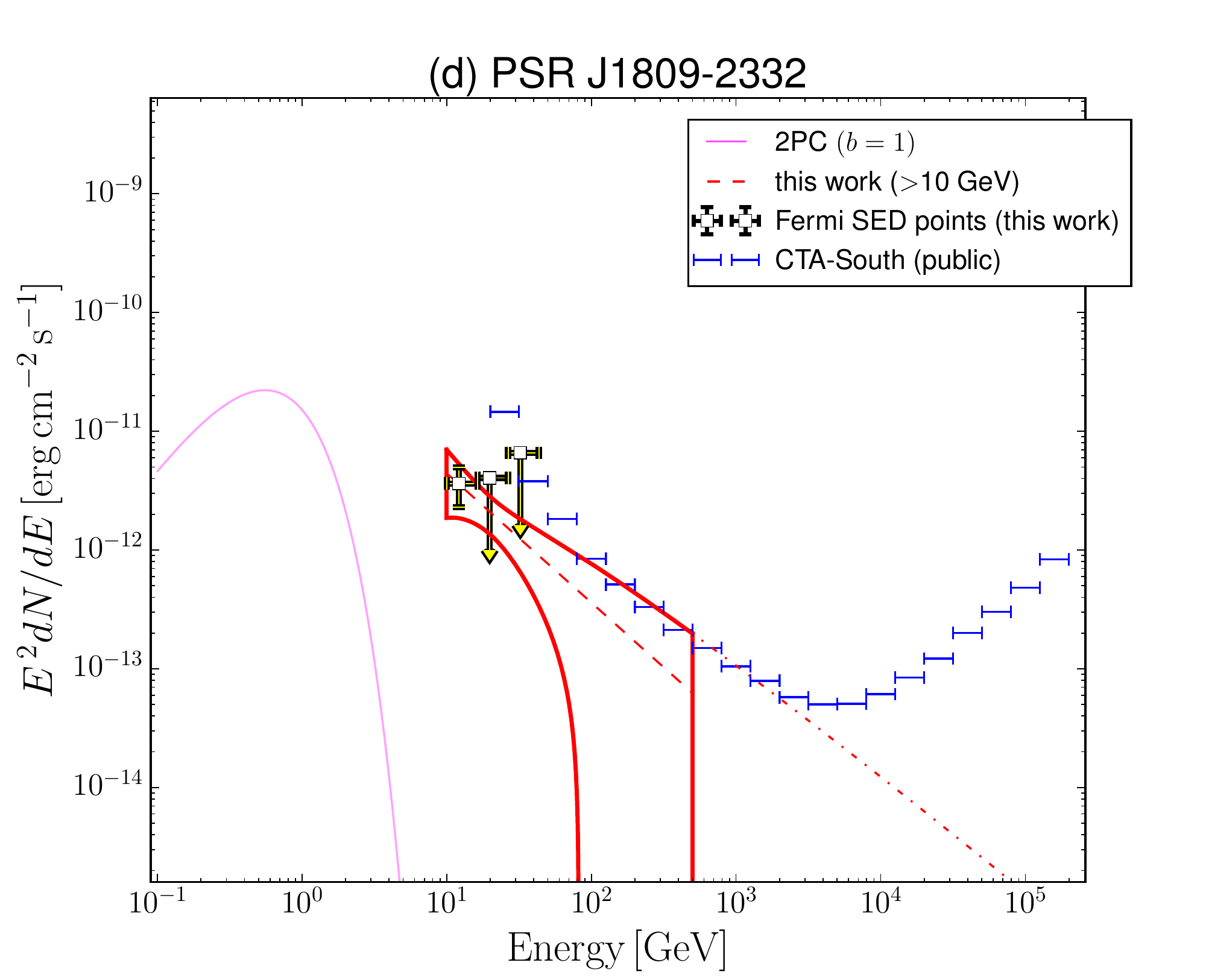} 
		\hfill
		\includegraphics[width=\pwnn\textwidth]{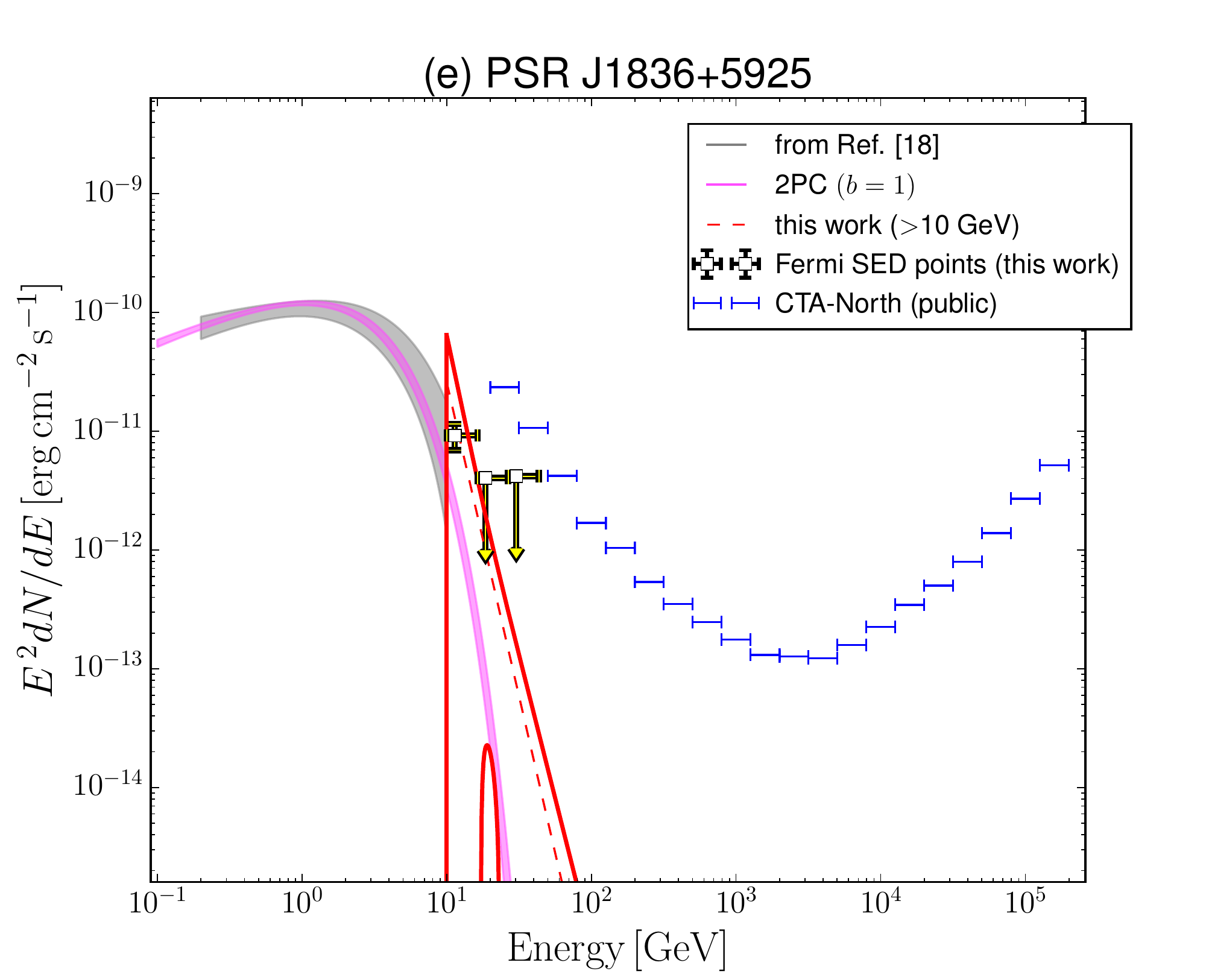}
	\end{minipage}

	\caption{Point sources off-peak (OP) spectral energy distributions above 10 GeV (red), compared with those from the literature (gray, cyan, see references in Table \ref{tab:res}), 3FGL (green, \citealt{Acero2015}), 2PC (magenta, \citealt{Abdo2013}) and 2FHL (blue crosses, \citealt{Ackermann2016}). The value of the parameter $b$ of the PLEC model used in 2PC is shown in brackets. The red dashed lines are the mean values of spectra from the high-energy spectral fits ($E$$>$10 GeV). The red dot-dashed lines show an extrapolation of the butterfly plots to the VHE range. For PSR J0633$+$1746 the arrows are the upper bounds of the corresponding spectra. Only the mean value of the uncertain 2PC spectrum of the OP emission of PSR J1809$-$2332 is shown in Fig. \ref{fig:PWN_Specs}d. The hatched areas in Figs. \ref{fig:PWN_Specs}a and \ref{fig:PWN_Specs}c are the VHE spectra from \citet{Aleksic2015} and \citet{Abramowski2012}, respectively. The Crab \citep{Aleksic2015} and Vela \citep{Abramowski2012} PWNe spectra, extrapolated down to 40 GeV are shown as black dot-dashed lines in Figs. \ref{fig:PWN_Specs}a and \ref{fig:PWN_Specs}c, respectively. Black dots show the \fermi-LAT spectral energy distribution (SED) in different energy bins (upper limits correspond to TS$<9$, see text for the details). Only statistical errors are considered in all butterfly plots and spectral energy distribution measurements. CTA 50 h sensitivity curves (blue bars) are taken from publicly available IRFs. Note, unlike Fig. \ref{fig:Fermi_Specs}, these curves are calculated considering only cosmic rays as the background and taking the duty cycles equal to 1.}
	\label{fig:PWN_Specs}
\end{figure*}

\subsection{Simulated CTA observations}\label{sec:4.3}
Using the results of the spectral fitting at energies $E$$>$10 GeV, we extrapolated the pulsar emission at VHE and simulated CTA observations of these objects. We consider observations with the northern or southern CTA installations (CTA-North and CTA-South). Simulations were performed including the isotropic background and the extrapolated VHE emission of the GDB  (if present, see Table \ref{tab:GDB_res}). The resulting significances $S$ of the pulsar point source detections in different energy bins ($E$$>$0.04, $E$$>$0.1, $E$$>$0.25 TeV) calculated for all events (phase average analysis, VHE-PhA) and considering the events from the reduced-background analysis\footnote{Reduced-background simulations with renormalized background are performed to mimic the results of the on-peak analysis (see Sect. \ref{subs:3.1}).} (VHE-RB) are listed in Table \ref{tab:Signif}. We obtained that the range of significances for each pulsar is dominated by the statistical uncertainties of its spectrum.

Residual and TS maps of the HE \fermi~pulsars for reduced-background simulations (VHE-RB) at energies $E$$>$0.1 TeV are shown in Figs. \ref{fig:VHE_resmaps} and \ref{fig:VHE_tsmaps}, respectively. These maps were obtained with \ctools, assuming the optimistic approximation of the pulsar spectrum reported in equation \ref{eq:7}. The Vela pulsar in Fig. \ref{fig:VHE_resmaps}e is shown with its PWN.

The spectral models of the OP emission in the Crab and Vela pulsars used in our VHE simulations are taken from \citet{Aleksic2015} and \citet{Abramowski2012}, respectively. OP components significantly detected in other pulsars (PSRs J0633$+$1746, J1809$-$2332 and J1836$+$5925) have rather uncertain spectral parameters above 10 GeV (see Table \ref{tab:res}) and their fluxes are below detectability in a 50 h CTA observation (see Fig. \ref{fig:PWN_Specs}). Although we included them in the VHE simulations, their contribution can be considered negligible.

PSRs J0007$+$7303, J1028$-$5819, J1048$-$5832, J1413$-$6205, J1836$+$5925 and J2229$+$6114 culminate at zenith angles above 30$^\circ$ (see Table \ref{tab:culm}) and, thus, low-energy CTA observations ($<$0.1 TeV) of these pulsars will not be efficient. The typical value of the energy threshold for high zenith angle observations assumed here is 0.1 TeV. Detection significances of these pulsars at energies $E$$>$0.04 TeV are listed in Tables \ref{tab:Signif} and \ref{tab:Tsig} for comparison.

As shown in Table \ref{tab:Signif} (column 3), the simulated CTA observations show significant detections\footnote{$S\ge5$ at least.} above 0.04 TeV for PSRs J0534$+$2201, J0614$-$3329, J0835$-$4510 and J1809$-$2332. Pulsars J0007$+$7302, J1028$-$5819 and J2229$+$6114 (in brackets in Table \ref{tab:Signif}) would be in principle detectable with CTA at low energies ($E$$>$ 0.04 TeV) if they were observed at a 20$^\circ$ zenith angle. 

Above 0.1 TeV, CTA will significantly detect PSRs J0534$+$2201, J0614$-$3329, J0835$-$4510, J1028$-$5819 and J2229$+$6114. In addition, pulsars J0007$+$7302, J1413$-$6205 and J1809$-$2332 will be detected in 50 h after off-peak background subtraction (Table \ref{tab:Signif}, column 7). A longer observing time is required for a significant detection of other pulsars with CTA above 0.1 TeV. PSRs J0534$+$2201, J0614$-$3329 and J0835$-$4510 can be detected at energies $E$$>$0.25 TeV in the phase-averaged analysis, while other 2 pulsars J1028$-$5819 and J2229$+$6114 turn out to be detectable in the reduced-background analysis. PSR J0614$-$3329 can be marginally detected above 1 TeV with $S=4.0^{+3.2}_{-1.2}$ in 50 h considering on-peak phase intervals (reduced-background analysis). Although in our analysis we considered $S$$=$5 as the detection threshold, we found that PSRs J0534$+$2201 and J2229$+$6114 will be also marginally detectable with CTA in 50 h at energies $E$$>$1 TeV (reduced-background analysis, performed assuming the optimistic approximation for the spectrum of each pulsar) with a significance $S$$=$4 and $S$$=$4.3, respectively. 

\begin{figure*}
	\centering
	\includegraphics[width=0.70\textwidth]{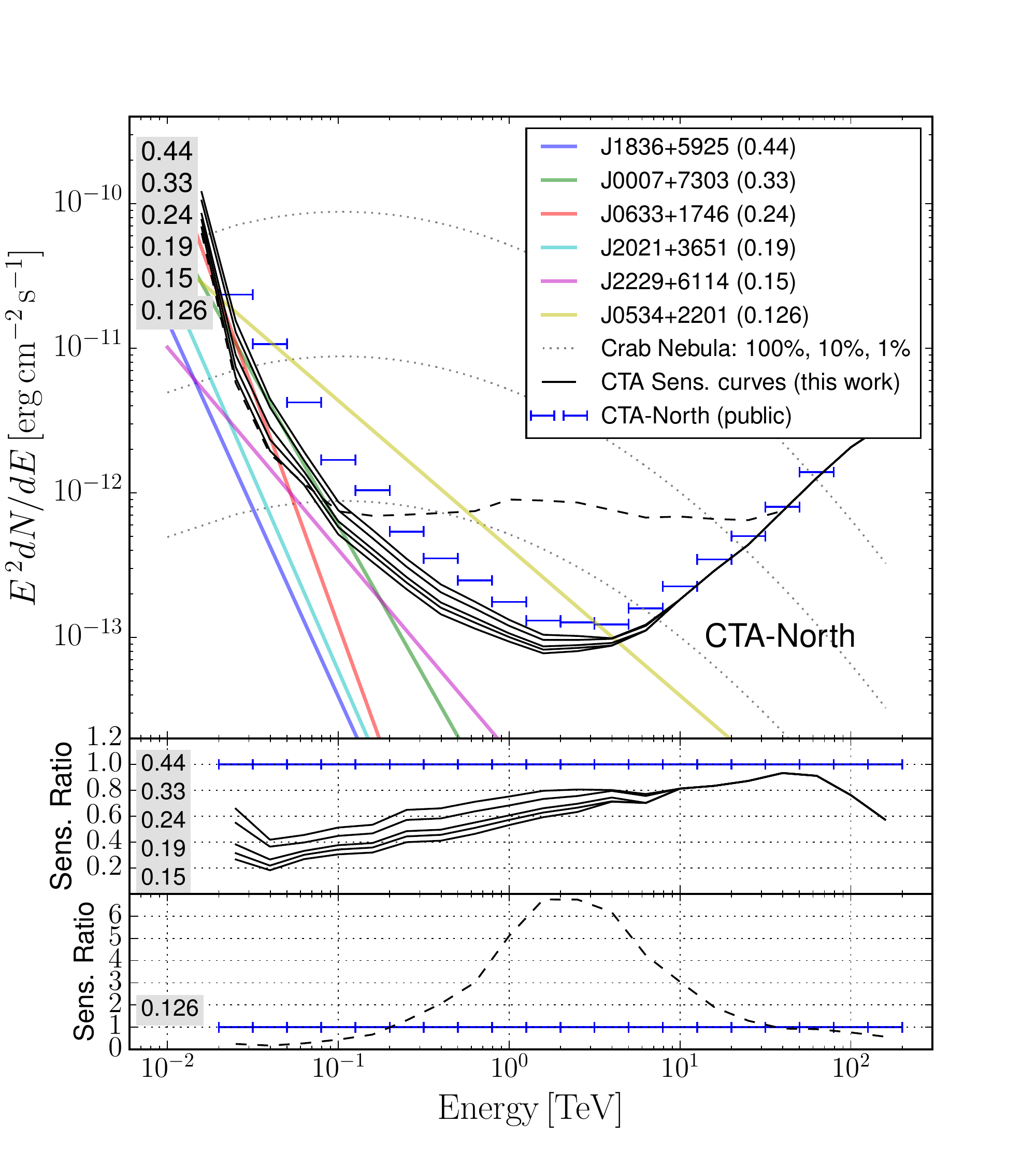}
	\caption{Sensitivity curves for a 50 h observation with the CTA-North array (black lines), calculated for the different duty cycles of the pulsars of our sample (shown on the top-left). The dashed black line corresponds to the sensitivity curve calculated for the Crab pulsar accounting for both its duty cycle and the OP emission of its nebula. The colored solid lines are the best fits of the pulsars spectra at energies $E$$>$10 GeV. The public 50 h CTA-North sensitivity curve (blue bars; duty cycle equal to 1) and the 100\%, 10\% and 1\% Crab nebula spectra (gray dotted lines, \citealt{Aleksic2015}) are shown for comparison. The bottom panels show the sensitivities normalized to the public 50 h CTA-North sensitivity curves.}
   	\label{fig:CTA-N}
\end{figure*}

\begin{figure*}
	\centering
	\includegraphics[width=0.70\textwidth]{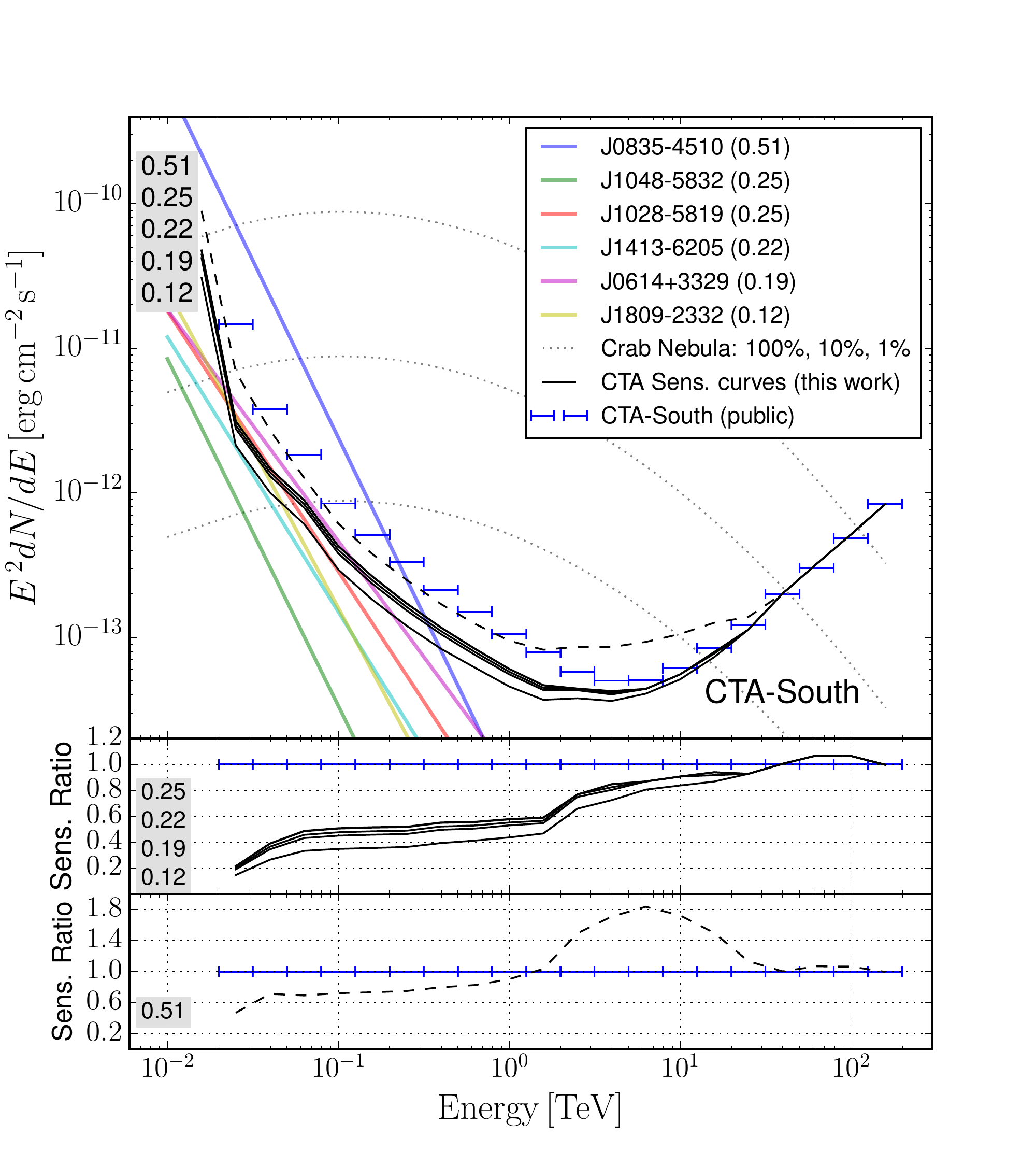}
	\caption{Same as Fig. \ref{fig:CTA-N} for the CTA-South array. The dashed black line corresponds to the sensitivity curve calculated for the Vela pulsar accounting for both its duty cycle and the OP emission of its nebula.}
   	\label{fig:CTA-S}
\end{figure*}

Individual CTA sensitivity curves, which account for different duty cycles and the possible contribution of OP components (see Sect. \ref{subs:3.2}), were calculated for 12 HE \fermi~pulsars. Result are shown in Figs. \ref{fig:CTA-N} and \ref{fig:CTA-S} along with the extrapolated power-law spectra of each pulsar. 

\begin{table*}
\caption{Significances $S$ of VHE pulsars detections during observations with the northern and southern CTA configurations for the phase-average (VHE-PhA) and the reduced-background (VHE-RB) analyses. The $S$-values are the square root of the corresponding test statistics (TS) in three energy bands ($E$$>$0.04, $E$$>$0.1 and $E$$>$0.25 TeV). $t_{\rm sim}$ is the duration of the observations. $\Delta\phi_{\rm on}$ is the on-peak phase interval.}
\label{tab:Signif} 
\centering
\begin{tabular}{l l | l l l l | l l l l l}
\hline\hline
Pulsar	& (Array $t_{\rm sim}$)\hspace{0.5cm} & | & \multicolumn{3}{|c|}{VHE-PhA} & | & \multicolumn{4}{|c}{VHE-RB}	\\
 		& & | & $E$$>$0.04 TeV & $E$$>$0.1 TeV & $E$$>$0.25 TeV & | & $E$$>$0.04 TeV & $E$$>$0.1 TeV & $E$$>$0.25 TeV & $\Delta\phi_{\rm on}$	\\
\hline
\textbf{J0007$+$7302}			& (North 50h)	& | & ($3.1^{+1.6}_{-1.4}$)	& $-$ 				& $-$ 				& | & ($7.1^{+3.4}_{-2.7}$) 	& $4.1^{+2.4}_{-2.7}$ 	& $-$				& 0.33	\\

J0534$+$2201$^a$ 			& (North 50h)	& | & $10.5^{+4.6}_{-3.6}$ 	& $7.2^{+1.8}_{-5.0}$ 	& $2.0^{+3.7}_{-2.0}$  	& | & $30.3^{+9.8}_{-8.2}$ 	& $18.7^{+8.1}_{-7.9}$	& $7.4^{+5.9}_{-4.3}$	& 0.126 	\\ 

J0614$-$3329					& (South 50h)	& | & $7.5^{+3.9}_{-5.0}$ 	& $5.0^{+4.2}_{-5.0}$ 	& $3.5^{+3.8}_{-3.5}$ 	& | & $14.9^{+7.5}_{-9.5}$ 	& $8.3^{+8.6}_{-8.3}$ 	& $5.0^{+7.0}_{-5.0}$ 	& 0.19	\\ 

J0633$+$1746$^b$			& (North 50h)	& | & $-$ 					& $-$ 				& $-$ 				& | & $-$ 				& $-$ 				& $-$					& 0.24	\\ 

J0835-4510$^c$				& (South 50h)	& | & $39.6^{+4.7}_{-5.0}$ 	& $14.9^{+2.0}_{-2.7}$ 	& $5.3^{+1.5}_{-0.2}$ 	& | & $52.8^{+6.5}_{-5.9}$ 	& $19.5^{+3.2}_{-3.2}$	& $6.2^{+1.5}_{-0.6}$ 	& 0.51	\\

\textbf{J1028$-$5819}			& (South 50h)	& | & ($4.9^{+2.9}_{-2.6}$) 	& $4.2^{+2.1}_{-4.2}$ 	& $-$ 				& | & ($9.6^{+6.1}_{-6.8}$) 	& $6.0^{+4.5}_{-4.9}$ 	& $5.3^{+2.5}_{-2.2}$ 	&  0.25 \\

\textbf{J1048$-$5832}			& (South 50h)	& | & ($-$) 				& $-$ 				& $-$ 				& | & ($-$) 				& $-$ 				& $-$  				& 0.25	\\

\textbf{J1413$-$6205}			& (South 50h)	& | & ($-$) 				& $2.4^{+1.4}_{-2.4}$& $-$ 				& | & ($3.9^{+3.3}_{-3.9}$)	& $2.1^{+3.1}_{-2.1}$ 	& $-$ 				& 0.22	\\ 

J1809$-$2332	$^b$			& (South 50h)	& | & $3.5^{+1.8}_{-1.3}$ 	& $-$ 				& $-$ 				& | & $5.1^{+5.6}_{-3.9}$ 	& $2.2^{+3.1}_{-2.2}$ 	& $-$ 				& 0.12	\\

\textbf{J1836$+$5925}$^b$		& (North 50h)	& | & ($-$) & $-$ & $-$ & | & ($-$) & $-$ & $-$  & 0.44	\\
J2021$+$3651				& (North 50h)	& | & $-$ & $-$ & $-$ & | & $-$ & $-$ & $-$  & 0.19	\\

\textbf{J2229$+$6114}			& (North 50h)	& | & ($2.2^{+3.0}_{-2.2}$) 	& $3.0^{+3.8}_{-2.0}$ 	& $-$ 				& | & ($5.9^{+7.4}_{-4.4}$) 	& $6.4^{+8.0}_{-6.4}$ 	& $5.8^{+5.0}_{-5.1}$  	& 0.15	\\ 
\hline

\multicolumn{11}{p{0.95\textwidth}}{\textbf{Notes.} Pulsars which will be observed at high zenith angles (above 30$^\circ$, i.e with high energy threshold $E$$>$0.1 TeV), are in \textbf{bold face}. Significances $S$ of these pulsars at energies $E$$>$0.04 TeV (in brackets) are shown for comparison. 
$^a$Simulations of the Crab pulsar observations performed assuming the log-parabola spectrum of the Crab nebula from \citet{Aleksic2015}.
$^b$Simulations of PSRs J0633$+$1746, J1809$-$2332 and J1836$+$5925 include the OP component found in the \fermi-analysis.
$^c$Simulations of the Vela pulsar observations performed assuming the power-law model with an exponential cut-off of the Vela PWN from \citet{Abramowski2012}. The spatial model for the extended Vela X emission is a uniform disk with radius of 1.2$^\circ$ centered at the coordinates $\alpha = 128.75^\circ$ and $\delta=-45.6^\circ$.
The lower/upper values of the significance $S$ are obtained using the optimistic/pessimistic approximations of the pulsar spectrum reported in equation \ref{eq:7}.}
\end{tabular}
\end{table*}

\newcommand*\cmw{0.24}
\begin{figure*}
	\includegraphics[width=\cmw\textwidth]{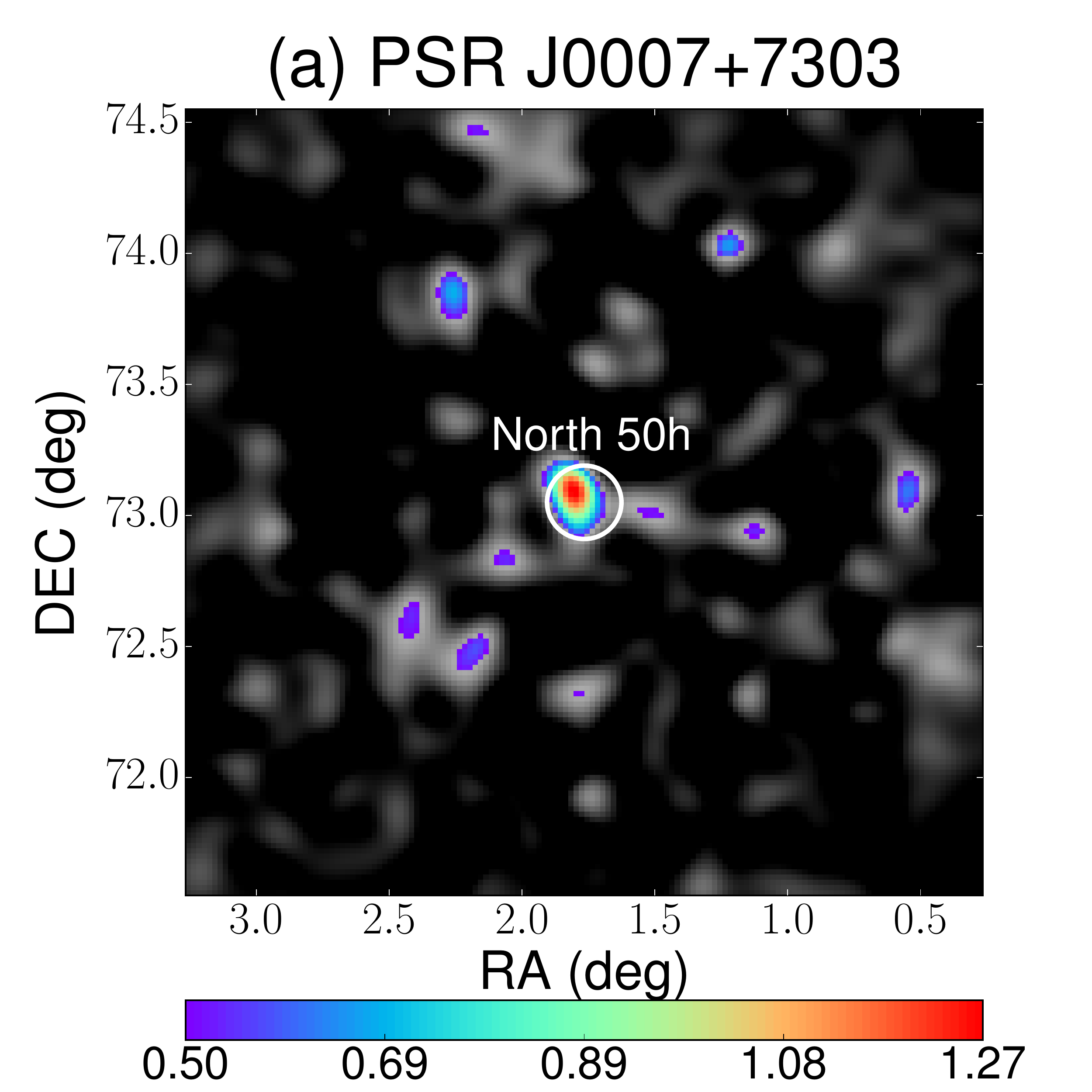}
	\hfill
	\includegraphics[width=\cmw\textwidth]{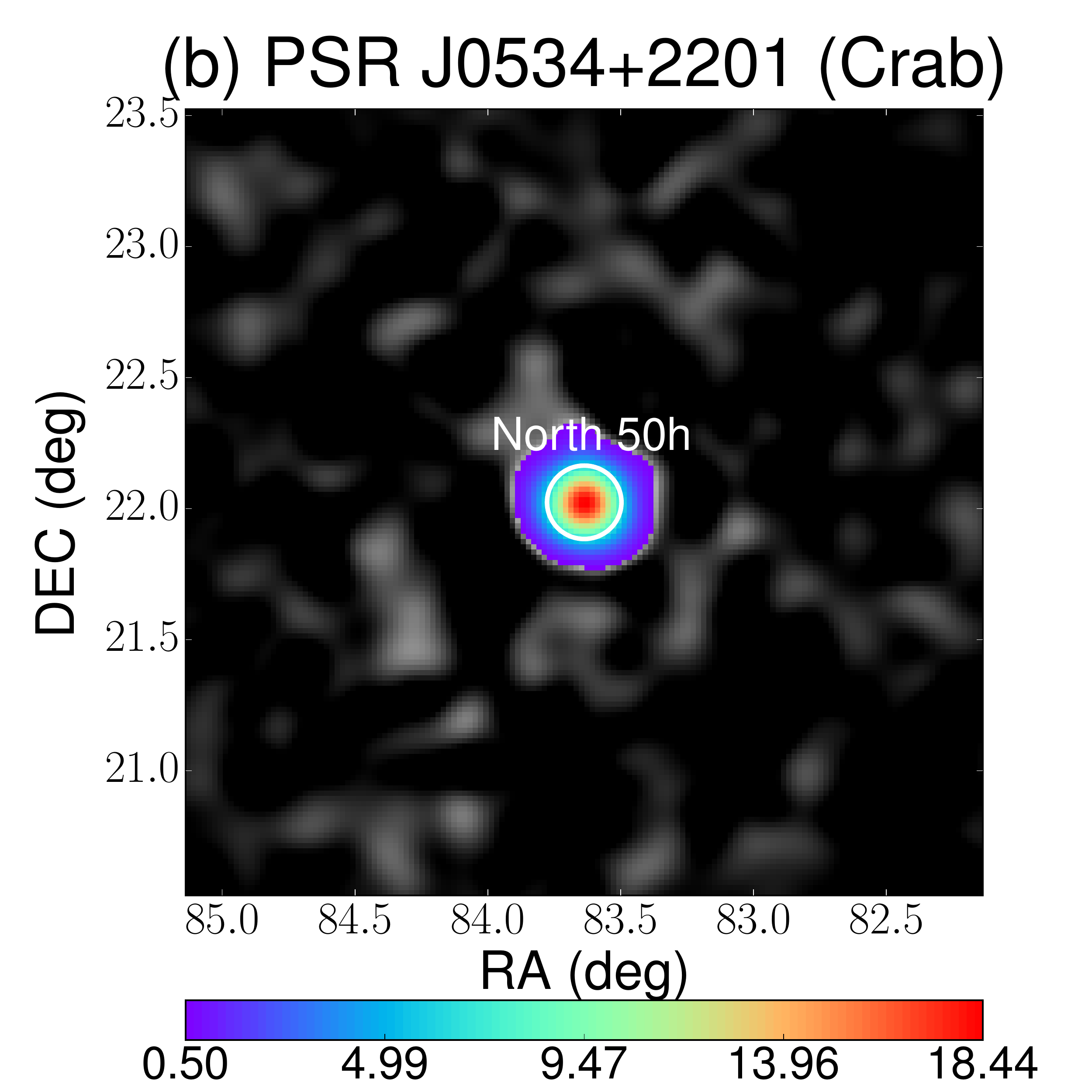} 
	\hfill
	\includegraphics[width=\cmw\textwidth]{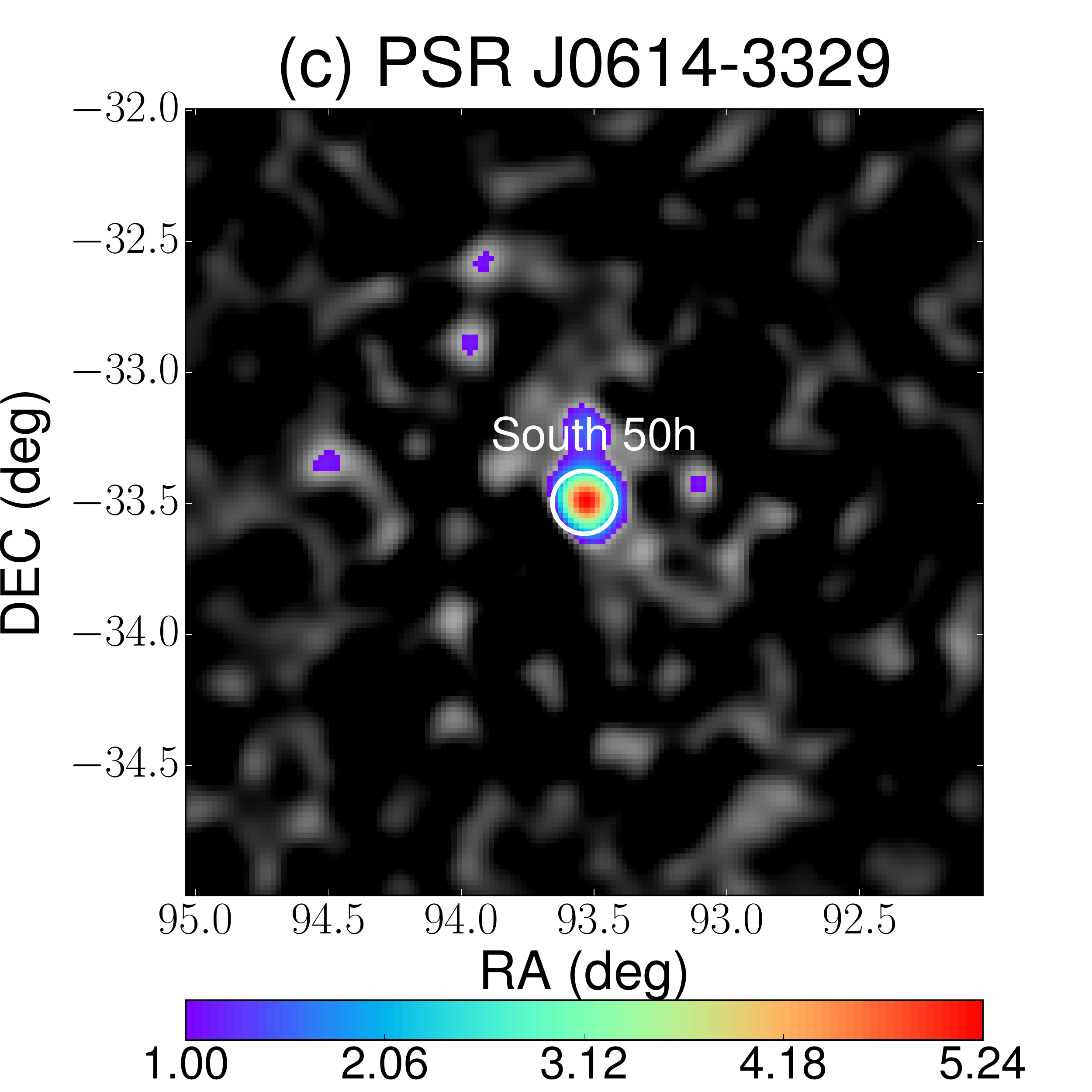}
	\hfill
	\includegraphics[width=\cmw\textwidth]{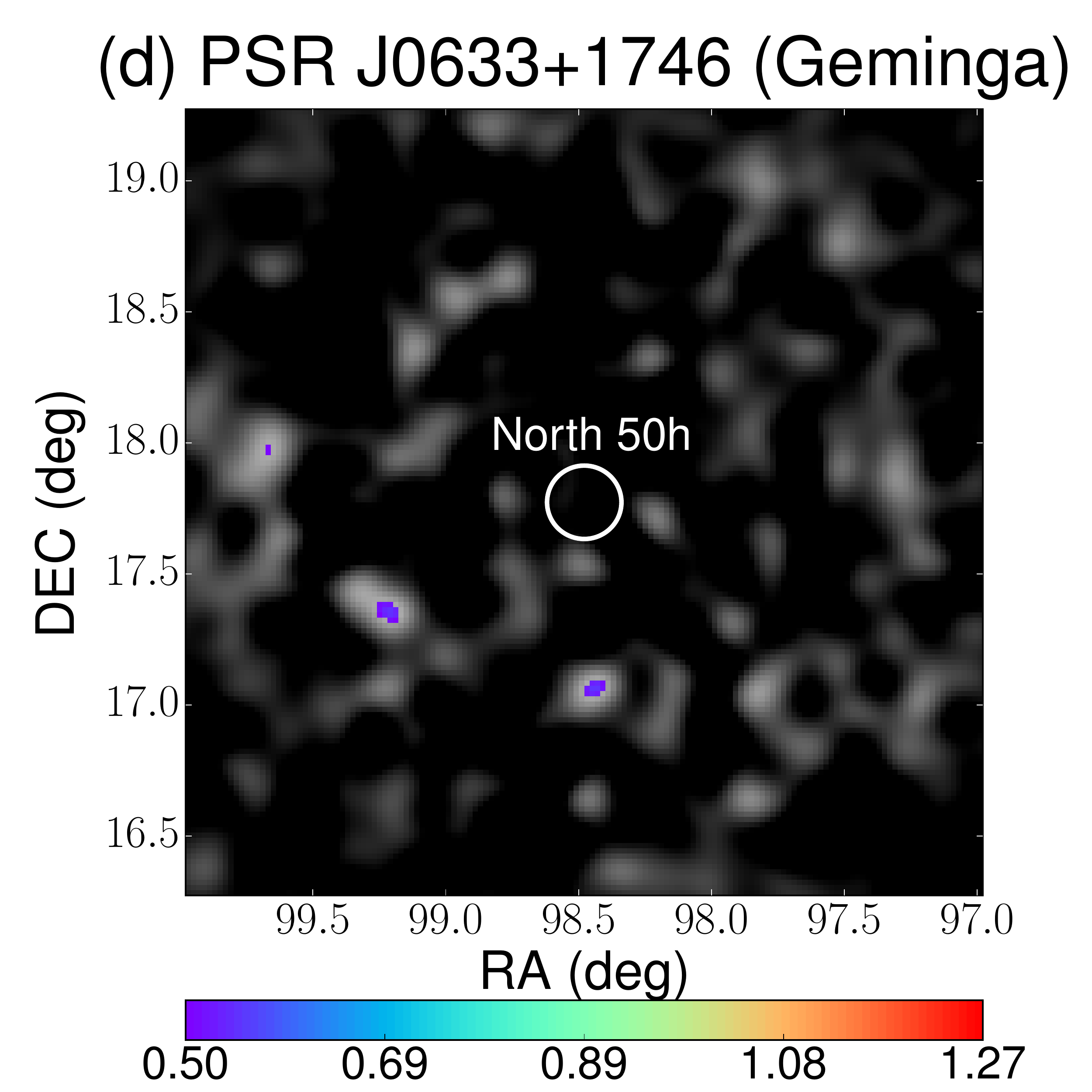} 
\\
	\includegraphics[width=\cmw\textwidth]{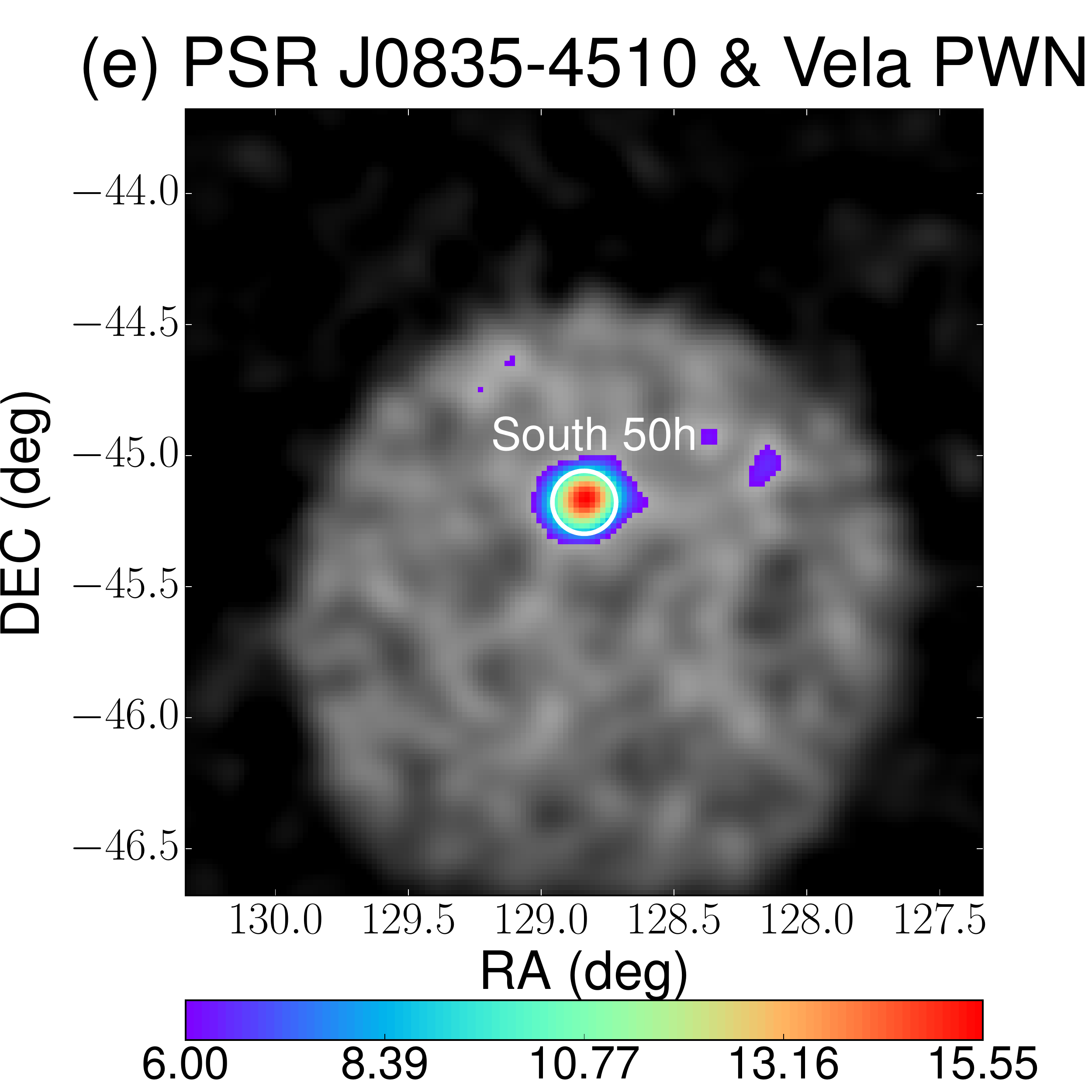}
	\hfill
	\includegraphics[width=\cmw\textwidth]{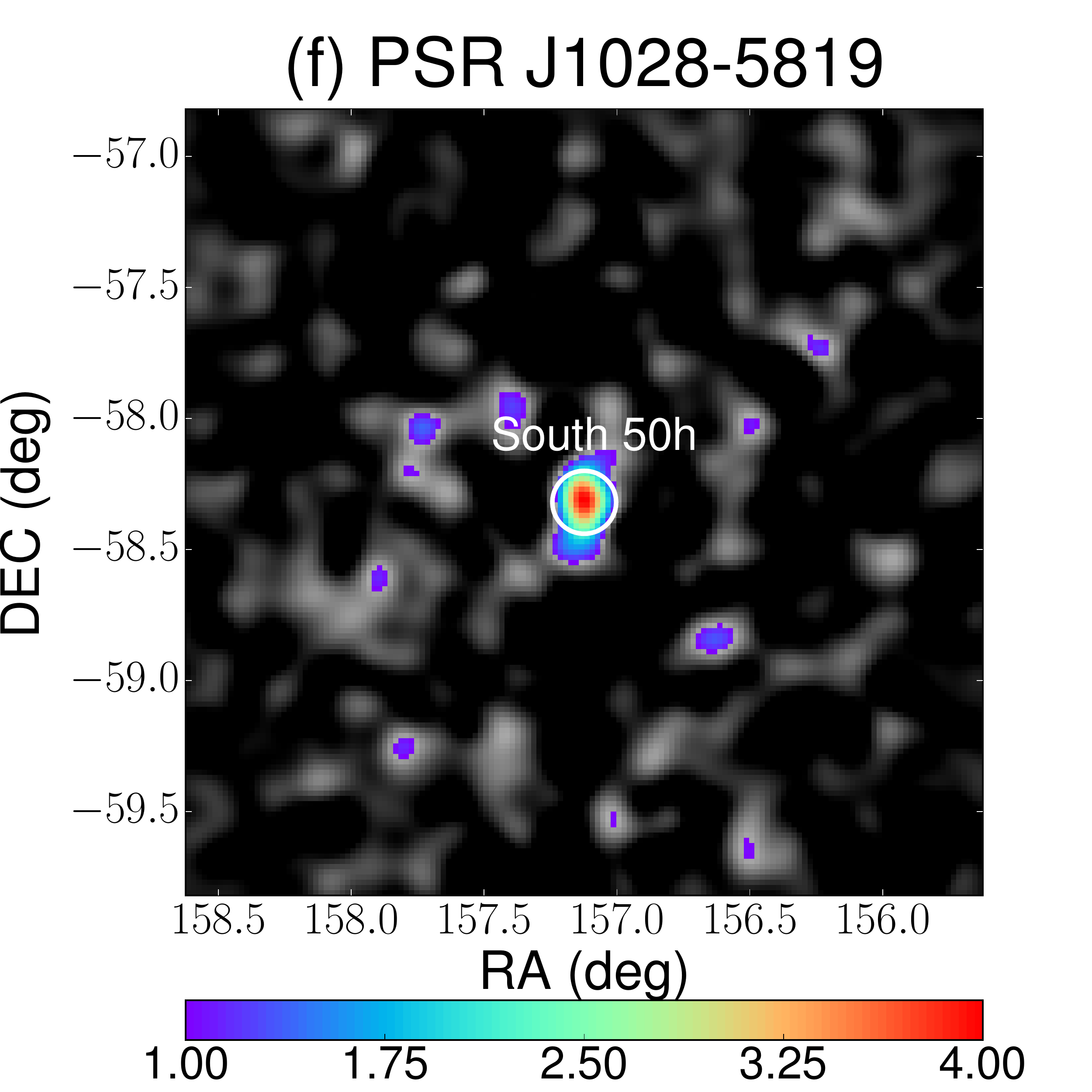}
	\hfill
	\includegraphics[width=\cmw\textwidth]{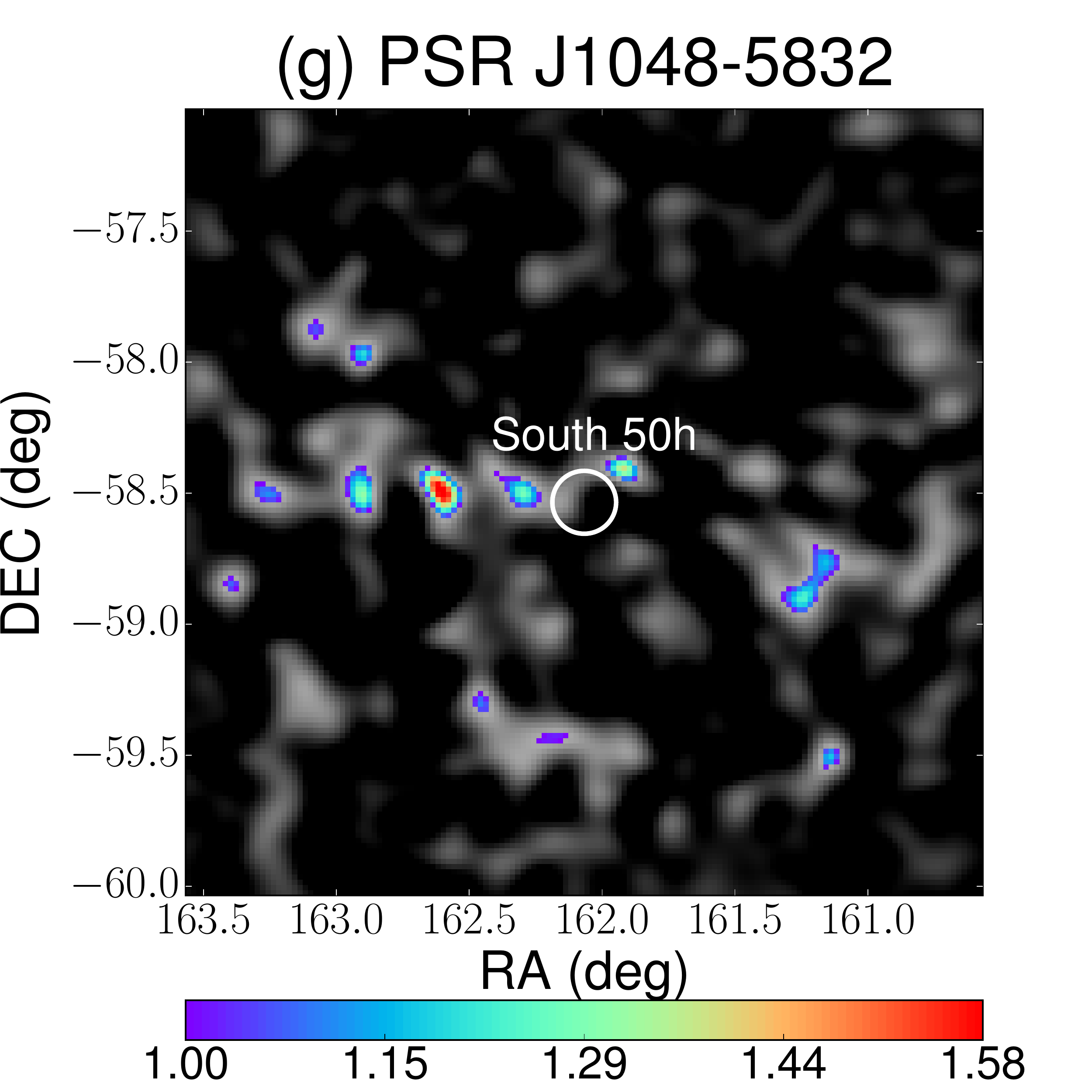} 
	\hfill
	\includegraphics[width=\cmw\textwidth]{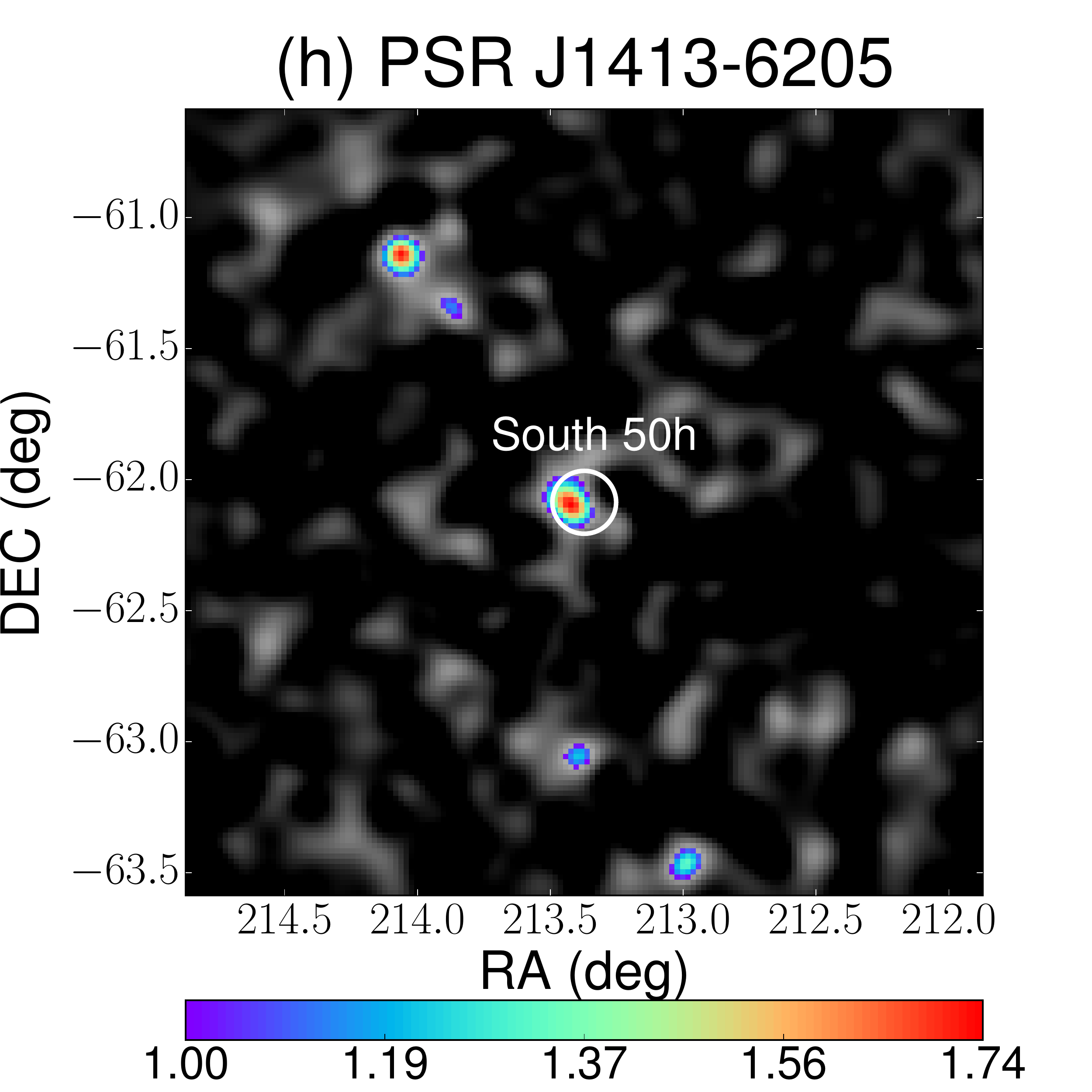} 
\\
	\includegraphics[width=\cmw\textwidth]{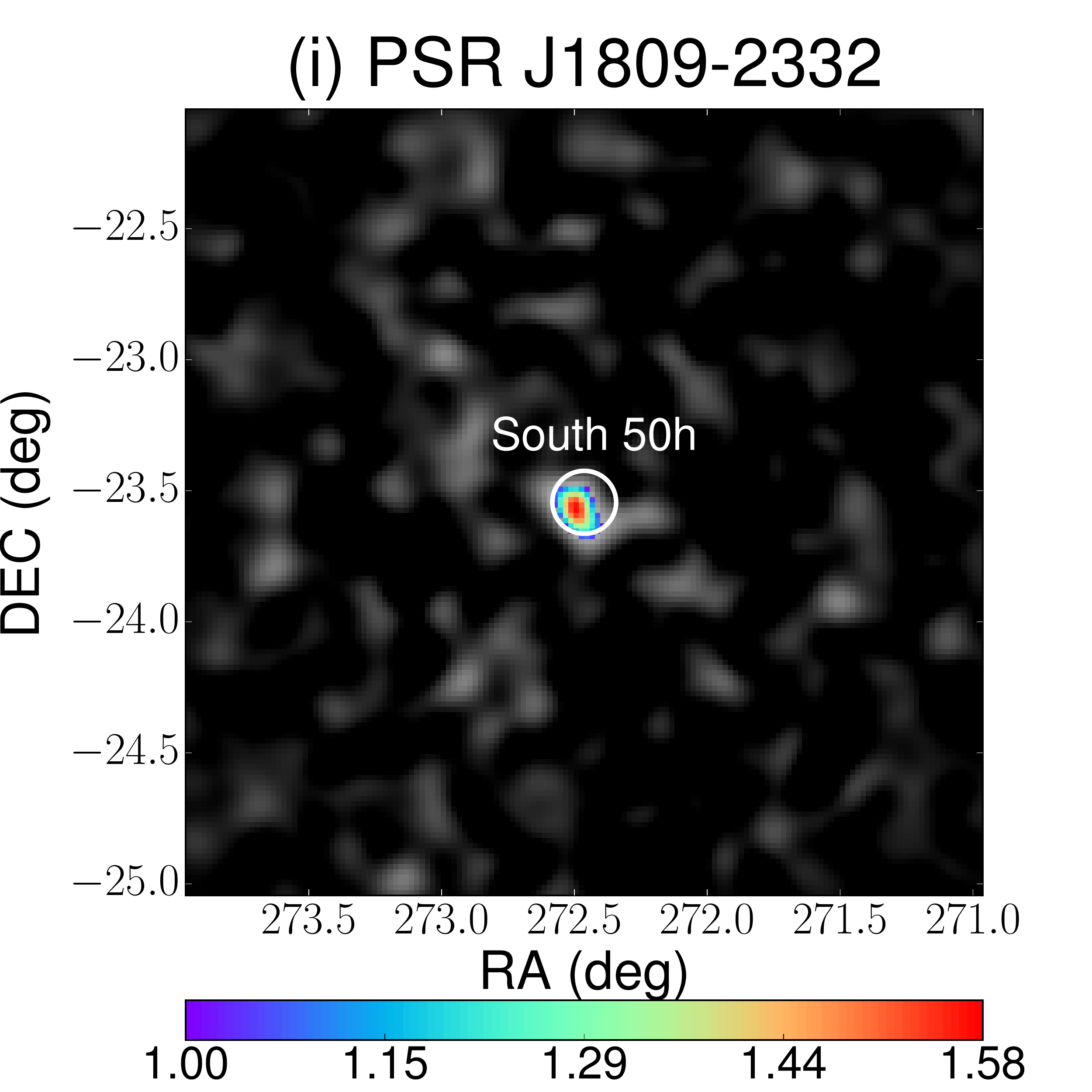}
	\hfill
	\includegraphics[width=\cmw\textwidth]{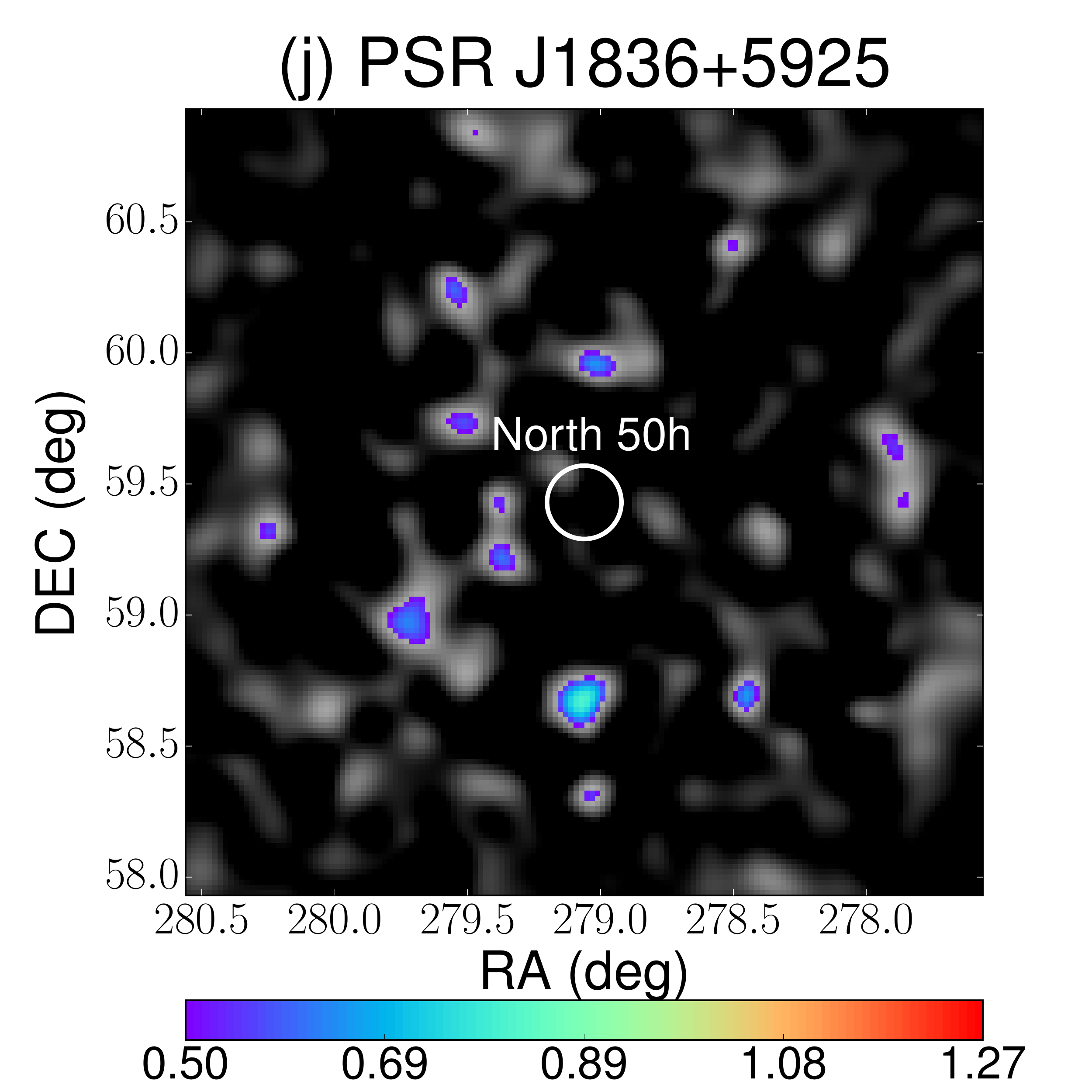}
	\hfill
	\includegraphics[width=\cmw\textwidth]{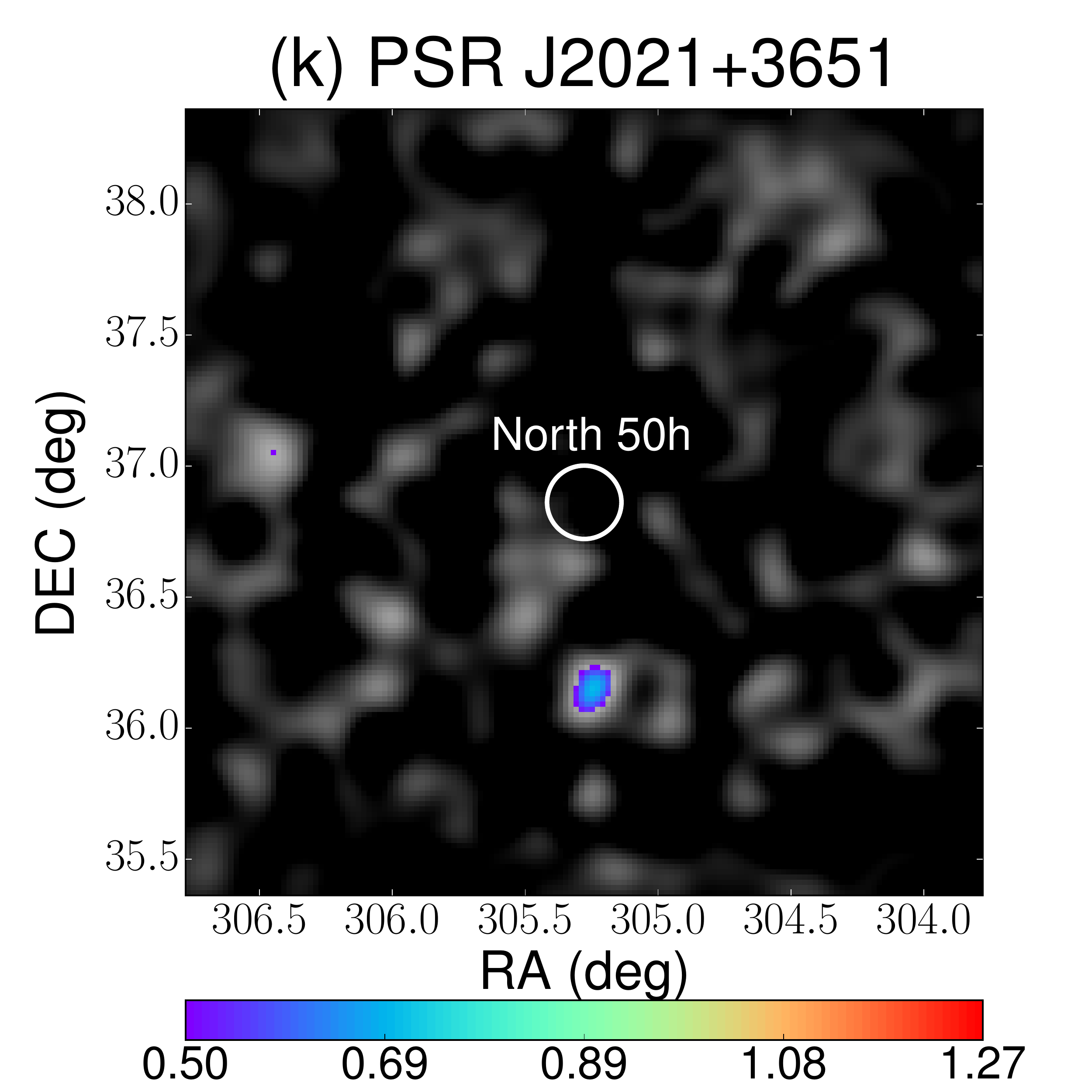}
	\hfill
	\includegraphics[width=\cmw\textwidth]{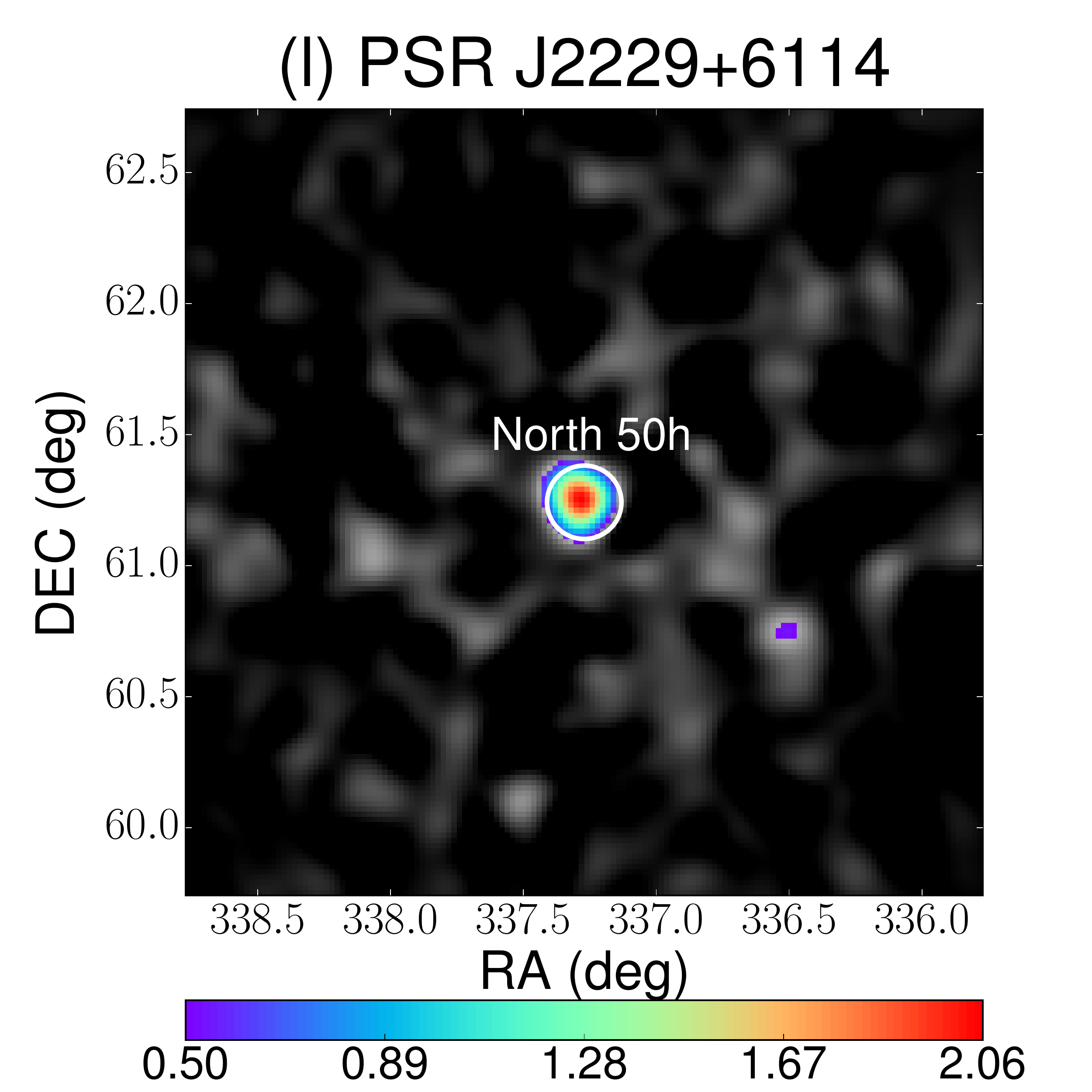} 

	\caption{VHE residual maps of HE \fermi~pulsars, simulated with \ctools~at energies $E$$>$0.1 TeV and considering only events from the on-peak phase intervals (reduced-background analysis, performed assuming the optimistic approximation for the spectrum of each pulsar). Each map covers a $3^{\circ}\times3^{\circ}$ area on the sky centered on the pulsars position. $x$-, $y$- axes are right ascension (RA) and declination (DEC) in degrees, respectively. Simulations are computed assuming a 50 h exposure of the central point source. All maps are smoothed with the 68\% of the PSF at 0.1 TeV, equal to 0.14$^\circ$ and 0.12$^\circ$ for the CTA-North and the CTA-South arrays, respectively. The size of these regions are shown as white circles. The color bars represent the brightness of each pixel (arbitrary units). The pixels with brightness below the background threshold are represented in gray, whereas the signal pixels (above the background threshold) are colored. PSR J0835$-$4510 in Fig. \ref{fig:VHE_resmaps}e is shown with its PWN.}
	\label{fig:VHE_resmaps}
\end{figure*}

\begin{figure*}
	\includegraphics[width=\cmw\textwidth]{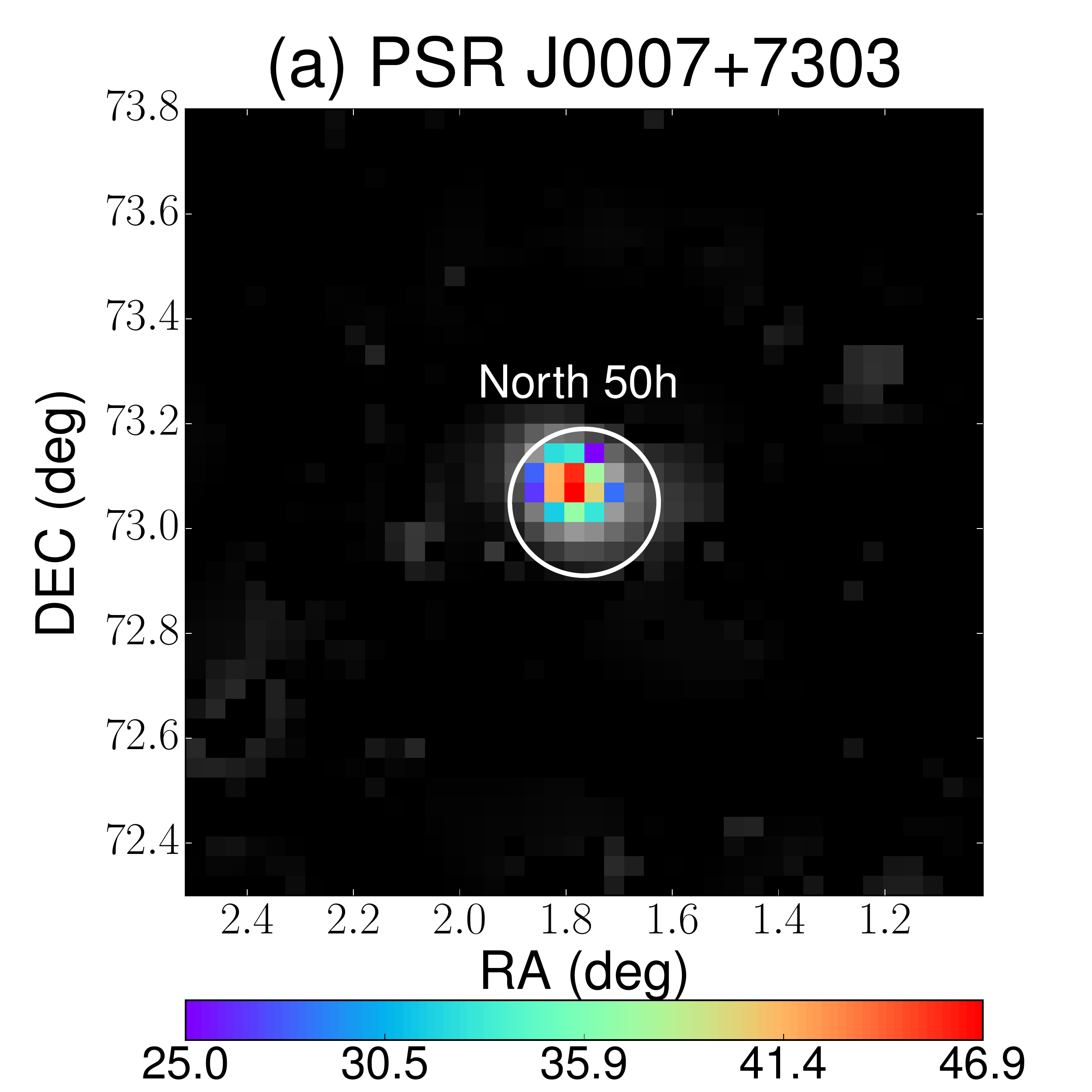}
	\hfill
	\includegraphics[width=\cmw\textwidth]{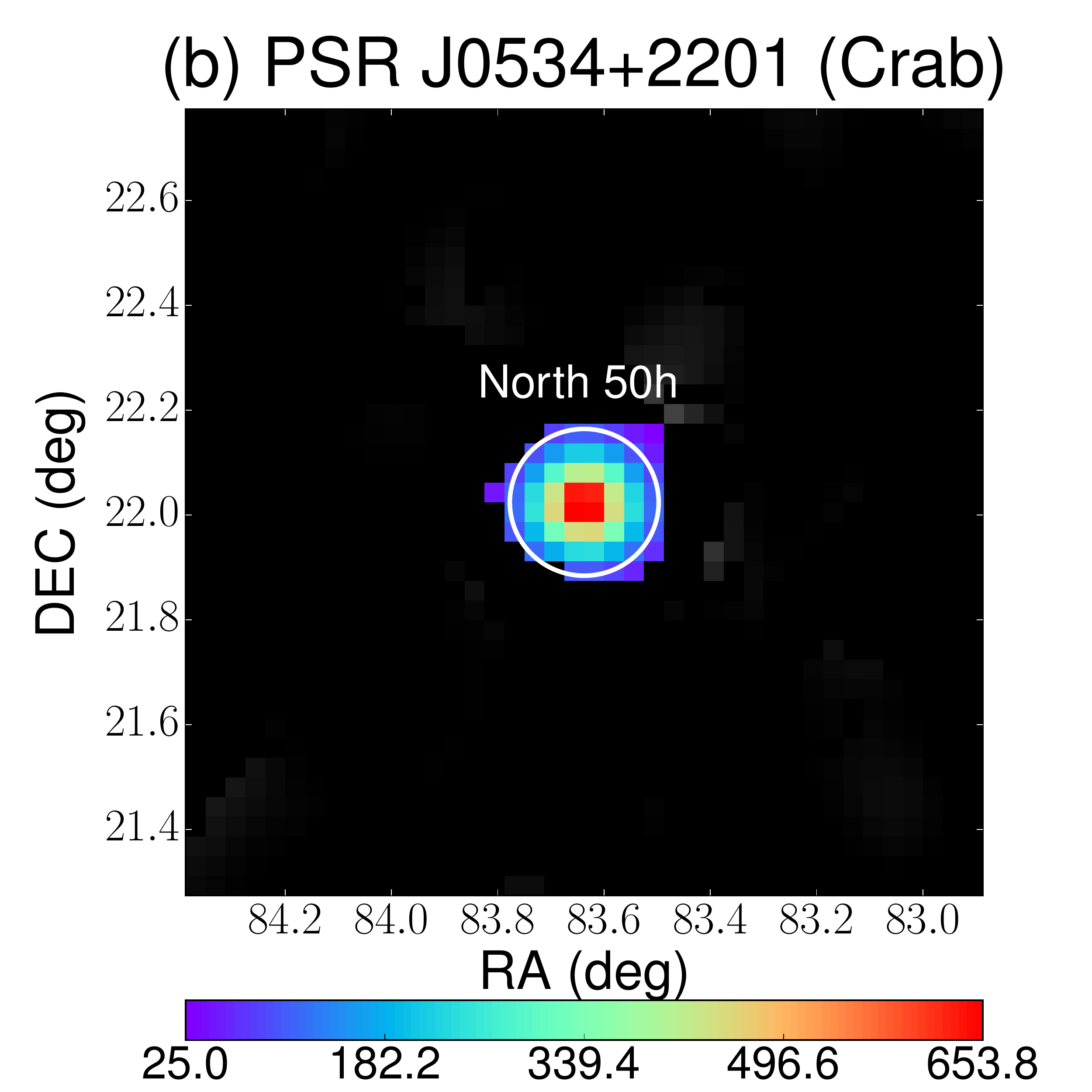} 
	\hfill
	\includegraphics[width=\cmw\textwidth]{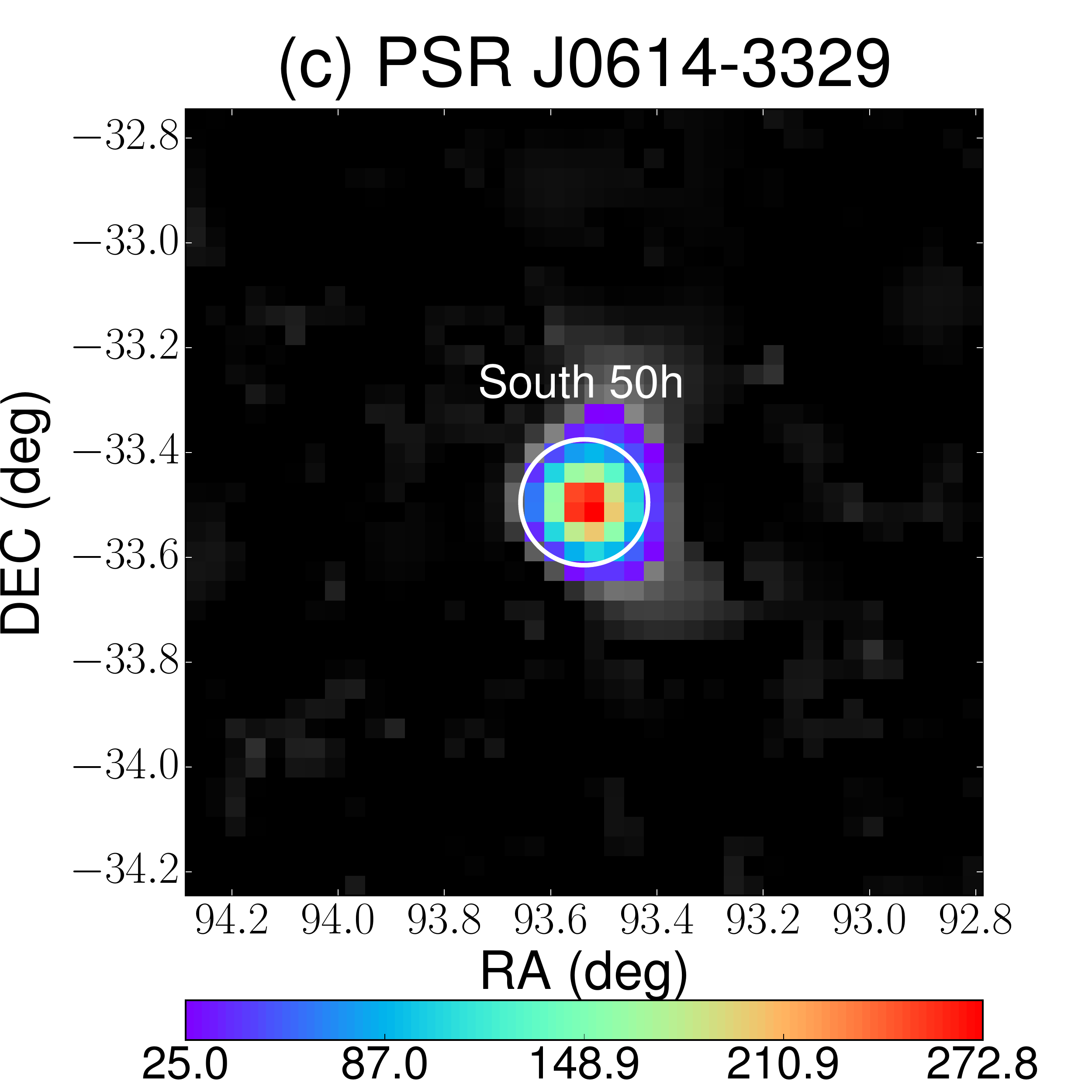}
	\hfill
	\includegraphics[width=\cmw\textwidth]{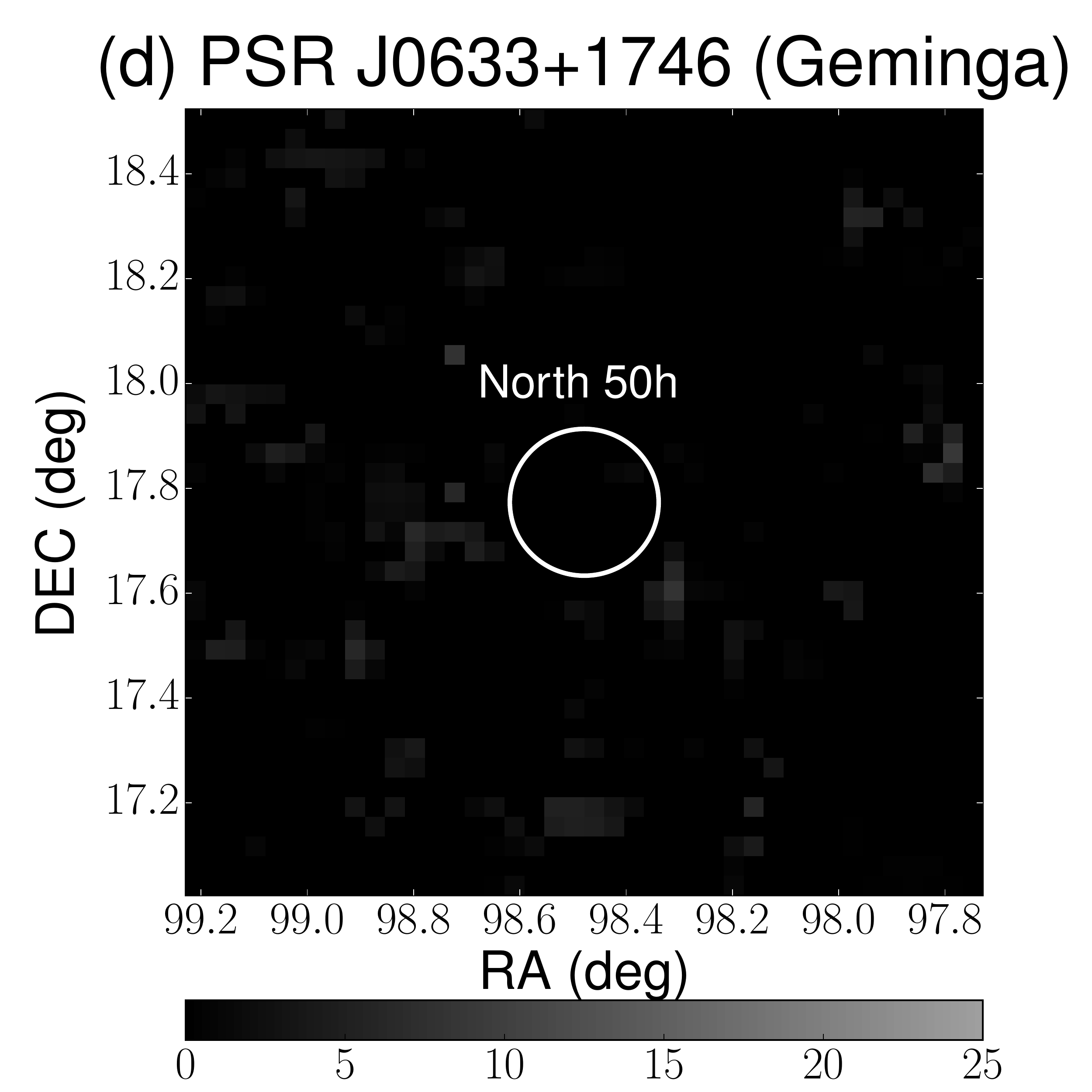}
\\
	\includegraphics[width=\cmw\textwidth]{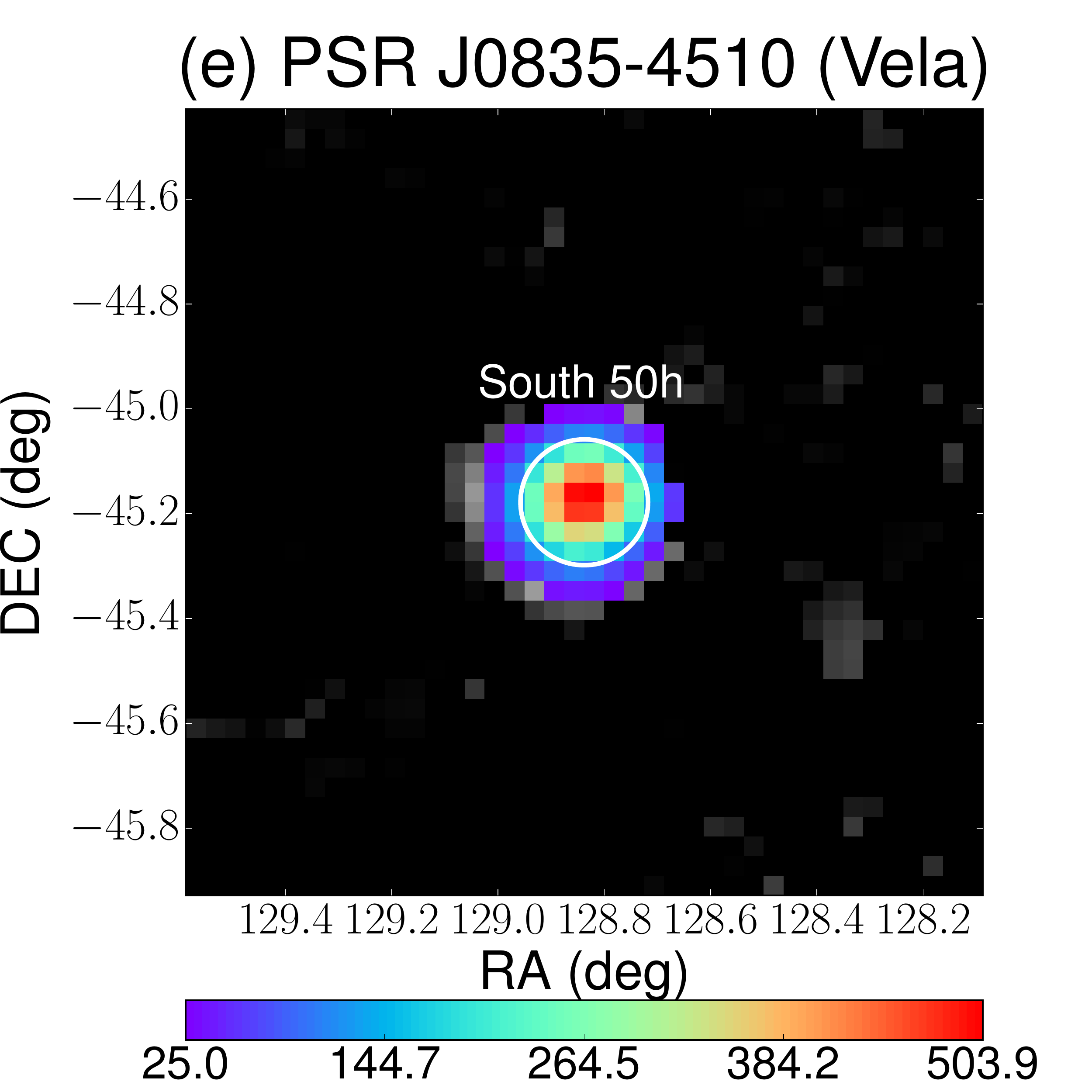}
	\hfill
	\includegraphics[width=\cmw\textwidth]{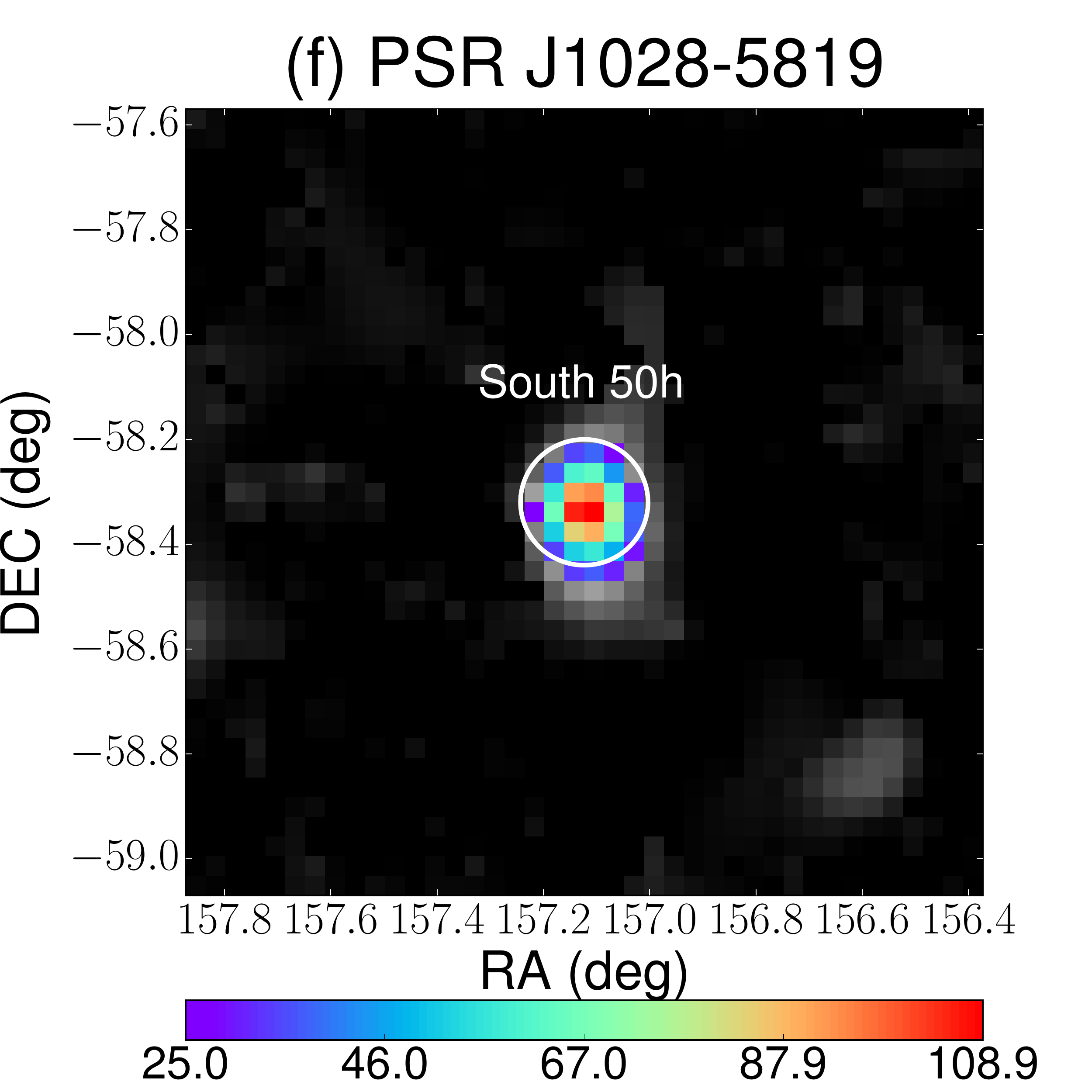}
	\hfill
	\includegraphics[width=\cmw\textwidth]{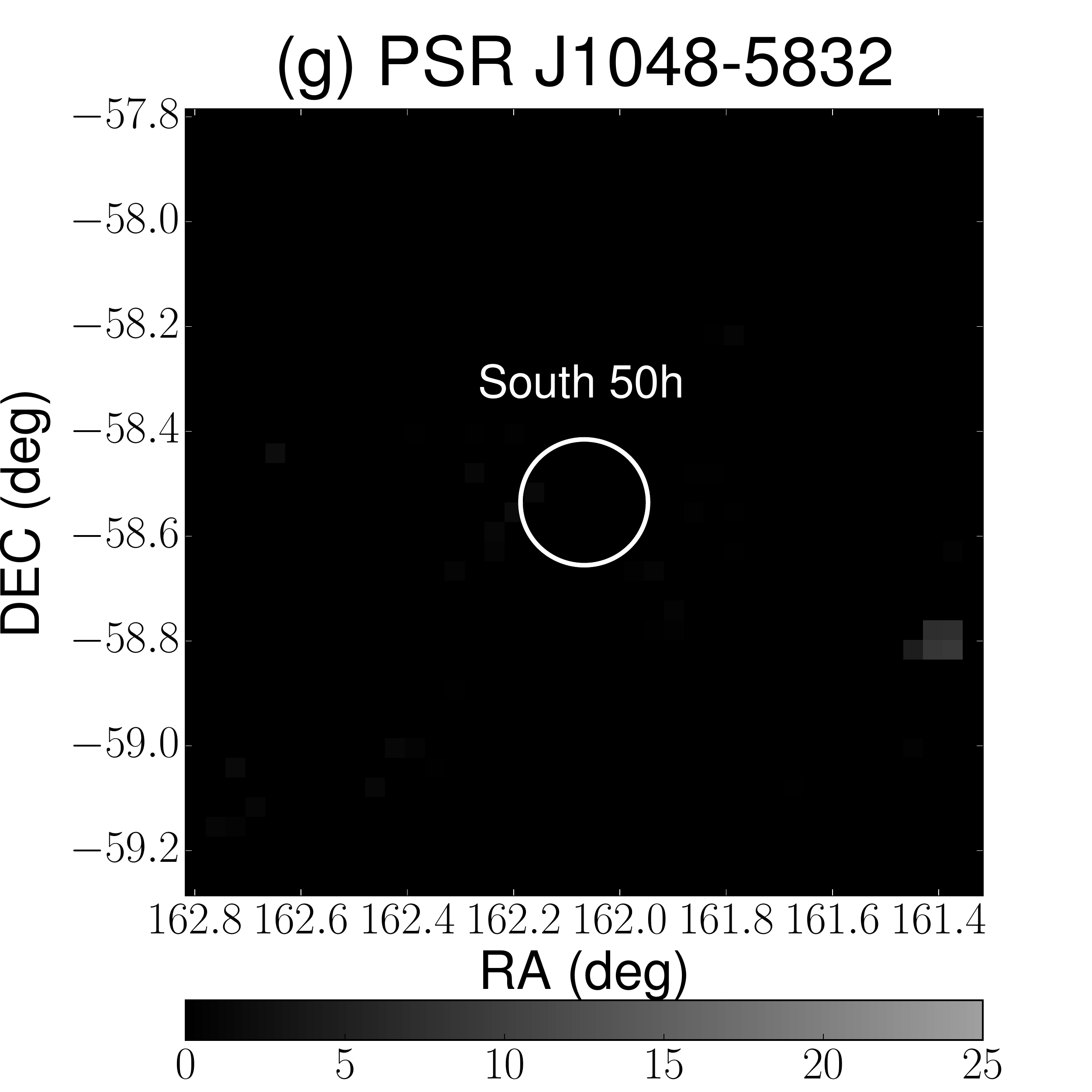} 
	\hfill
	\includegraphics[width=\cmw\textwidth]{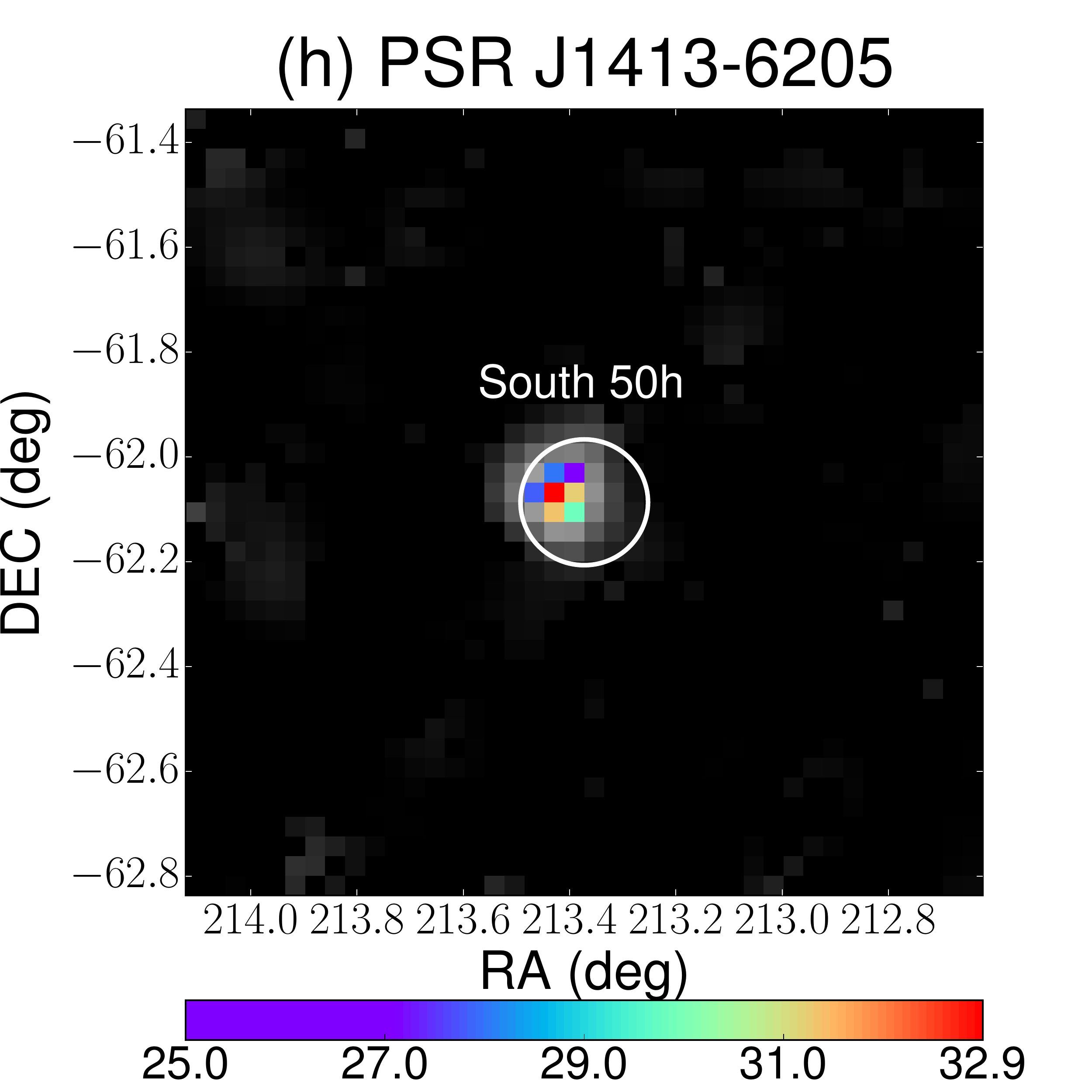}
\\
	\includegraphics[width=\cmw\textwidth]{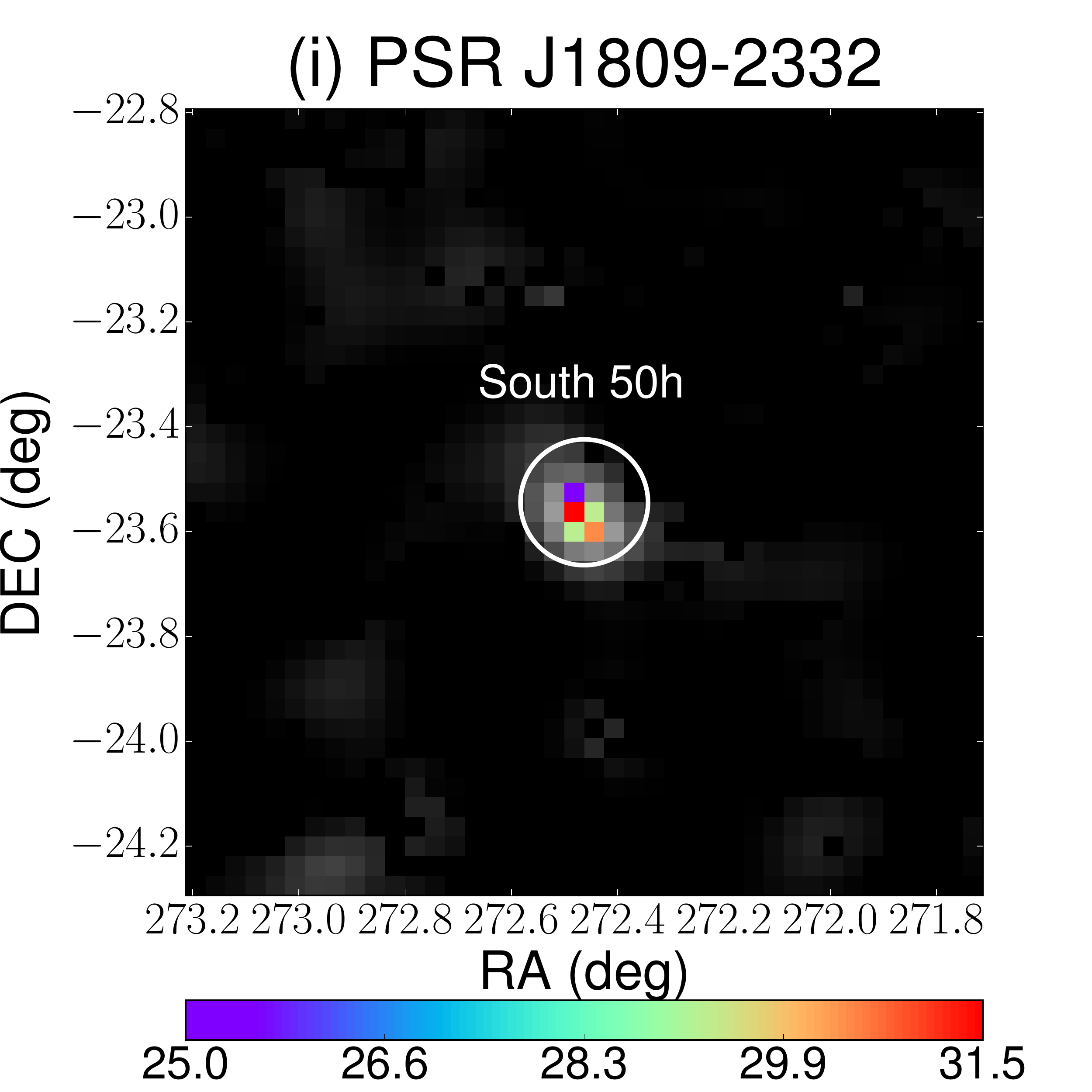}
	\hfill
	\includegraphics[width=\cmw\textwidth]{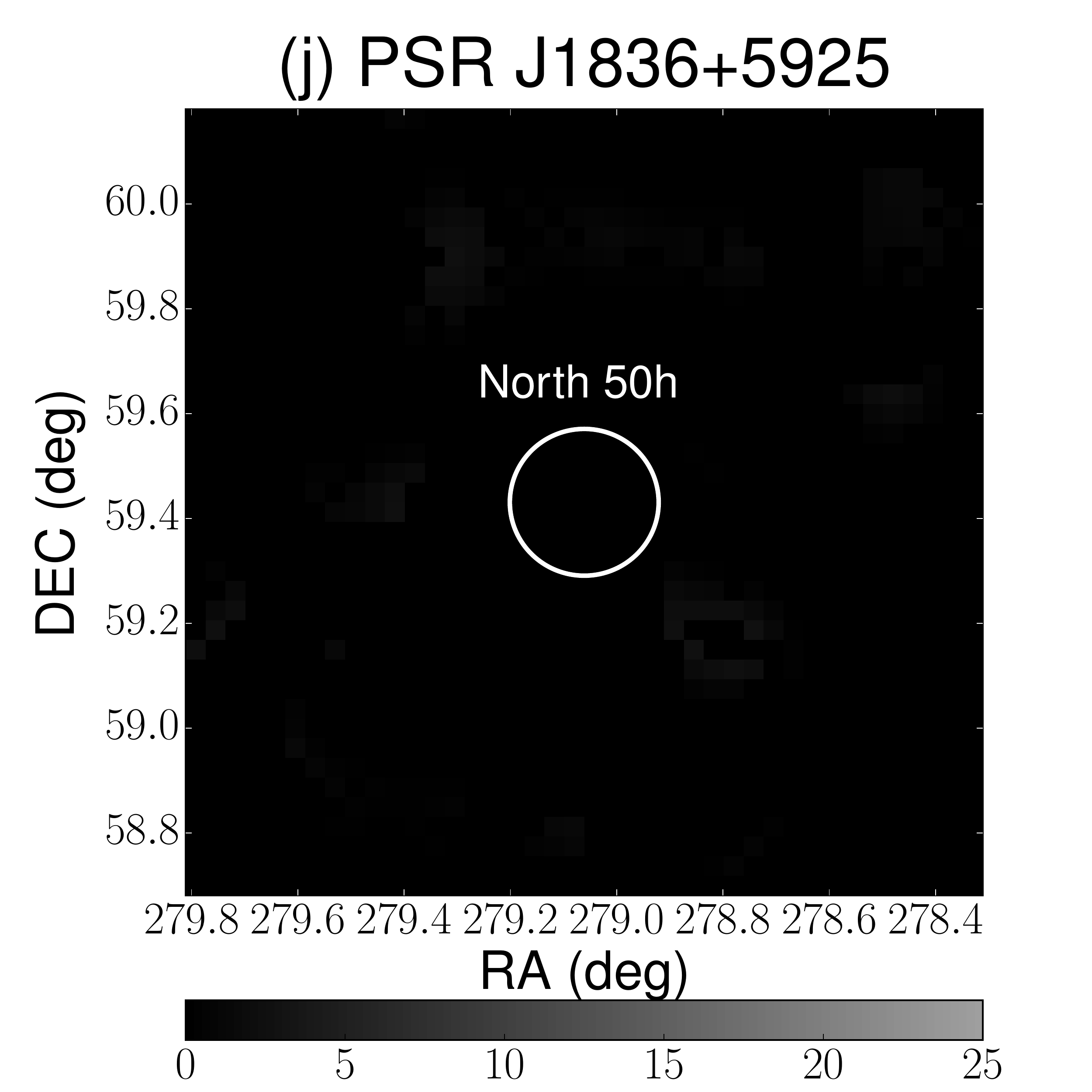}
	\hfill
	\includegraphics[width=\cmw\textwidth]{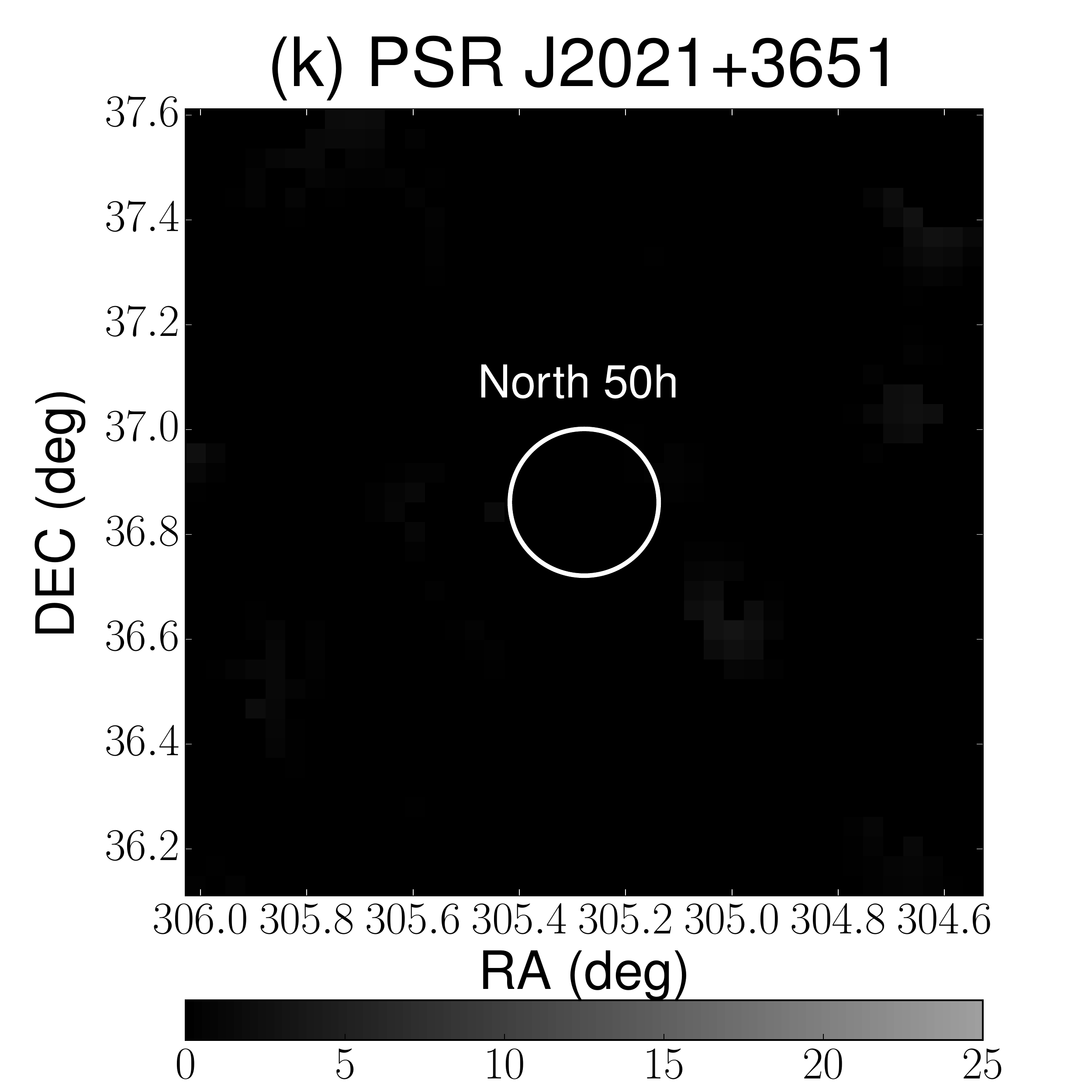}
	\hfill
	\includegraphics[width=\cmw\textwidth]{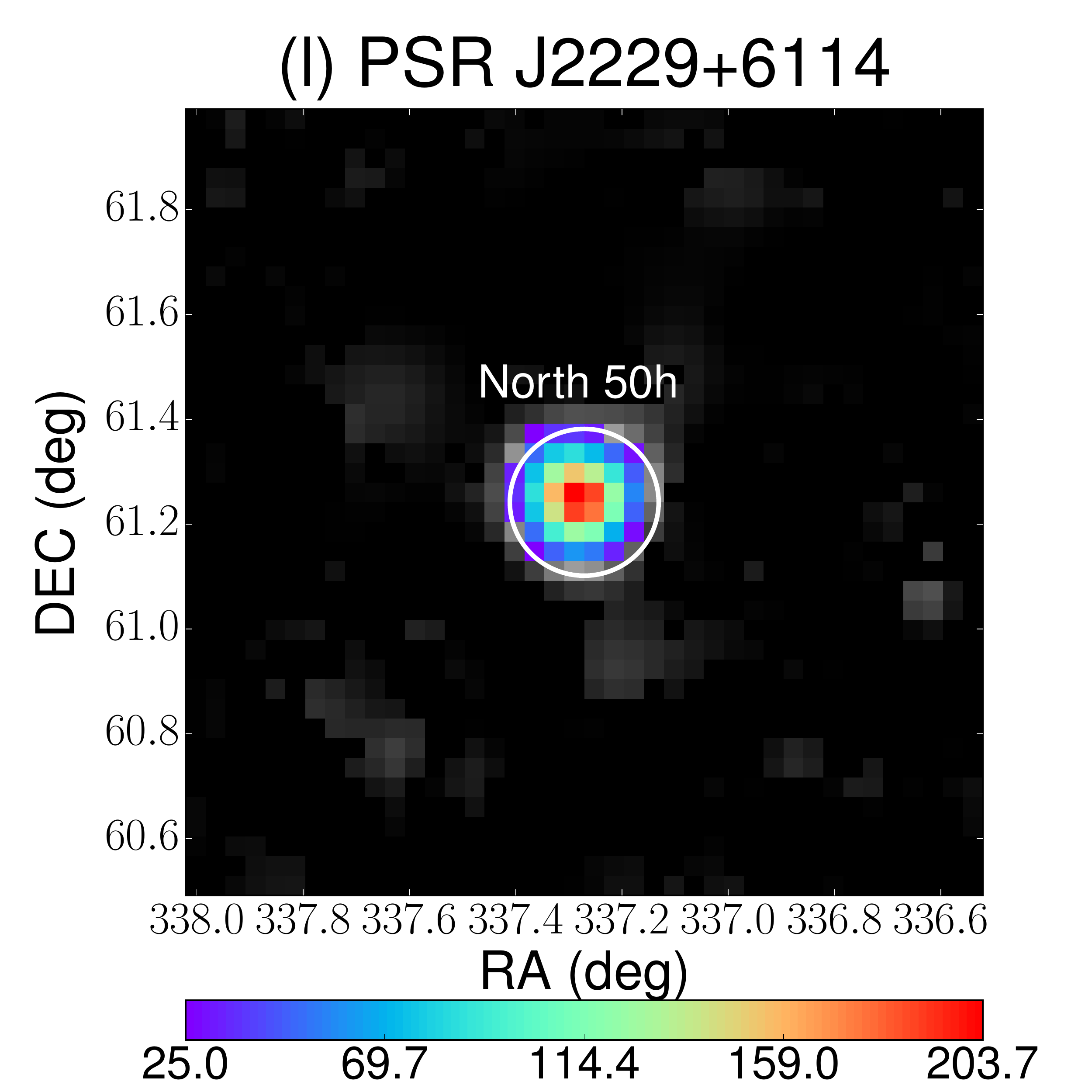}

	\caption{VHE TS maps of HE \fermi~pulsars, simulated with \ctools~at energies $E$$>$0.1 TeV and considering only events from the on-peak phase intervals (reduced-background analysis, performed assuming the optimistic approximation for the spectrum of each pulsar). Each map covers a $1.5^{\circ}\times1.5^{\circ}$ area on the sky centered on the pulsars position. $x$-, $y$- axes are right ascension (RA) and declination (DEC) in degrees, respectively. Simulations are computed assuming a 50 h exposure of the central point source. The size of CTA PSFs is shown as white circles. TS ($\equiv S^2$) below 25 is marked in gray and above 25 in color (scale shown at the bottom of each panel).}
	\label{fig:VHE_tsmaps}
\end{figure*}


\section{Discussion}\label{sec:5}

We presented an analysis of 12 HE \fermi~pulsars at high energies ($E$$>$10 GeV). We performed the spectral fitting of both the on- and off- peak emission and compared the results of our \fermi-analysis with those from the literature (Table \ref{tab:res}) and from the Fermi catalogs (3FGL, 2PC, 1FHL, 2FHL and 3FHL). While the pulsars spectra from the \fermi~catalogs refer to both the peak and bridge emission, our analysis focuses only on the peak emission (without the bridge). Since the bridge emission is rather faint at energies $E$$>$10 GeV, in all cases the spectra obtained here are consistent with those from the catalogs. We compared the Crab pulsar spectrum from our analysis with that from \citet{Ansoldi2016}, who report a less energetic spectrum of the on-peak emission (both peaks) in the energy range from 10 GeV to 1.5 TeV (cmp. red and hatched butterfly plots in Fig. \ref{fig:Fermi_Specs}b). This is due to the broader on-peak phase interval of the Crab pulsar used here ($\Delta\phi_{\rm on}=0.126$) as compared to that adopted by \citet{Ansoldi2016}: $\Delta\phi_{\rm M}=0.088$.

We then simulated CTA observations of our sample of 12 HE \fermi~pulsars. We calculated the significances of pulsar detection, assuming that the VHE pulsar spectrum is an extrapolation of the high-energy power-law \fermi~spectral fit above 10 GeV. We note that our estimates of the CTA detectability can be considered as optimistic, since a cut-off may be present in the pulsars VHE emission. So far, there are only two pulsars, for which significant pulsations have been detected with ground-based Cherenkov telescopes -- the Crab pulsar (MAGIC,  \citealt{Aleksic2012,Aleksic2014_2,Ansoldi2016}; VERITAS, \citealt{Aliu2011}) and the Vela pulsar (\hess\footnote{\url{http://www.mpg.de/8287998/velar-pulsar}}). According to our calculations CTA will probably detect $\sim$5-8 HE \fermi~pulsars in 50 h in the VHE range ($E$$>$0.1 TeV). Although the angular resolution of CTA improves with energy\footnote{For CTA-South the 68\% of the PSF radius at 0.04, 0.1, 0.25 and 1 TeV is 0.19$^\circ$, 0.12$^\circ$, 0.09$^\circ$ and 0.05$^\circ$, respectively.} \citep[see e.g.][]{Bernlohr2013}, which tends to lead to an improvement in the detection significance because of the decrease in the isotropic background contamination, only 3 pulsars (J0534$+$2201, J0614$-$3329 and J0835$-$4510, with rather flat spectra and high fluxes; see Table \ref{tab:res}) are bright enough to be detectable with CTA above 0.25 TeV. PSR J2229$+$6114 has a spectral index smaller than that of PSR J0614$-$3329, but it is not significantly detected because of the lower flux. 

In addition to the phase-averaged point source detection, we calculated also the $S$-values for the reduced-background simulations (see columns 6--8 in Table \ref{tab:Signif}). This choice decreases the background contamination from the off-peak phase intervals and increases the signal-to-noise ratio, leading to higher detection significances as compared to the phase-averaged ones. Different gamma-ray observations \citep{Abdo2010,Aliu2011,Aleksic2012,Ackermann2013,Aleksic2014_2,Ansoldi2016} showed that there is a trend of narrowing of the pulsars peaks with energy. Therefore, in this respect our results can be considered as rather conservative. Clearly, in order to perform this type of measurements a knowledge of the accurate pulsars ephemerides is required, which can be obtained from simultaneous low-energy  observations (e.g. radio, optical, X-rays or gamma-rays). Considering the events from the reduced-background analysis, we obtain significant detections at energies $E$$>$0.25 TeV also for pulsars J1028$-$5819 and J2229$+$6114. Above 1 TeV, PSRs J0614$-$3329 and J2229$+$6114 are sufficiently bright to be marginally detectable. Despite the harder high-energy spectrum ($\gamma=3.0$), a simulation of a 50 h observation of the Crab pulsar above 1 TeV also yields a marginal detection ($S$$=$4) because of the contamination from the bright point-like (in gamma-rays) Crab nebula. This result is consistent with the simulations presented by \citet{Ona2013}. It also demonstrates an improvement of the CTA performance compared with the currently operating IACTs. For example, the MAGIC telescope detected the interpulse (P2) of the Crab pulsar at a significance ``3.5-sigma'' above 0.95 TeV with a much larger observing time of $\sim$320 h \citep{Ansoldi2016}. 

The two independent analyses presented here (Sects. \ref{subs:3.1} and \ref{subs:3.2}) show consistent results (see Fig. \ref{fig:Fermi_Specs} and Table \ref{tab:Signif}). Only the extrapolated spectrum of PSR J0614$-$3329 at energies $E$$>$1 TeV turns out to be below the CTA sensitivity curve (Fig. \ref{fig:Fermi_Specs}c). However, as described in  Sect. \ref{subs:3.2}, we require 10 photons for a significant detection of the source. Such a condition is not satisfied in the \ctools~sensitivity calculation. Taking all this into account, the probability of detecting this pulsar at energies $E$$>$1 TeV is marginal. 

In Figs. \ref{fig:CTA-N} and \ref{fig:CTA-S} we compare the CTA sensitivity curves calculated for a 50 h observation and for different duty cycles of the HE \fermi~pulsars (from $\sim$0.1 up to $\sim$0.5). The improvement of the CTA  sensitivities obtained for the pulsars as compared to that with duty cycle equal to 1 can be as high as a factor of $\sim$2 for both CTA-North and CTA-South at energies $E$$\lesssim$1 TeV. Above $\sim$10 TeV, where the sensitivity of CTA is signal limited, the background suppression due to different duty cycles does not play an important role and the difference between sensitivity curves becomes negligible. Moreover, significant variation from the public sensitivity curve can occur if other sources -- e.g. surrounding PWNe -- provide an additional contribution to the background. As shown in Figs. \ref{fig:CTA-N} and \ref{fig:CTA-S} (dashed lines), for the Crab pulsar the PWN contribution becomes significant at intermediate energies (0.2-30 TeV) while, for the Vela pulsar, at higher energies (1-30 TeV). At lowest energies ($E$$\lesssim$0.1 TeV) the PWN contribution is less crucial since the CTA sensitivity is limited manly by the background systematics (for details see e.g. \citealt{Bernlohr2013}).

As mentioned in Sect. \ref{subs:3.2}, we used the following criteria to simulate the CTA sensitivity curves for each pulsar from our sample: 
\begin{enumerate}
	\item significance of 5 sigma in each energy bin, 
	\item number of excess events ($N_{\rm ex}$) larger than 10 in each energy bin, 
	\item number of excess events larger than 5\% of the background rate ($N_{\rm bg}$) in each energy bin.
\end{enumerate}
The ratio $\xi=N_{\rm ex}/(5\%\,N_{\rm bg})$ mentioned in criterion (iii) scales linearly with the duty cycle. At low energies ($E$$\lesssim$0.1 TeV), where the background systematics are significant, the denominator is large and the criterion $\xi>1$ becomes more stringent than the other two. On the other hand, at energies $E$$\gtrsim$10 TeV the background is not larger than a few counts per energy bin and, therefore, the more stringent criterion is (ii). In this case, as the number of excess events $N_{\rm ex}$ scales linearly with observing time, this requirement does not depend on the duty cycle. At intermediate energies, the CTA sensitivity is determined mainly by criterion (i). The significance then scales with the square root of the observing time and with the inverse of the square root of the duty cycle.

Analyzing the \fermi-LAT data at energies above 10 GeV, we found significant OP components in PSRs J0633$+$1746, J1809$-$2332 and J1836$+$5925, which most probably have magnetospheric origin (see e.g. \citealt{Abdo2010_5,Abdo2010_4,Abdo2013}). However, recently TeV emission from the extended PWN has been possibly detected near PSR J0633$+$1746 with the HAWC instrument \citep{Baughman2015arx}. Further investigations of these sources will be crucial to clarify the mechanism of the OP emission in these pulsars. 

We also obtained significant OP components in the Crab and Vela pulsars, which correspond to the surrounding nebulae. The spectrum of the Crab nebula obtained here from the \fermi-LAT analysis is consistent with that from 2FHL. However, as shown in Fig. \ref{fig:PWN_Specs}a, above 0.3 TeV our power-law spectral fit it harder than the log-parabola model obtained fitting the VHE MAGIC data \citep{Aleksic2015}. The spectrum of the Vela PWN obtained here shows a different spectral slope at energies $E$$>$50 GeV as compared to that from 2FHL, because in the latter, the larger dataset of was considered ($\sim$6.67 yr of \fermi-LAT data; see Fig. \ref{fig:PWN_Specs}c). Our extrapolated power-law spectrum is also not consistent with the exponential cut-off measured with \hess~\citep{Abramowski2012}.

For the Crab and Vela OP components we adopted VHE spectra that are different from those obtained in our \fermi-LAT off-peak analysis. For the Crab nebula, extrapolating our power-law spectral fit to energies $E$$>$0.3 TeV would lead to overestimate the actual VHE flux of the Crab nebula, measured with MAGIC \citep{Aleksic2015}. In our VHE simulations we, therefore, adopted the log-parabola model of \citet{Aleksic2015}. For the Vela PWN, in the simulations we adopted the \hess~spectrum extrapolated down to 0.04 TeV. To estimate the systematics at low energies (0.04$-$0.25 TeV) modeled by the \hess~spectrum, we performed other simulations of the Vela X region in the energy range 0.04-0.25 TeV using two different spectral models: (i) one with the spectrum of Vela PWN resulting from our analysis and (ii) another one with the \hess~spectrum. Using the \hess~spectrum of Vela PWN, we infer systematic uncertainties of the Vela pulsar spectral parameters of the order of $\sim$10\% for the normalization factor and $\sim$2\% for the spectral index. This is comparable with the corresponding statistical errors. The relative difference in the TS values of the Vela pulsar calculated in these two different cases is $\Delta {\rm TS}/{\rm TS}\sim5\%$. Similar estimates were performed in the range 0.04-0.75 TeV using the spectrum of the Vela PWN from 2FHL (blue line in Fig. \ref{fig:PWN_Specs}c). Performing simulations with the \hess~spectrum instead of that from 2FHL leads to a (systematic) fractional difference in the Vela pulsar detection significance $\Delta {\rm TS}/{\rm TS}\sim17\%$.

Our results are strongly dependent on the assumed spectral behavior of the pulsars above few hundreds GeV. We compare the extrapolated VHE spectral slope of the Crab pulsar ($\gamma=3.0\pm0.2$) with the most recent MAGIC results. \citet{Ansoldi2016} performed a joint spectral analysis of the \fermi-LAT and MAGIC data, fitting separately the emission of the main pulse P1 and the interpulse P2 with a power-law model. We estimated the spectral index $\gamma_{\rm F+M}$ of the power law that fits the total emission of both peaks (P1$+$P2)\footnote{The calculations were performed using the \texttt{JAGS} program from \url{http://mcmc-jags.sourceforge.net/}.}, obtaining $\gamma_{\rm F+M}=3.2$ with a relative error of $\sim3\%$. We conclude that the VHE spectrum of the Crab pulsar assumed in our simulations is consistent with that detected with MAGIC in the energy range 10 GeV -- 1.5 TeV \citep{Ansoldi2016}.

Within the framework of 1FHL, \citet{Ackermann2013} proposed a criterion for the selection of good candidates for future VHE investigations. The criterion is as follows: (i) the source is significantly detected ($N_{\rm sigma}\gtrsim3$) at energies $E$$>$30 GeV within 1FHL, (ii) the spectral index above 30 GeV is smaller than 3 and (iii) the flux above 50 GeV is $E$$>$10$^{-11}$ photons cm$^{-2}$ s$^{-1}$. None of the HE \fermi~pulsars, except those which have been already detected at VHE, passed these criteria. \citet{Ona2013} estimated the number of gamma-ray pulsars detectable with CTA, assuming a power-law tail with spectral index  $\gamma=3.52$ above the cut-off energy for the 46 gamma-ray pulsars from the First \fermi-LAT Catalog of Gamma-ray Pulsars \citep{Abdo2010_1PC}. They found that about 20 pulsars might be detected in less than 50 h of dedicated observations with the baseline CTA configuration under such assumptions. We found a lower number of detectable pulsars, because our sample is smaller (the 12 most energetic \fermi~pulsars) and the high-energy spectra of most of our pulsars, obtained from the \fermi-LAT analysis, are steeper ($\gamma>3.52$).

We estimated also the exposure times required for a significant detection of the point-like emission from pulsars of our sample above 0.04 and 0.1 TeV in the phase-averaged analysis. The OP emission was considered as background. Results for J0007$+$7303, J0534$+$2201, J0614$-$3329, J0835$-$4510, J1028$-$5819 and J2229$+$6114 are shown in Table \ref{tab:Tsig}. All other pulsars will require unrealistically long observing times for a significant ($S=5$) detection with CTA. The Crab pulsar will require a longer exposure to be detected than the Vela pulsar, since the contamination from the Vela PWN is lower than that from the Crab nebula below a few TeV and the Vela pulsar is brighter below 70 GeV.

\begin{table}
\caption{Exposure time $t_{\rm obs}$ (in hours) required for a significant ($S=5$) detection of the point-like emission from some HE \fermi~pulsars above 0.04 and 0.1 TeV with CTA (phase-averaged analysis, OP emission is considered as background).}
\label{tab:Tsig}
\centering
\begin{tabular}{l c c}
\hline\hline
Pulsar~\hspace{1cm} &\hspace{0.2cm}$E$$>$0.04 TeV\hspace{0.3cm} &\hspace{0.3cm}$E$$>$0.1 TeV\hspace{0.3cm} \\
\hline
\textbf{J0007$+$7303}	& (75 h)	& 300 h	\\
J0534$+$2201$^a$	& 6 h 	& 30 h	\\
J0614$-$3329			& 25 h	& 50 h	\\
J0835$-$4510$^b$	& 1 h 	& 6 h 	\\
\textbf{J1028$-$5819}	& (50 h)	& 70 h	\\
\textbf{J2229$+$6114}	& (250 h)	& 210 h	\\
\hline
\multicolumn{3}{p{0.45\textwidth}}{\textbf{Notes.} Pulsars which will be observed at high zenith angles (above 30$^\circ$, i.e with high energy threshold $E$$>$0.1 TeV), are in \textbf{bold face}. Exposure time required for a significant detection of these pulsars above 0.04 TeV (in brackets) are shown for comparison.
$^a$Simulations of the Crab pulsar observations are performed using the log-parabola spectrum of the Crab nebula from \citet{Aleksic2015}.
$^b$Simulations of the Vela pulsar observations are performed using the power-law model with an exponential cut-off of the Vela PWN from \citet{Abramowski2012}. The spatial model for the extended Vela X emission is a uniform disk with radius of 1.2$^\circ$.}
\end{tabular}
\end{table}

\section{Conclusions}\label{sec:6}
We presented the prospects of pulsars detection at VHE with the next-generation IACT instrument -- CTA. We simulated 50 h observations of each pulsar, considering such relatively short exposure times reasonable for a significant detection of the most energetic \fermi~pulsars. In our most optimistic scenario CTA will detect 5-8 gamma-ray pulsars from our sample at very high energies ($E$$>$0.1 TeV). We also found that up to 5 pulsars are sufficiently bright for a significant VHE detection above 0.25 TeV with CTA. PSRs J0534$+$2201, J0614$-$3329 and J2229$+$6114 may possibly be detectable also above 1 TeV in 50 h. These conclusions are based on the assumption that there is no high-energy cut-off in the spectra of HE \fermi~pulsars. Observations of these pulsars in the energy range from 0.1 up to several TeV will help us to clarify whether the VHE component discovered in the Crab pulsar is intrinsic also to other gamma-ray pulsars.

\section*{Acknowledgements} 
This research made use of \ctools, a community-developed analysis package for IACT data. \ctools~is based on GammaLib, a community-developed toolbox for the high-level analysis of astronomical gamma-ray data. This research made use also of Matplotlib \citep{Matplotlib2007} and of Astropy, a community-developed core Python package for Astronomy \citep{Astropy2013}. The work is supported by the Italian Ministry of Education, University, and Research (MIUR) with funds specifically assigned to the Italian National Institute of Astrophysics (INAF) for the Cherenkov Telescope Array (CTA), and by the Italian Ministry of Economic Development (MISE) within the ``Astronomia Industriale'' program. We acknowledge support from the Brazilian Funding Agency FAPESP (Grant 2013/10559-5) and from the South African Department of Science and Technology through Funding Agreement 0227/2014 for the South African Gamma-Ray Astronomy Programme. We acknowledge support from the agencies and organizations listed under Funding Agencies at \url{http://www.cta-observatory.org/}, from the University of Padova and from the Hakubi project of the Kyoto University. This work is partially supported by CISAS-University of Padova under the project ``Technological development and scientific exploitation of Aqueye+ and Iqueye for High Time Resolution Astronomy''. 

This paper has gone through internal review by the CTA Consortium.

\bibliographystyle{mnras}

\bsp	
\label{lastpage}
\end{document}